\pdfoutput=1
\documentclass[showpacs,aps,pre,reprint,floatfix]{revtex4-1}
\usepackage{graphicx}
\usepackage{dcolumn}
\usepackage{bm}
\usepackage{hyperref}
\usepackage{color}

\begin{document}
\title{Interface Dynamics of Immiscible Two-Phase Lattice-Gas Cellular Automata: A Model with Random Dynamic Scatterers and Quenched Disorder 
in Two Dimensions}
\author{R. M. Azevedo}
\author{R. R. Montenegro-Filho}
\author{M. D. Coutinho-Filho}
\affiliation{Laborat\'{o}rio de F\'{i}sica Te\'{o}rica e Computacional, Departamento de F\'{i}sica, Universidade Federal de Pernambuco, 50760-901 Recife-PE, Brasil}
\date{\today}

\begin{abstract}
We use a lattice gas cellular automata model in the presence of random \textit{dynamic scattering sites} and quenched disorder in the two-phase immiscible model
with the aim of producing an interface dynamics similar to that observed in Hele-Shaw cells. The dynamics of the interface is studied as one fluid displaces the
other in a clean lattice and in a lattice with quenched disorder. For the clean system, if the fluid with a lower viscosity displaces
the other, we show that the model exhibits the Saffman-Taylor instability phenomenon, whose features are in very good agreement with those observed in real (viscous) fluids. 
In the system with quenched disorder,
we obtain estimates for the growth and roughening exponents of the interface width in two cases: viscosity-matched fluids and
the case of unstable interface. The first case is shown to be in the same universality
class of the random deposition model with surface relaxation. 
Moreover, while the early-time dynamics of the interface behaves similarly, viscous fingers develop in the second case
with the subsequent production of bubbles in the context of a complex dynamics. 
We also identify the Hurst exponent of the subdiffusive fractional Brownian motion associated with the interface,
from which we derive its fractal dimension and the universality classes related to a percolation process. 

\end{abstract}

\maketitle

\section{Introduction}

The dynamics of Newtonian fluids is governed by the Navier-Stokes equations \cite{landau,*spurk}. In fact,
a number of distinct flow behavior has been identified, such as laminar or turbulent, depending on boundary conditions and the Reynolds number, $\text{Re}=VL/\nu$, where $V$ ($L$) is some characteristic velocity (length) and $\nu$ is the kinematic viscosity of the fluid.
A special case of wide interest is that of fluid dynamics in porous media \cite{Sahimi_Review,*sahimi2012flow,bear1972}.

In a microscopically disordered media the flow is very complex and its macroscopic behavior can be homogeneous or
heterogeneous \cite{Sahimi_Review,*sahimi2012flow}: in the former case, the system displays size-independent transport properties,
while in the latter the system is better described by a position-dependent permeability \cite{PhysRevE.66.056307,tarta}.
Further, at very low Reynolds number with the characteristic length estimated from, e. g., the average size of
the solid obstacles, a proper volume average of the porous media strongly indicates, in agreement with experimental observation, that
the single-phase flow is well described by Darcy's law \cite{Sahimi_Review,*sahimi2012flow,bear1972}. In addition, if dispersion (diffusion) effects are relevant, an advection-dispersion (convection-diffusion) equation must be included in the description of a variety 
of phenomena, such as fluid flow and
solute transport \cite{tartaprl}, and miscible viscous fingering \cite{Homsy,mishracomsol,*mishrapf,*mishrapre,dewit}.
On the other hand, as Re increases, violations of Darcy's law have been reported due to inertial effects, described by the Forchheimer
equation \cite{Chen199172,soares99}, and very interesting non-Newtonian behavior, with power-law permeability and data collapse in a
broad range of Reynolds conditions \cite{soares2009}. Very recently, the stability analysis of two-phase buoyancy-driven flow, in the presence of a capillary transition zone,
was investigated in detail \cite{PhysRevE.87.033009}.

A large body of work on fluid dynamics in porous media \cite{Sahimi_Review,*sahimi2012flow,joekar,*NAG:NAG898} clearly indicates
that several powerful continuum and discrete approaches have been put forward to treat many cases of interest. In the continuum version both analytical
and direct numerical integration of the pertinent dynamic equations have proved very efficient \cite{bear1972,joekar,*NAG:NAG898,PhysRevE.87.033009,tarta,tartaprl,mishracomsol,*mishrapf,*mishrapre,soares99,soares2009}, while several
concepts from percolation, growth models, random walk and fractal geometry have been very useful in dealing with
discrete models \cite{Sahimi_Review,*sahimi2012flow,feder,Barabasi}, which are alternative approaches heavily based on numerical simulations
and supplemented by continuum stochastic equations \cite{Barabasi}. Among the discrete models widely used in the literature, we shall emphasize the 
lattice gas cellular automata.

Lattice gas cellular automata (LGCA) is a family of computational models whose particles have velocities defined by a discrete set and
collide with each other on sites of an hexagonal lattice; the resulting velocity configuration obeys mass and momentum conservation
\cite{PhysRevLett.56.1505,*Frisch_et_al,Wolfram_article, Rothman_book}.
Making use of the Chapman-Enskog expansion, it was demonstrated that in two \cite{PhysRevLett.56.1505,*Frisch_et_al,Wolfram_article, Rothman_book} and three \cite{0295-5075-2-4-006} dimensions (2D and 3D) the average values of the
pertinent variables in a macroscopic scale satisfies the
incompressible Navier-Stokes equations; the 3D case requires minor additional considerations.
More recently, 2D-LGCA flows on curved surfaces with dynamical
geometry were also proposed \cite{PhysRevE.82.046705,*Love13062011}, with motivation in describing a variety of phenomena.

Challenging aspects in the description of flow in porous media are associated
with the complex geometry of the pore space and the region near the fluid-fluid interface in the multiphase case.
The LGCA model handles the no-slip boundary condition in the fluid-solid
interface in a very simple manner through the local bounce-back rule, which is easily implemented even for the complex geometries arising from the pore space \cite{rothman96}.
In particular, the LGCA single phase flow in 2D porous media satisfies Darcy's \cite{roth} and Forchheimer's \cite{Chen199172} laws, and 
the permeability of real systems can be estimated \cite{Chen199172,Humby20021955}
from images of their microgeometry. We also remark that, in a clean channel, Poiseuille flow is observed \cite{roth}.
Two-phase flows in 2D are also modeled through the LGCA by the use of colored particles and an additional collision-color conservation law
\cite{Rothman_Keller}. This model leads to a well-defined interface between the two fluids, with a finite surface tension,
thereby allowing the observation of Laplace's law and the phenomenon of phase separation \cite{Rothman_Keller}.
Beyond that, it was shown \cite{rothf,*PhysRevE.53.16} that this model captures the expected interface hydrodynamic fluctuations
associated with capillary waves, and exhibits dynamic scaling exponents \cite{Barabasi} in agreement with those of the
Kardar-Parisi-Zhang equation \cite{PhysRevLett.56,[{See also, }] [{ and references therein}]PhysRevE.85.010601}, apart from logarithmic corrections for the largest systems studied
\cite{rothf,*PhysRevE.53.16}. The macroscopic properties of two-phase immiscible fluids
in 3D porous media were described \cite{FLM:13517} using lattices built from rock's microgeometry. Further, the flow of
two-phase immiscible fluids in the presence of an amphiphile species in 2D \cite{PhysRevE.62.2898} and 3D \cite{PhysRevE.64.061302,*Love2003340}
porous media were also considered.

We also stress that much attention has been devoted to the development of the lattice-Boltzmann method (LBM) \cite{PhysRevLett.61.2332,revlbm}. This variant of the
LGCA method \cite{Rothman_book} can be simplified through a linearized collision operator \cite{higuera,extralbm}, 
both in 2D and 3D. The LBM was also used in the study of single-\cite{0295-5075-10-5-008,cancelliere:2085} and 
two-phase \cite{rothman95,PhysRevE.53.743,PhysRevE.67.046304,PhysRevE.72.026705,Boek20102305} flows   
in artificial and real 3D porous media; as well as in the study of single-phase flow in multiscale 
porous media \cite{PhysRevE.66.056307}. Further, the LBM was adapted \cite{PhysRevE.66.036304} to model single-phase flow in a 2D porous medium,
in which case the action of the solid space is introduced as linear (Darcy's law) and non-linear (Forchheimer's law) effective body-force terms
in the generalized Navier-Stokes equations \cite{Nithiarasu19973955}. On the other hand, viscous finger phenomenon in 2D channels was studied
through the LBM for two miscible fluids \cite{rako} and two immiscible fluids with zero \cite{rako} and finite \cite{langaas} surface tension. 
We also mention that the LBM was adapted to describe the hydrodynamics of 3D liquid crystals \cite{15082004}, and very recently used 
\cite{PhysRevLett.110.026001,*PhysRevLett.110.048303} to study the flow of a nematic liquid crystal in microfluidic channels.

In the following we digress on the statistical properties and the dynamics of an interface separating two immiscible fluids 
flowing in a Hele-Shaw cell, or in 2D and 3D systems modeling these cells, which is the main subject of this work. 
The dynamics of growing interfaces has stimulated intensive theoretical and experimental investigations motivated by its own
challenging fundamental aspects and potential technological applications as well \cite{Barabasi,feder}. In this context, the
displacement of a fluid by another during the flow in a porous medium has been considered as a representative example in the study
of the dynamics of roughening interfaces \cite{Barabasi,feder}. The main experimental tool in such studies is the 
Hele-Shaw cell \cite{Proc.R.Soc.,chuoke},
which mimics the porous media and have unveiled a rich variety of phenomena.
The cell is formed by two plates separated by a little gap of size $b$. The velocity of a viscous single-phase flow through the gap, averaged through the direction
perpendicular to the plates, satisfies Darcy's law with permeability $b^2/12$. For an immiscible two-phase flow in a horizontal 
cell, if the less viscous fluid displaces the more viscous one, the 
interface is unstable, and a linear stability analysis predict 
the occurrence of one-finger or multifinger patterns \cite{Proc.R.Soc., chuoke}. More recently, very detailed experimental aspects of this 
phenomenon \cite{maher,PhysRevA.33.794,*jfmlib,*PhysRevA.36.1894,*tabeling1989experimental,RevModPhys.58}, and theoretical studies based on 
vortex-in-cell method \cite{aref}, random-walk techniques and methods from complex analysis \cite{RevModPhys.58}, have been put forward; which also include 
phase-field model \cite{PhysRevE.60.1734}, potential flow analysis \cite{FLM:389437} of radial fingering \cite{PhysRevA.39.5863,kim:074106}
and rigorous arguments \cite{PhysRevA.44.6490,*PhysRevLett.80.2113,*PhysRevLett.84.5106,*FLM:85633,*Crowdy01112006}.

In the context of flow in clean Hele-Shaw cells, a second variant of the LGCA method was formulated \cite{Hayot1,notagrosfils} shortly after the original 
proposal \cite{PhysRevLett.56.1505,*Frisch_et_al,Wolfram_article}, in  
which case the Darcy's law  obtains without the need of the detailed microscopic geometry
of the medium. In fact, through the introduction of random scattering sites, in a stochastic context, the viscous dissipation of the single-phase flow in 
a Hele-Shaw cell was investigated using this modified LGCA method.
On the other hand, in the proposed model for immiscible fluids \cite{Hayot2,*Hayot1989277}, the authors used \textit{dynamic random scattering sites} (a time-dependent
approach) and find that the interface fluctuations exhibit typical random-walk behavior, with a roughness exponent 1/2; 
however, the fluctuations are not associated with hydrodynamic capillary waves, since the kinetic rules \cite{Hayot2,*Hayot1989277}, and
other assumptions, differ from those of Ref. \cite{Rothman_Keller}.
Notwithstanding, Saffman-Taylor fingering instability \cite{Proc.R.Soc.,chuoke} is observed through this modeling \cite{Hayot89,*Hayot199164}:  
the interface is unstable either if the fluids differ in pressure or 
in the density of dynamic random scattering sites.
On the other hand, we mention that a two-phase model similar to that of Ref. \cite{Rothman_Keller}, 
but with only local update rules \cite{Somers199139} and with the inclusion of stochastic scattering sites, 
was suggested \cite{lutsko} to describe multifractal properties of fingering patterns
associated with the Saffman-Taylor instability in a clean Hele-Shaw cell. 
The viscosity contrast was introduced through the use of distinct collision rules for each fluid.
The authors find that for high surface tension the Hausdorff (fractal) dimension is in the interval (1.41 - 1.46), while for moderate surface 
tension the fractal dimension is found to be 1.75 (see discussion below).

We now discuss highlights of viscous-fingering phenomena in Hele-Shaw cells with quenched disorder.
In the case of a circular cell, fluid invasion at the center of the cell exhibits fractal fingering
structure for high capillary number $\text{Ca}=\mu V/\sigma$, where $\mu=\rho\nu$ is the dynamic viscosity, $\rho$ is the density of the fluid, and $\sigma$
is the interface tension \cite{PhysRevLett.55.2688,*PhysRevA.36.318}. On theoretical grounds, for simplicity, the viscosity of the invading fluid is considered
negligible (infinity viscosity ratio), thus allowing a map \cite{PhysRevLett.52.1621,PhysRevLett.55.1892} of the growing front dynamics onto the
diffusion-limited aggregation (DLA) process \cite{witten1,*witten2},
characterized by a fractal dimension $D_f \approx 1.71$. 
Notwithstanding, the
measured \cite{PhysRevLett.55.2688,*PhysRevA.36.318} 
fractal dimension of the fingering structure for high Ca, $D_f=$ (1.62 - 1.64)$\pm 0.04$, 
is very close to the fractal dimension of the percolation cluster backbone  \cite{PhysRevLett.53.1121,*law}, $D_f=(1.61\text{ - }1.62)\pm0.02$;
it also satisfies the inequality, $1.3<D_{DLA}^{eff}<D_{VF}^{eff}<1.7$, 
suggested \cite{PhysRevLett.57.1875} to be valid above the percolation threshold in the context of the analogy between DLA, percolation 
and viscous fingering.
However, at low Ca the fractal structure is described
by invasion percolation with trapping \cite{wilkinson,PhysRevLett.54.2226}, $D_f\approx 1.82$, which should be compared with the fractal 
dimension of ordinary \cite{stauffer} or invasion \cite{wilkinson,dias,liacir} percolation at the threshold: $D_f=91/48\approx 1.89$. 
We also remark that the diffusion front of random walkers on 2D lattices is a
fractal object of dimension $D_f=1.75$, in close connection with the percolation cluster hull \cite{sapoval,*PhysRevB.32.6053}; 
in 3D lattices the front is a dense object with $D=3$ \cite{PhysRevLett.57.3195}. 
Further, in rectangular cells, two-case studies have been reported \cite{stokes}: 
(i) water (wetting fluid) displacing oil (nonwetting fluid); (ii) oil (nonwetting fluid) displacing water and glycerol (wetting fluid),
with viscosity ratio $M=\mu_2/\mu_1\approx 200$ in the two cases. The authors find that in the first case (imbibition process, with the finger 
width $\lambda\sim \text{Ca}^{-\frac{1}{2}}$), 
the finger width is much larger than the pore size (viscous fingering), whereas in the second case (drainage process), the finger
width has the same order of the pore size (capillary fingering). 
In this context, a dynamical transition has
been identified \cite{PhysRevLett.60.2042}. On the other hand, self-affine (anisotropic) fractal interfaces \cite{feder,mandelbrot} from immiscible displacement
in porous media were observed when the displacing fluid is more viscous and more effectively wets the medium (e. g., water or glycerol displacing
air in a Hele-Shaw cell filled with glass beads). In these experiments
the scale-dependent roughness, $W(L)=AL^\alpha$, where $W(L)$ is the interface width, $\alpha\approx 0.73$ is the roughness exponent, 
and $A\sim \text{Ca}^{-\frac{1}{2}}$ \cite{rubio}, while large variations with time and flow conditions
suggest a more complex dependence on Ca \cite{PhysRevLett.69}, with the effective value of $\alpha$ varying over a wide range (0.65 - 0.91).
In the latter experimental study, connection with the dynamics of random-field Ising model was proposed, although in this case a crossover
from $\alpha=0.8$ at small scales to  $\alpha=\frac{1}{2}$ at larger scales was found
\cite{PhysRevLett.71.2074}, in agreement with experiments on wetting invasion \cite{0305-4470241,aurora}. We remark that, in the context of fractal growth
in hydrodynamic dispersion through random porous media, direct integration of the advection-diffusion and linear stokes equations shows
that the following steady flows are obtained \cite{PhysRevE.50.335} according to the value
of the P\'eclet number, $\text{Pe}=\frac{VL}{D_m}$, where $D_m$ is the diffusion constant: (i) for $\text{Pe}=\infty$, the fractal morphology of the spreading
dye (tracer) is characterized by a fractal dimension close to $D_{DLA}$; (ii) for $\text{Pe}=0$, a self-affine fractal interface is identified
with scaling consistent with a growth process as predicted by the KPZ equation \cite{PhysRevLett.56}; (iii) for finite Pe, the behavior is characterized by
the competition between advection and diffusion effects. Lastly, a series of very detailed experimental studies in Hele-Shaw cells with disorder have 
investigated the interface scaling and fractal properties under combined viscous, gravity, and capillary effects \cite{lenormand,PhysRevE.55.2969,PhysRevE.66.05,PhysRevLett.102.074502,*PhysRevE.80.036308,vadose, [{
We also mention a recent LBM investigation of immiscible fingering on a 2D lattice with regularly spaced circular obstacles: }] [{}] dong2011}. 
In particular, for immiscible displacement of viscosity-matched fluids in 2D porous media \cite{PhysRevE.55.2969}, the fractal structures of the front are characterized by 
using the box counting algorithm, which implies $D_f=1.33 \pm 0.05$, consistent with the fractal dimension of the external perimeter in invasion 
percolation \cite{PhysRevLett.67.584}, and the density-density correlation function (data collapse for several values of Ca), which
implies $D_f=2-\alpha=1.42 \pm 0.05$ and $\beta=0.8 \pm 0.3$, where $\beta$ is the growth exponent ($W_L\sim t^\beta$).
In addition, the authors find that the fractal dimension of the structure left behind the front is $D_f=1.85$, consistent with invasion percolation 
with trapping. Further, a detailed analysis of the two-phase flow of immiscible fluids in disordered porous media allowed the authors \cite{vadose} to propose a 
quite unified view of the macroscopic transport properties. Indeed, the characteristic crossover scales between fractal regimes, i. e., from 
capillary to viscous fingering, are discussed using box counting, where the box side size $l$ ranges from the cell width $L_y$ (cell extent $L_x$) down to the pixel size. They thus find: $D = 1.00$ for
$l>L_y$, $D_f = 1.60$ for $L_y>l>a/\text{Ca}$, and $D_f = 1.83$ for $l<a/\text{Ca}$, where $a$ is the pore size. We stress that the two mentioned fractal 
dimensions are consistent with those of 2D percolation cluster backbone and invasion percolation with trapping, respectively.

The main goal of our work is to study the interface dynamics between two immiscible fluids in a Hele-Shaw cell geometry in clean systems and
in systems with quenched disorder; both cases are investigated through the LGCA model \cite{Rothman_Keller} in the presence of 
\textit{dynamic random scattering sites} \cite{Hayot89,*Hayot199164}. 
In fact, our approach combines the best features of the LGCA model, namely, its microscopic kinetic rules \cite{Rothman_Keller}, interface
hydrodynamic fluctuations effects \cite{rothf,*PhysRevE.53.16}, and dynamic random scattering sites  \cite{Hayot89,*Hayot199164}, thereby allowing a nice description of the Saffman-Taylor instability
in a clean Hele-Shaw cell, in very good agreement with experimental observations \cite{Proc.R.Soc.,chuoke,RevModPhys.58,PhysRevA.33.794,*jfmlib,*PhysRevA.36.1894,*tabeling1989experimental}. In addition,
the inclusion of quenched disorder allows us to consider the competition between viscous and capillary effects at the pore level, thus causing the roughening
interface to exhibit very interesting self-affine features.

The work is organized as follows, in Section II we briefly review the LGCA models from which we build the two phase LGCA with scatterers.
In Sec. III we present our results for the interface dynamics in a clean system, exhibiting the Saffman-Taylor instability, and in a system
with quenched disorder, in which case we also study scaling behavior, critical exponents, and fractal dimension of the interface.
Finally, a discussion on the main results and concluding remarks are presented in Section IV.

\section{LGCA Models}
\subsection{Single Phase LGCA}

The basic model reviewed in this section is named the
random-collision 7-velocity \cite{Rothman_book}. Time and space are discrete and the particles can lie only on the sites of a regular triangular lattice.
Moreover, its velocities can assume only seven possible values, one is null (particle at rest) and six have magnitude of one lattice unit (unit of length) per time step (unit of time) 
in the direction of the lattice edges, $\mathbf{c}_i = (\cos(i\pi/6),\sin(i\pi/6))$ with $i = 1, \ldots,6$, where a lattice unit is the distance between two sites.
Not more than one particle per site in one direction is allowed. In each time step the outcome of particle collisions are such that
total momentum and particle number are conserved at each site of the lattice,
after each collision the particles hop to a neighboring site in accord with its velocity.
In a collision, the model takes into account all possible randomly
changes in the velocity allowed by the conservation laws; the collision is local, i.e., depends only on particles at the same site.

In this model, it is natural to define the following hydrodynamic variables:
\begin{equation}
\overline{\rho}(\mathbf{x},t) = \sum_{i=0}^6{N_i(\mathbf{x},t}),
\label{density}
\end{equation}
the density of particles per site at position $\mathbf{x}$ and time $t$;
\begin{equation}
\overline{\mathbf{g}}(\mathbf{x},t)= \sum_{i=1}^6 {N_i(\mathbf{x},t) \mathbf{c}_i},
\label{momentum}
\end{equation}
the momentum density per site, where $N_i$ is the
probability of a particle at position $\bm{x}$ and time
$t$ to have a velocity in the direction $i$, considering $N_0$ in Eqs. (\ref{density}) and (\ref{momentum}) as the probability of a particle to be at rest. These
probabilities are obtained by making space-time averages. Densities per unit area,
$\rho$ and $\mathbf{g}$, are given by
\begin{equation}
\rho=\overline{\rho}/(\sqrt{3}/2)\text{ and }\mathbf{g}=\overline{\mathbf{g}}/(\sqrt{3}/2)
\end{equation}

In systems not far from equilibrium, with
smooth variations in space and time, Chapman-Enskog expansions allows one to obtain the following dynamic equation
\cite{PhysRevLett.56.1505,*Frisch_et_al,Wolfram_article,Rothman_book}:
\begin{equation}  \label{eq_macroscopica}
\frac{\partial \mathbf{u}}{\partial t} + g(\overline{\rho})(\mathbf{u}\cdot \nabla )%
\mathbf{u} = -\frac{1}{\rho}\nabla p + \nu\nabla ^{2}\mathbf{u} + \frac{1}{\rho}\mathbf{f},
\end{equation}
where $\mathbf{u}$ is the average flow velocity, $\nu$ is the kinematic viscosity, $\mathbf{f}$ is the body force per unit area and
\begin{equation}  \label{ro_FHP-II}
g(\overline{\rho}) = \frac{14}{24} \frac{7 - 2\overline{\rho}}{7-\overline{\rho}}.
\end{equation}

Boundary conditions are easily implemented \cite{PhysRevLett.56.1505,*Frisch_et_al,Wolfram_article,Rothman_book} and
no-slip boundary condition is introduced by bouncing-back the particles when they arrive at sites marked as solid sites.
In our simulations the force is introduced by increasing
the momentum at a randomly chosen site, if such a change is possible,
until the total desired momentum increase is reached in each time step.

\subsection{Immiscible Two Phase LGCA}

The Immiscible Lattice Gas (ILG) was introduced by Rothman and Keller \cite{Rothman_Keller} and mimics
the dynamics of two immiscible fluids in two dimensions. This model is similar to what we have described above except that there are
two kinds of particles: red and blue. As before, we allow not more than one particle with the same velocity at the same site, even if
they have different colors. In the propagation, the particle maintains its identity and in a collision, in addition to the conservation of
momentum and particle number, the number of particles of each color is also conserved at each site. Nonetheless, not all collisions are allowed; in fact, they depend on the neighborhood and in the way that particles of the same color tend to cluster. More specifically, the site configuration could be represented by 14 Boolean variables, $(r_0, b_0, \ldots, r_6,b_6)$ each one representing the presence
of a red or blue particle in the direction $i$ (again, the direction 0 is associated with a particle at rest). The effect of the neighborhood on the
result of a collision at the site $\mathbf{x}$ is introduced through the definition of the vector quantity \textit{color gradient}:
\begin{equation}  \label{gradiente_de_cor}
\mathbf{F}(\mathbf{x}) = \sum_i \mathbf{c_i} \sum_j[r_j(\mathbf{x} + \mathbf{%
c_i}) - b_j(\mathbf{x} + \mathbf{c_i})];
\end{equation}
the vector \textit{color flux}:
\begin{equation}  \label{fluxo_de_cor}
\mathbf{q}[\mathbf{r}(\mathbf{x}),\mathbf{b}(\mathbf{x})]) = \sum_i \mathbf{%
c_i} \big[ r_i(\mathbf{x}) - b_i(\mathbf{x}) \big],
\end{equation}
and by considering that the resultant site configuration maximizes the quantity:
\begin{equation}  \label{maximiza}
\mathbf{q}(\mathbf{r^{\prime }},\mathbf{b^{\prime }}) \cdot \mathbf{F},
\end{equation}
and conserves the number of each kind of particle and the total momentum. In the above equation, prime indicates the configuration after a collision. If there are more
than one configuration that maximizes Eq. (\ref{maximiza}), one of them is randomly chosen.

Through these rules, an initially mixed system will separate into regions with only red or blue particles, depending upon the density of particles and the relative
concentration of color \cite{Rothman_book,Rothman_Keller}.
In general, if the density is above 2 particles per site and the relative concentration is not too high or too low, the separation occurs.
In a phase separated system it is verified by simulations that there is a discontinuity in pressure which obeys Laplace's law \cite{Rothman_book,Rothman_Keller}:
\begin{equation}  \label{Laplace}
p_1 - p_2 = \sigma \kappa, \qquad
\end{equation}
where $p_1$ and $p_2$ are the pressure in regions with different colors, $\sigma$ is the surface tension and $\kappa$ is the curvature of the interface.
The value of $\sigma$ varies with the density of particles and for the density used in this paper it is approximately 0.37.
In regions where there is just one kind of particle, the dynamics is the same as that of the 7-random-velocity model, but in the interface
the dynamics changes.

\subsection{Single Phase LGCA with Scatterers}
\begin{figure}
\includegraphics[width=.45\textwidth]{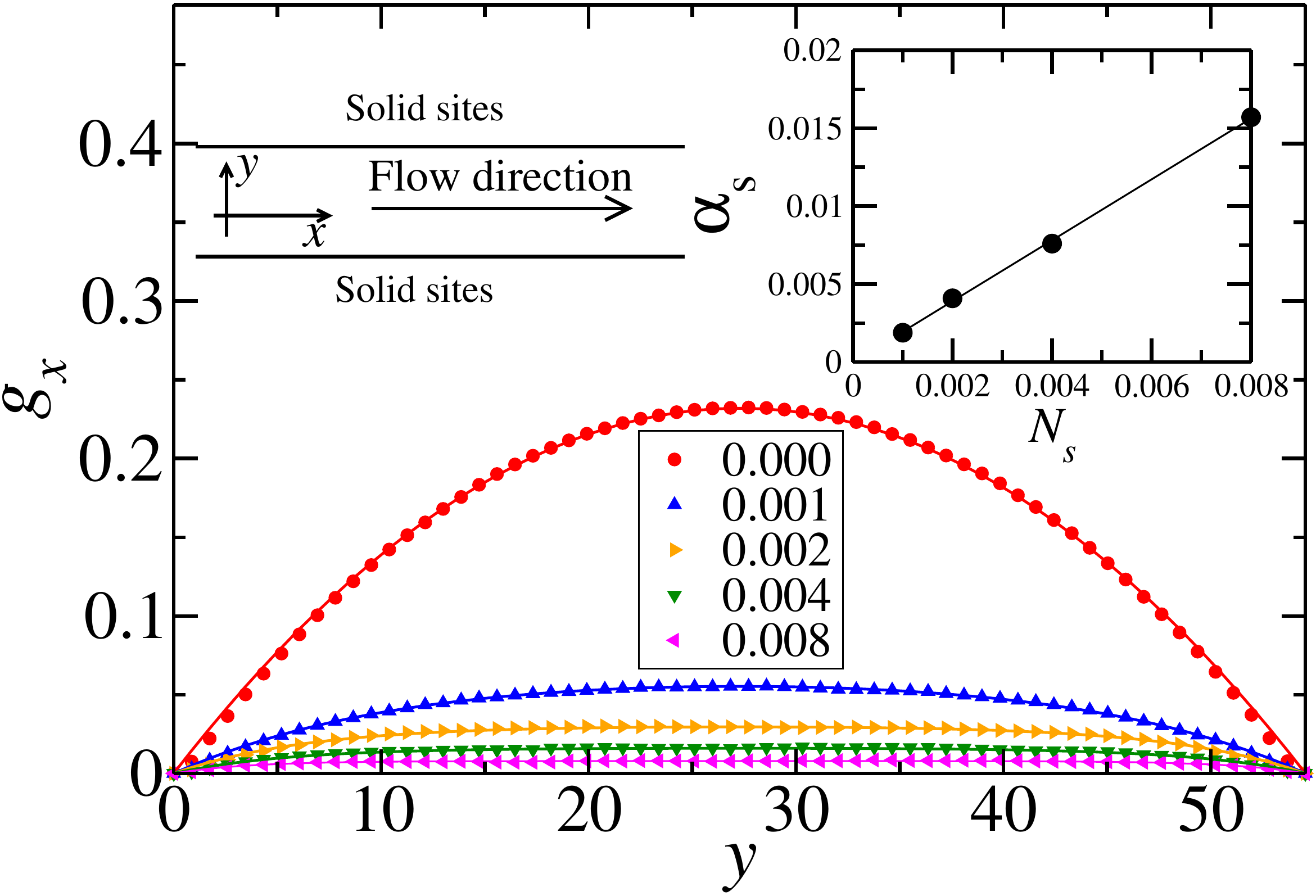}
\caption{(Color online) Profile of the x-component of the momentum per site, $\overline{g}_x$, for the single phase LGCA with scatterers in a flow with the same boundary condition as the Poiseuille flow (indicated
in the figure). The density of scatterers, $N_s$, is indicated, the density per site is $\overline{\rho}=3.5$ and the body force per site is
$\overline{f}=1/(256\times 64)\approx 6.1\times 10^{-5}$. Solid lines are fittings of the data to Eq. (\ref{poiseuille}). Inset: dissipative factor $\alpha_s$ as a function
of $N_s$ from the presented profiles.}
\label{ScFig}
\end{figure}
This model was introduced by Hayot \textit{et al.} \cite{Hayot1} with the goal of reproducing the fluid dynamics ruled by Darcy's law. The scatterers provide
momentum dissipation similarly to the solid region of porous media or the plates of Hele-Shaw cells. Here the scatterers are introduced in the
random-collision 7-velocity model. In each time step a fixed number of sites is chosen randomly
as solid scattering sites, thus inverting the velocity of the scattered particles and introducing a damping term in Eq. (\ref{eq_macroscopica}).
The number of scattering sites must be very low compared with the number of particles in the system.
Furthermore, if the density is chosen to be 3.5 particles per site $g(\overline{\rho}) = 0 $, thereby eliminating the advective term in Eq. (\ref{eq_macroscopica}).
In this manner, we obtain \cite{Hayot1} the following steady state equation

\begin{equation}  \label{NS_H}
\alpha_s \mathbf{u} = -\frac{1}{\rho}\nabla p + \nu\nabla ^{2}\mathbf{u} +
\frac{1}{\rho} \textbf{f},
\end{equation}
where $\alpha_s$ is positive and proportional to the density of scatterers \cite{Hayot1}.
For a lattice gas in the $xy$ plane subject to a force density $f$ and a constant pressure ($p$) gradient
also in the $x$ direction, Eq. (\ref{NS_H}) reads:
\begin{equation}  \label{NS_Hx}
\alpha_s u_x = -\frac{1}{\rho}\frac{dp}{dx} + \nu\nabla ^{2}u_x +
\frac{1}{\rho} {f},
\end{equation}
The effect of the dissipative term can be understood by comparing the velocity profile for the boundary conditions of a
Poiseuille flow
with the parabolic profile which is obtained in the absence of this term. For this boundary condition the fluid is
confined between two lines of solid sites at $y=0$ and $y=L$, which implies $u_x(0)=0$ and $u_x(L)=0$, due to non-slip boundary conditions in the
solid sites, and the following solution to Eq. \ref{NS_Hx} \cite{Hayot1}:
\begin{equation}
 u_x(y)=\frac{1}{\alpha_s\rho}(-\frac{dp}{dx}+f)\left\{1-\frac{\cosh[r(y-(L/2))]}{\cosh(rL/2)}\right\},
\label{poiseuille}
\end{equation}
with $r=\sqrt{\alpha_s/\nu}$.

In Fig. \ref{ScFig} we present the velocity profile for a simulation \footnote{The codes we use 
were built from \url{http://www.lmm.jussieu.fr/\~zaleski/latgas.html}; see Ref. \cite{Rothman_book}.} of the single phase LGCA model
with scatterers, adopting periodic boundary conditions (PBC) in the $x$ direction,
for the indicated values of density of scatterers, $N_s$, homogeneous density of force per site $\overline{f}$, as well as data for a simulation
without scatterers. Since the applied force is homogeneous, the density ($\rho$) and the pressure ($p$) remain
constant on average ($\frac{dp}{dx}=0$); therefore, the data of the simulations are well fitted by Eq. (\ref{poiseuille}). As we can see, the presence of scatterers flattens the velocity profile in comparison with the parabolic one, except near the walls \cite{Hayot1,PhysRevE.66.036304, notagrosfils}. 
For $N_s=0$, we obtain $\nu=0.196$, which is in accordance with reported results \cite{Rothman_book} for the value of $\nu$ for the random-collision 7-velocity model. Following the procedure described in Ref. \cite{Hayot1}, we obtain $\nu=0.197, 0.176, 0.207 \text{ and } 0.196$ for $N_s=0.001,0.002,0.004\text{ and }0.008$,
respectively. Also, we find the dissipative factor $\alpha_s=1.95N_s$, which is near the theoretical estimate from the microscopic model through the
use of the Boltzmann approximation \cite{Hayot1}: $\alpha_s=2N_s$, but greater than the value observed for the 6-velocity model \cite{Hayot1}: $\alpha_s\approx1.1N_s$.
\begin{figure}
\includegraphics[width=.45\textwidth]{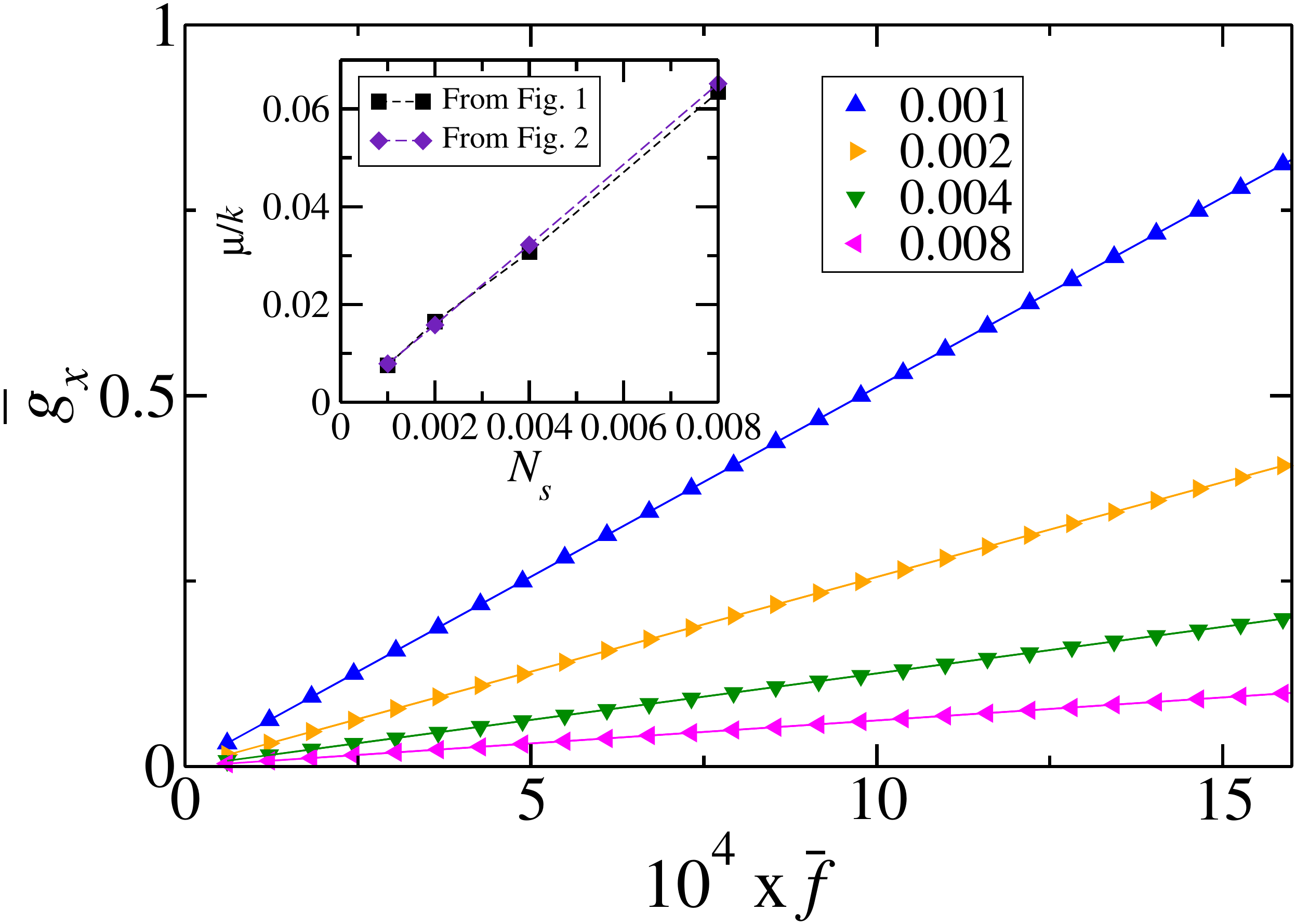}
\caption{(Color online) Component $x$ of the momentum per site, $\overline{g}_x$, as a function of the body force per site, $\overline{f}$, for $\overline{\rho}=3.5$ and
periodic boundary conditions in the $x$ and $y$ directions, for the values of $N_s$ indicated. Solid lines are fittings to a Darcy's law equation form, Eq. (\ref{eqcap}).
Inset: (From Fig. 1) $\mu/k=2\alpha_s \overline{\rho}/\sqrt{3}$ as a function of $N_s$ calculated from Eq. (\ref{coeficiente}) and the estimated value of $\alpha_s$ from Fig. 1, }
\label{Darcy}
\end{figure}

In the region where the velocity is flat, the Laplacian term in Eq. (\ref{NS_Hx}) can be neglected, and this equation reduces to Darcy's law 
\cite{Hayot1,PhysRevE.66.036304, notagrosfils}:
\begin{equation}
u_x = \frac{k}{\mu}\left(-\frac{dp}{dx} + f\right),
\label{Darcy_H}
\end{equation}
where $\mu=\nu\rho$ is the dynamic viscosity of the fluid and $k$ is the permeability of the porous medium; for fluid flow in a Hele-Shaw cell, $k=b^2/12$, where
$b$ is the gap between the two plates of the cell. This implies
\begin{equation}
\frac{k}{\mu}\equiv\frac{1}{\alpha_s\rho},
\label{coeficiente}
\end{equation}
which allows a physical interpretation of the underlying role played by the scatterers in the model.
For homogeneous pressure, Eq. (\ref{Darcy_H}) can also be written in terms of the momentum and force densities
per site:
\begin{equation}
 \overline{g}_x = \frac{k}{\nu}\overline{f},
 \label{eqcap}
\end{equation}
In Fig. \ref{Darcy} we present the simulation of a flow on a lattice
without solid sites and with periodic boundary conditions in the $x$ and $y$ directions and without any obstructed sites. 
The simulations were performed using 10$^6$ time steps
and the results for the average momentum are obtained discarding the first 20000 time steps. A linear relation between $f$ and $g_x$ or $u_x$ is obtained. The values of
$\frac{\mu}{k}=\alpha_s\rho$ shown in the inset of this figure are consistent with those obtained from the simulation in a channel
with $\alpha_s\approx2N_s$, in agreement with the estimate from the Boltzmann approximation \cite{Hayot1}.

Therefore, the approach with scatterers is quite remarkable since it allows the unified description of Hele-Shaw cell experiments with
fluid dynamic viscosity tuned by $\mu=(\frac{b^2}{12})\alpha_s\rho$, or the flow in a general 2D porous media, which may include quenched disorder effects,
with an average homogeneous permeability $k$ fixed by Eq. (\ref{coeficiente}).

An interesting situation is the possibility of introducing scattering sites in the ILG model, in particular for red and blue fluids
having distinct scattering rates, which can be interpreted as two fluids with different viscosities. This situation can be realized and shall be discussed in next
section for an homogeneous lattice and a lattice with quenched disorder.

\section{Immiscible LGCA with Dynamic Scatterers}
This model is built by introducing random dynamic scattering sites in the ILG model.
In regions of the lattice where there is just one kind of fluid, the flow dynamics is the same as the LGCA model with scatterers and the
change occurs only on the interface. In a static situation, simulations as in Ref. \cite{Rothman_Keller} confirm that Laplace's law holds
and the surface tension has a value
nearly equal to the one found in the ILG model ($\approx 0.37$). In particular, if a flow is established in a system with distinct scattering
rates for the red and blue fluids, we
can observe a Saffman-Taylor-like instability \cite{Proc.R.Soc.,chuoke} in the interface, as described below.

\subsection{Interface Instabilities}
\begin{figure}
\includegraphics[width=.45\textwidth]{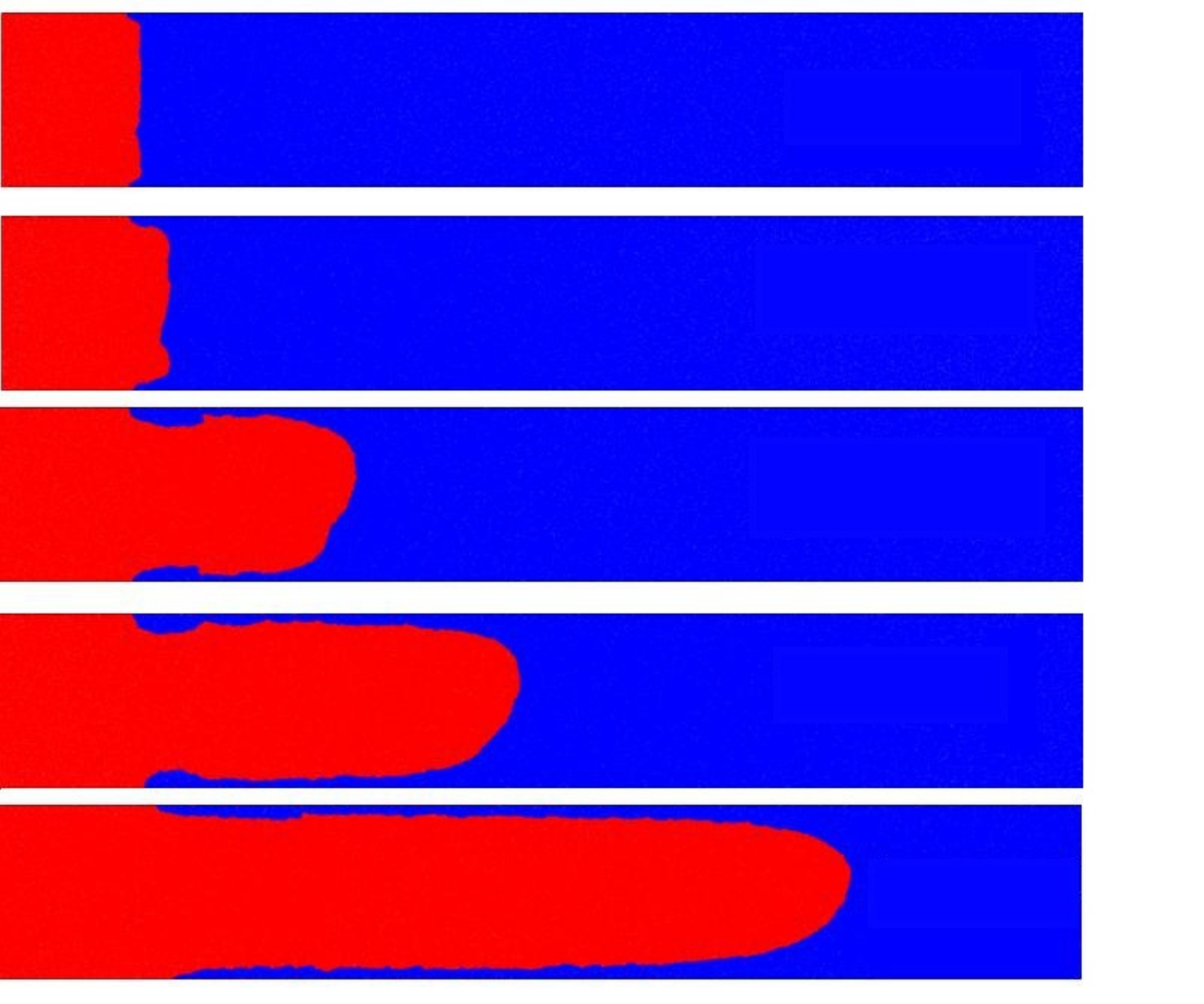}
\caption{(Color online) Snapshots of a simulation in which the red (light gray) fluid with a density of 0.001 scatterers per site displaces the blue fluid, which has a density
of 0.008 scatterers per site. The forcing rate is 0.0001 and acts from left to right. The width of the system is 500 $\sqrt{3}/2$. Simulation time
step increases from top to bottom.}
\label{ST1}
\end{figure}
\begin{figure}
\includegraphics[width=.45\textwidth]{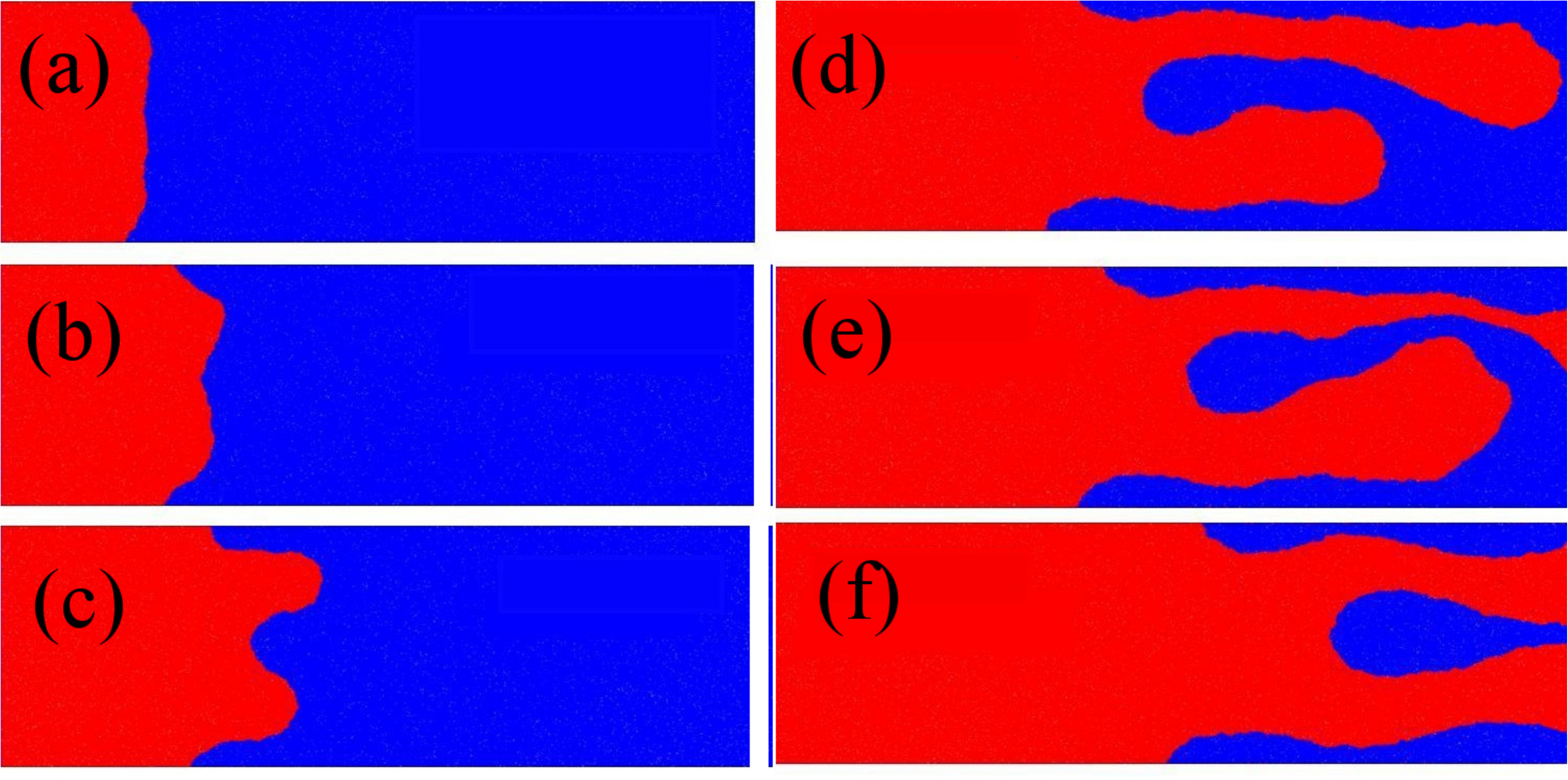}
\caption{(Color online) Snapshots of a simulation in which the red (light gray) fluid with a density of 0.001 scatterers per site displaces the blue fluid, which has a density
of 0.008 scatterers per site. The forcing rate is 0.0005 and acts from left to right. The width of the system is 500 $\sqrt{3}/2$. Simulation time
step increases from (a) to (f).}
\label{ST2}
\end{figure}
\begin{figure}
\includegraphics[width=.45\textwidth]{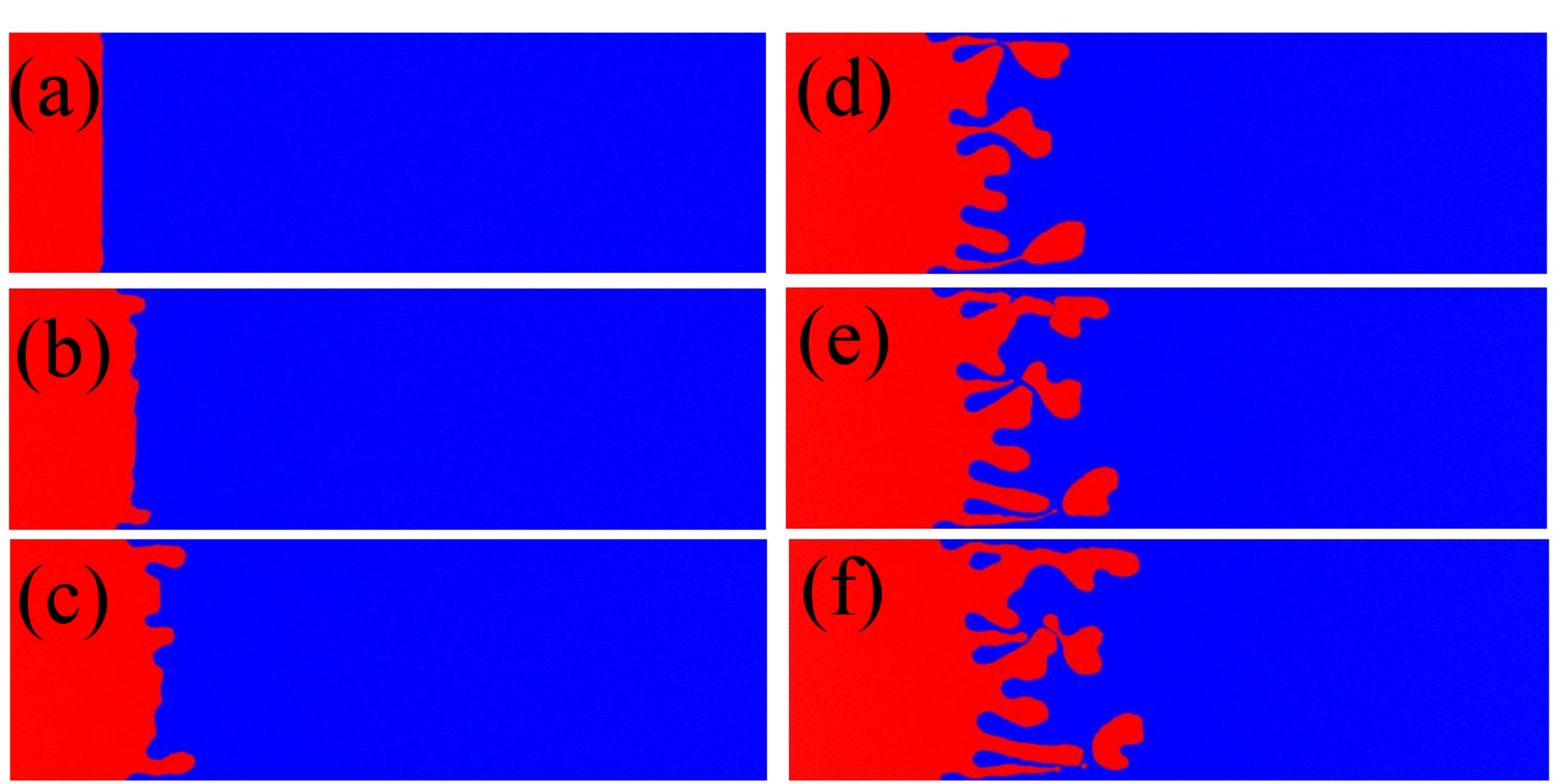}% fig319.eps Here is how to import EPS art
\caption{(Color online) Snapshots of a simulation in which the red (light gray) fluid with a density of 0.001 scatterers per site displaces the blue fluid, which has a density
of 0.008 scatterers per site. The forcing rate is 0.0005 and acts from left to right. The width of the system is 2000 $\sqrt{3}/2$. Simulation
time step increases from (a) to (f).}
\label{ST3}
\end{figure}
The dynamics of the interface between two immiscible fluids as one displaces the other in a Hele-Shaw cell can be discussed by considering the
average velocity of flow and the relation between the viscosities of the two fluids \cite{Proc.R.Soc.,chuoke,maher,RevModPhys.58,PhysRevA.33.794,*jfmlib,*PhysRevA.36.1894,*tabeling1989experimental}.
Three general behaviors for the interface are observed as
these parameters are varied: (i) If a more viscous fluid displaces a less viscous one or a less viscous displaces a more viscous with a low
average velocity, the interface is flat; (ii) if a less viscous fluid displaces a more viscous one with intermediate values of the average
velocity, a single stable finger is observed; (iii) for sufficiently higher values of the average velocity a multifinger dynamics sets in.
If fluid 2 displaces fluid 1 in a cell with width $L_y$ the finger patterns observed are governed by two dimensionless parameters \cite{maher,aref}:
the viscosity contrast
\begin{equation}
 A=\frac{\frac{\mu_{1}}{k_{1}}-\frac{\mu_{2}}{k_{2}}}{\frac{\mu_{1}}{k_{1}}+\frac{\mu_{2}}{k_{2}}};
 \label{Aparameter}
\end{equation}
and a dimensionless surface tension
\begin{equation}
  B=\frac{2\sigma}{L_y^2}\frac{1}{(\frac{\mu_{1}}{k_{1}}-\frac{\mu_{2}}{k_{2}})V},
  \label{Bparameter}
\end{equation}
where $V$ is the average velocity of the flow. Further, from a linear stability analysis of the problem, it is possible to show that the
perturbation with
the wavelength \cite{feder,chuoke}
\begin{equation}
 \lambda_m=2\pi L_y\sqrt{\frac{3}{2}}\sqrt{B}=2\pi\sqrt{\frac{3\sigma}{(\frac{\mu_{1}}{k_{1}}-\frac{\mu_{2}}{k_{2}})V}}
 \label{lambda}
\end{equation}
presents the maximum growth rate; however, all perturbations with $\lambda>\lambda_m/\sqrt{3}$ are unstable. In the case of a Hele-Shaw cell,
with $\mu_2<<\mu_1$, $\lambda_m\approx \pi b /\sqrt{\text{Ca}}$, where $\text{Ca}=\mu V / \sigma$ is the capillary number.

Turning our attention to the ILGS, with the red fluid displacing the blue fluid, we observe similar regimes to those discussed above and the parameters $A$, $B$ and
$\lambda_m$ can be written as
\begin{equation}
 A=\frac{N_{s,b}-N_{s,r}}{N_{s,b}+N_{s,r}};
\end{equation}
\begin{equation}
  B=\frac{2\sigma}{L_y^2}\frac{1}{1.95(N_{s,b}-N_{s,r})\rho V};
\end{equation}
\begin{equation}
 \lambda_m=2\pi L_y\sqrt{\frac{3}{2}}\sqrt{B}=2\pi\sqrt{\frac{3\sigma}{1.95(N_{s,b}-N_{s,r})\rho V}},
\end{equation}
where we have used $(\mu/k)=\alpha_s\rho$, with $\alpha_s=1.95N_s$, in Eqs. (\ref{Aparameter}), (\ref{Bparameter}) and (\ref{lambda});
moreover the subscripts $r$ and $b$ refer to the red and blue fluids, respectively.

First, if the scattering rate of the red fluid is higher than that of the blue fluid, the interface is flat as in (i) above. If the red fluid has
a lower scattering rate, we observe the three situations exemplified in Figs. \ref{ST1}, \ref{ST2} and \ref{ST3}, which are obtained in a lattice
under the following conditions: the first and last lines are made of solid sites; the density of both fluids is 3.5 particles per site, periodic
boundary conditions are applied in the horizontal direction with the color of the particle changing when it crosses the boundary sites, as made in \cite{Rothman_book} to
simulate a fluid invasion process. The viscosity contrast for the three simulations is $A=0.78$. 
For a finite forcing rate, the velocity is not constant and depends on the fraction of each fluid;
its average is calculated by considering the time needed to the most advanced point of the interface to meet the right boundary of the lattice.
In particular, in Fig. \ref{ST1} we show a simulation with a low forcing rate (=0.0001) and an average velocity
$V\approx 4.9\times 10^{-3}$ presenting the formation of one finger, which is our
version of the so-called Saffman-Taylor instability \cite{RevModPhys.58}. The value of $\lambda_m\approx400$ implies $L_y/\lambda_m\approx 1.1$,
in agreement with the simulation. For a higher forcing rate (=0.0005), and an average velocity $V\approx 2.2\times10^{-2}$, we observe in
Figs. \ref{ST2} and \ref{ST3} a multifinger dynamics for the interface between the two lattice gases, as in regime (iii) above, and in
excellent agreement with the observed two-fluid dynamics in a Hele-Shaw cell \cite{PhysRevA.33.794,*jfmlib,*PhysRevA.36.1894,*tabeling1989experimental}
for the two-finger (Fig. \ref{ST2}) and multifinger (Fig. \ref{ST3}) cases. Indeed, in these two cases, $\lambda_m\approx190$ implying $L_y/\lambda_m\approx 2.3$ ($\approx 9.1$) for the simulation presented in Fig. \ref{ST2} (Fig. \ref{ST3}). Moreover, the agreement between our predictions for the
number of fingers, shape and time evolution, as shown in Figs. \ref{ST1}, \ref{ST2} and \ref{ST3}, is remarkable \cite{maher,PhysRevA.33.794,*jfmlib,*PhysRevA.36.1894,*tabeling1989experimental,RevModPhys.58}. We also emphasize that the finger dynamics is quite complex; indeed,
as time increases, some fingers can give rise to both bubble formation and finger amalgamation [see Figs. \ref{ST3} (e)-(f)]. 
The mentioned agreement is, in fact, the real reliability test of our approach.

While we do not provide a rigorous mapping between the immiscible two-fluid flow in a Hele-Shaw cell and
the ILGS model, the simulations discussed above suggest the close relation between the two systems. 

\subsection{Interface dynamics with quenched disorder}
\begin{figure}
\includegraphics[width=.4\textwidth]{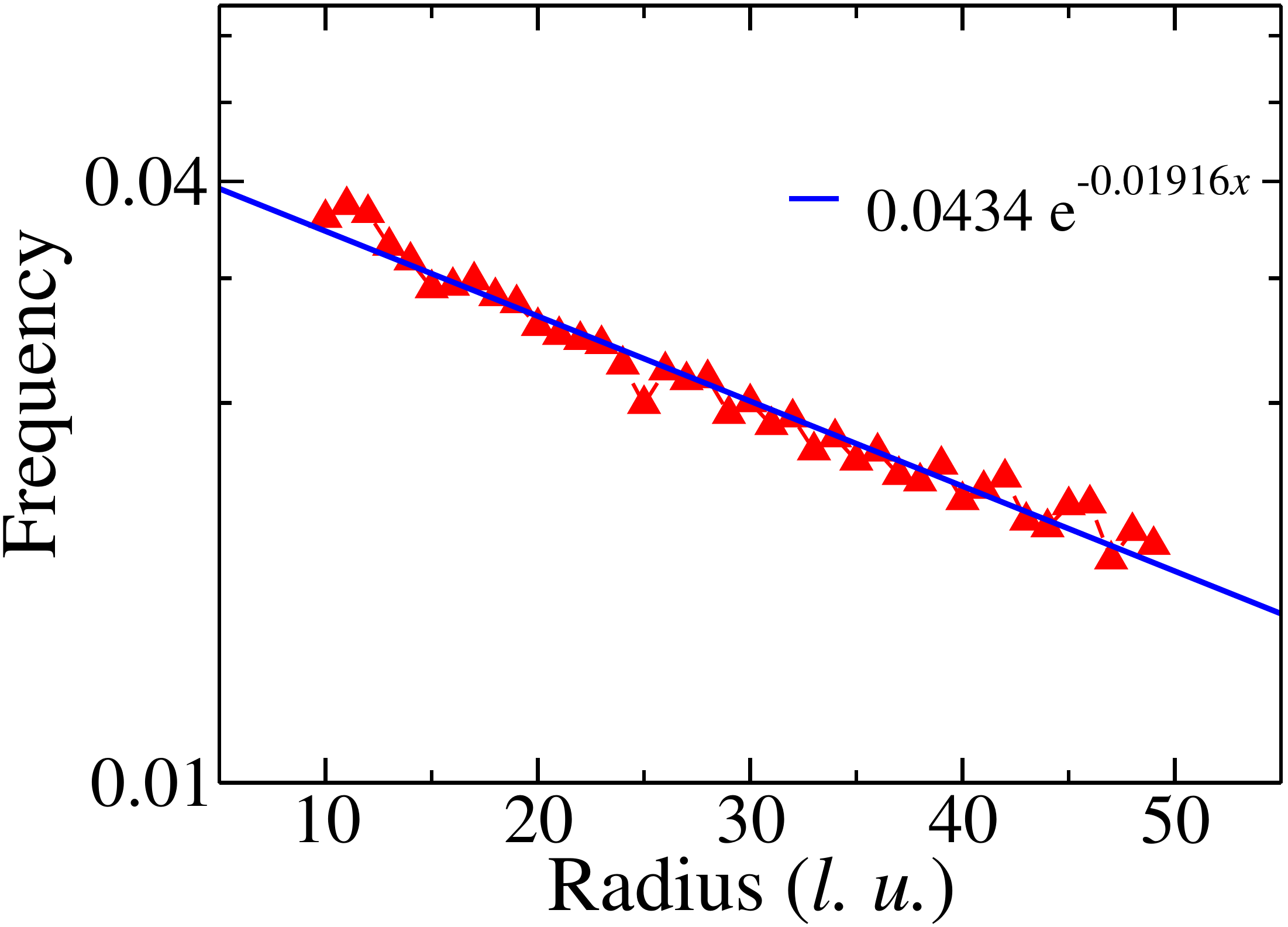}% Here is how to import EPS art
\caption{(Color online) Distribution of beads radii. The data is obtained averaging each frequency trough 100
realization of initial configuration and is well fitted by a exponential distribution.}
\label{radii}
\end{figure}
\begin{figure}
\includegraphics[width=.45\textwidth]{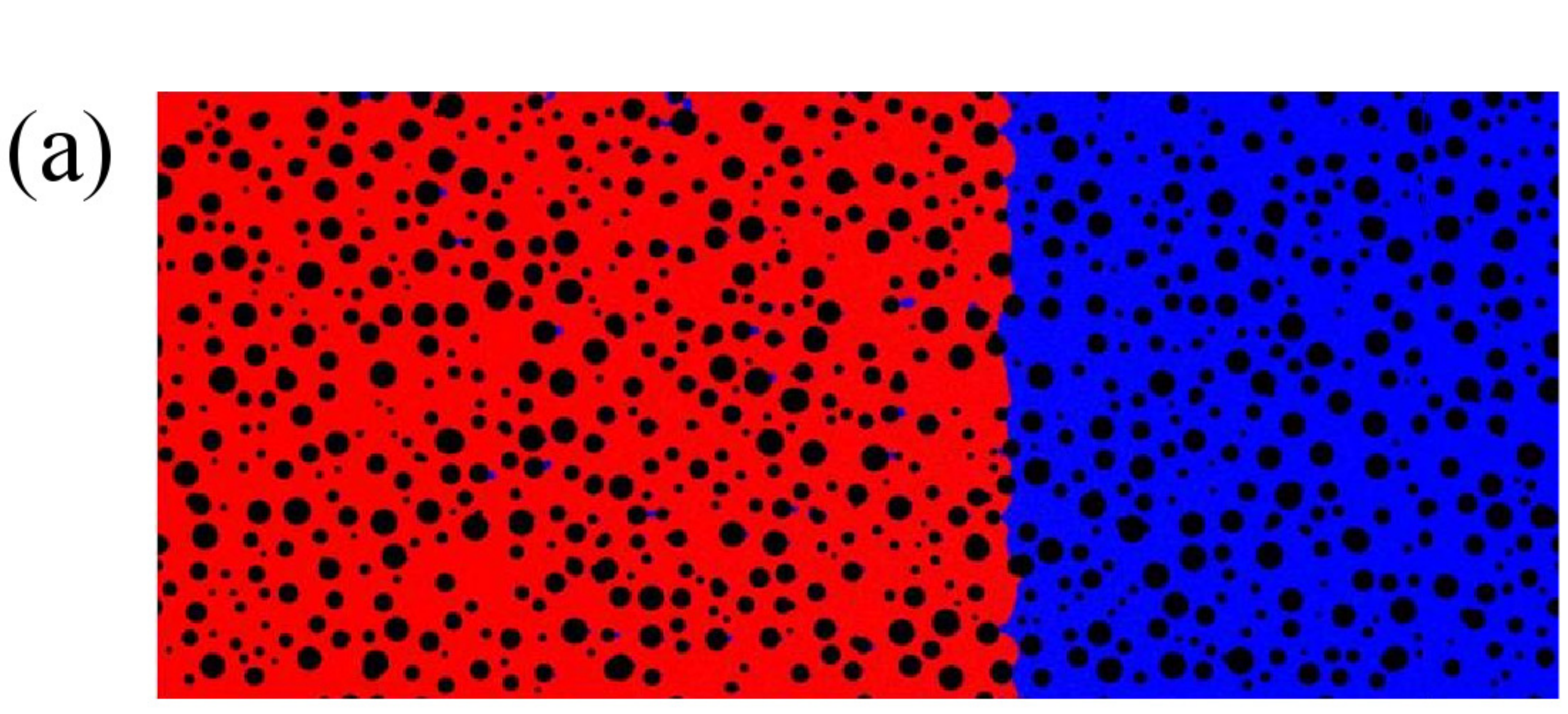}\newline
\includegraphics[width=.45\textwidth]{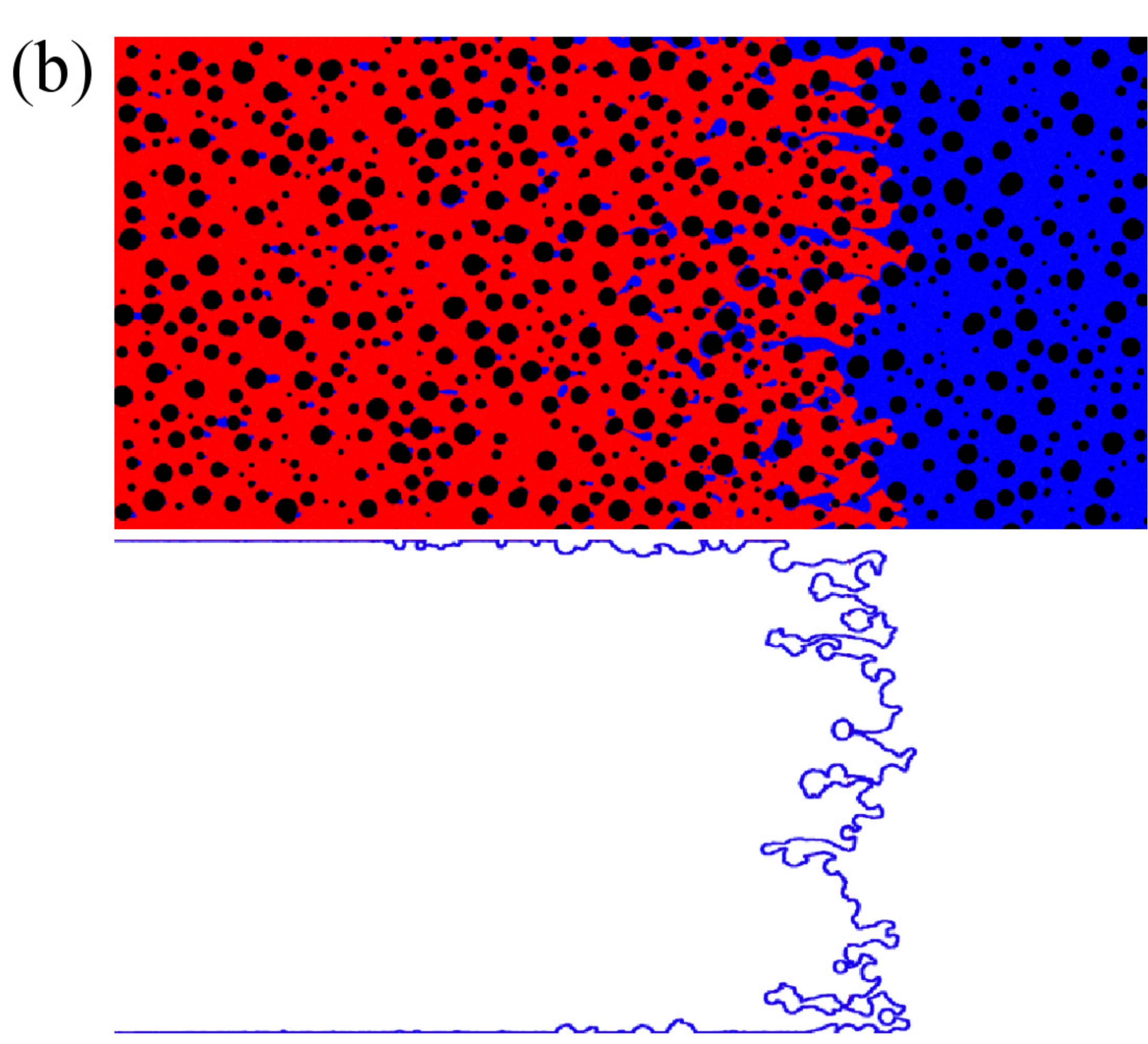}\newline
\includegraphics[width=.45\textwidth]{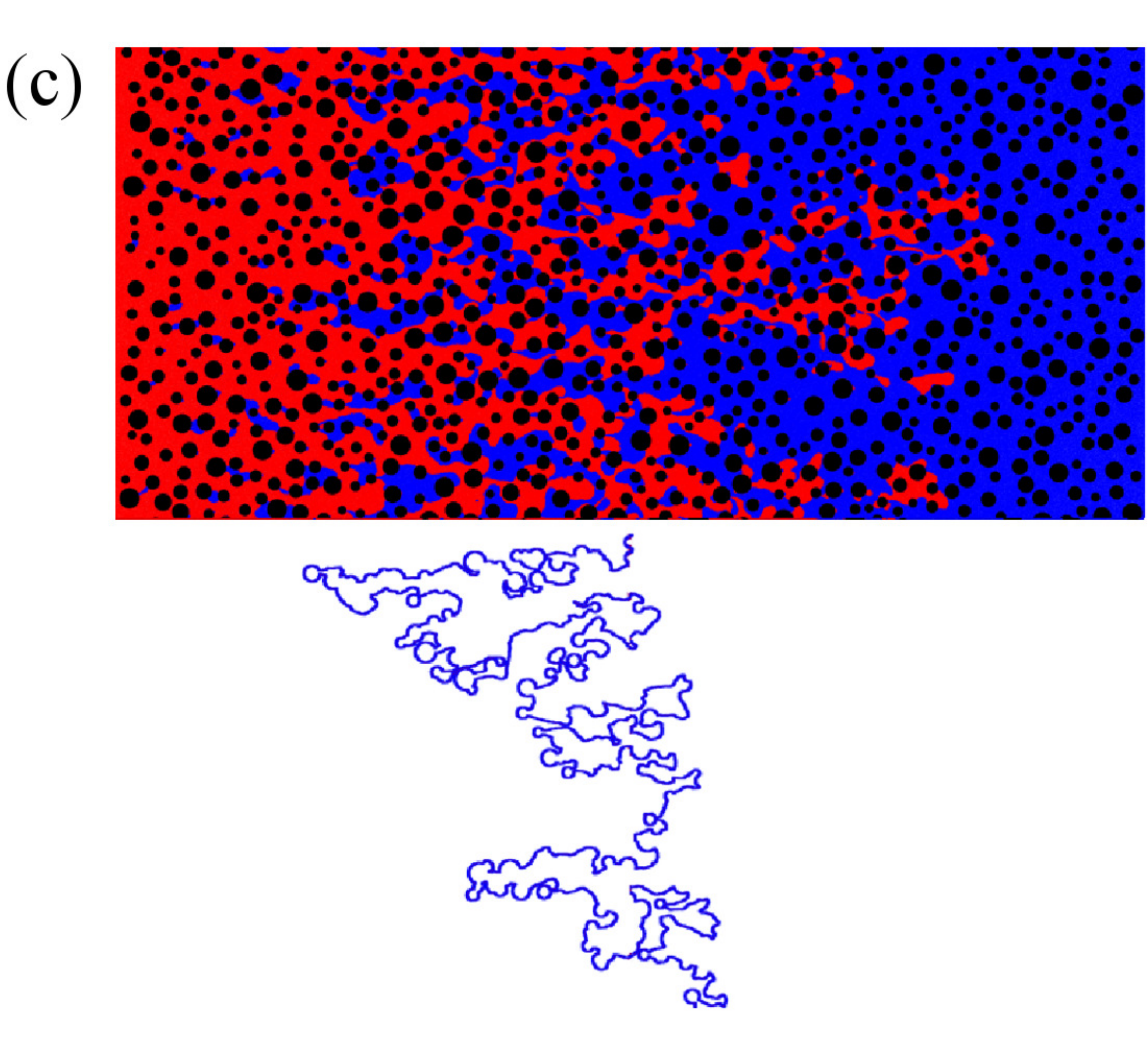}
\caption{(Color online) Snapshots from three simulations with distinct parameters of the displacement of the blue fluid
by the red fluid in a disordered medium, the scattering rate of the red (light gray) [blue] fluid is 0.008 [0.001] in (a),
0.001 [0.001] in (b) and 0.001 [0.008] in (c). In (b) and (c) just below the image of the snapshot we present the
interface between the two fluids. The size of the systems is 5000 x 2500 $\sqrt{3}/2$ and the
forcing rate is 3.0 x 10$^{-4}$. In (b) the sites at the first and the last horizontal lines are solids.}
\label{simulacaoFoto}
\end{figure}
The dynamics of the interface between two fluids were observed in a great variety of experiments in Hele-Shaw cells with some kind
of disorder \cite{vadose,PhysRevE.80.036308,lenormand,PhysRevLett.54.2226,PhysRevE.55.2969,PhysRevE.66.05,Barabasi}. The disorder can be introduced by the spatial variation of the gap between
the plates of the cell \cite{aurora} or by the introduction of randomly distributed circular obstacles \cite{rubio,PhysRevLett.71.2074}. In
both cases the dynamics of the interface may obey scaling laws, but the values of the exponents are usually distinct \cite{aurora,PhysRevE.55.2969}.
Discarding the effect of gravity, the interface is unstable to finger formation if the defending fluid is the higher viscous one, while
viscous forces stabilize the interface in the opposite case. In the unstable situation, the finger patterns can be controlled predominately by capillary
forces (acting in the scale of the pores), or by viscous forces, in which case crossover length scales between the pole-scale structure and macroscopic
scales can be defined \cite{vadose}.

Inspired by these experiments, we consider now the simulation of ILGS model for the displacement of the blue fluid by the
red fluid in a system with a random distribution of circular beads (see Fig. \ref{simulacaoFoto}). We use no-slip boundary conditions at solid sites,
periodic boundary conditions in the vertical and horizontal directions, but in the horizontal direction the color of the particle changes when it crosses the boundary sites. The average density of
particles is equal for the two fluids and has the same value for all simulations:
3.5 particles per site. We do not consider effects of wettability in the system, although it is possible to introduce it in the cellular automata model
\cite{Rothman_book}. Thus, in the unstable situation studied below, the fingers are controlled by viscous forces at large scales
and capillary effects at small scales.

The bead-filling algorithm is as follows: first, we choose the total amount of sites to be solid sites,
then we choose randomly (with an uniform distribution) the position of the bead center and its radius (between 20 and 50 lattice units).
If the bead is sufficiently distant from any other solid site (20 lattice units is chosen),
the new bead is accepted and the total amount of solid sites increases. This process is iterated
until the pre-defined total fraction of solid sites is reached or exceeded. This algorithm leads to the exponential distribution of radii
shown in Fig. \ref{radii}, which is well fitted by the Poisson distribution and implies an average radius equal to 25 lattice units
and an average porosity (the void fraction of the system) equal to 0.81.

\begin{figure}[!htb]
\includegraphics*[width=0.4\textwidth]{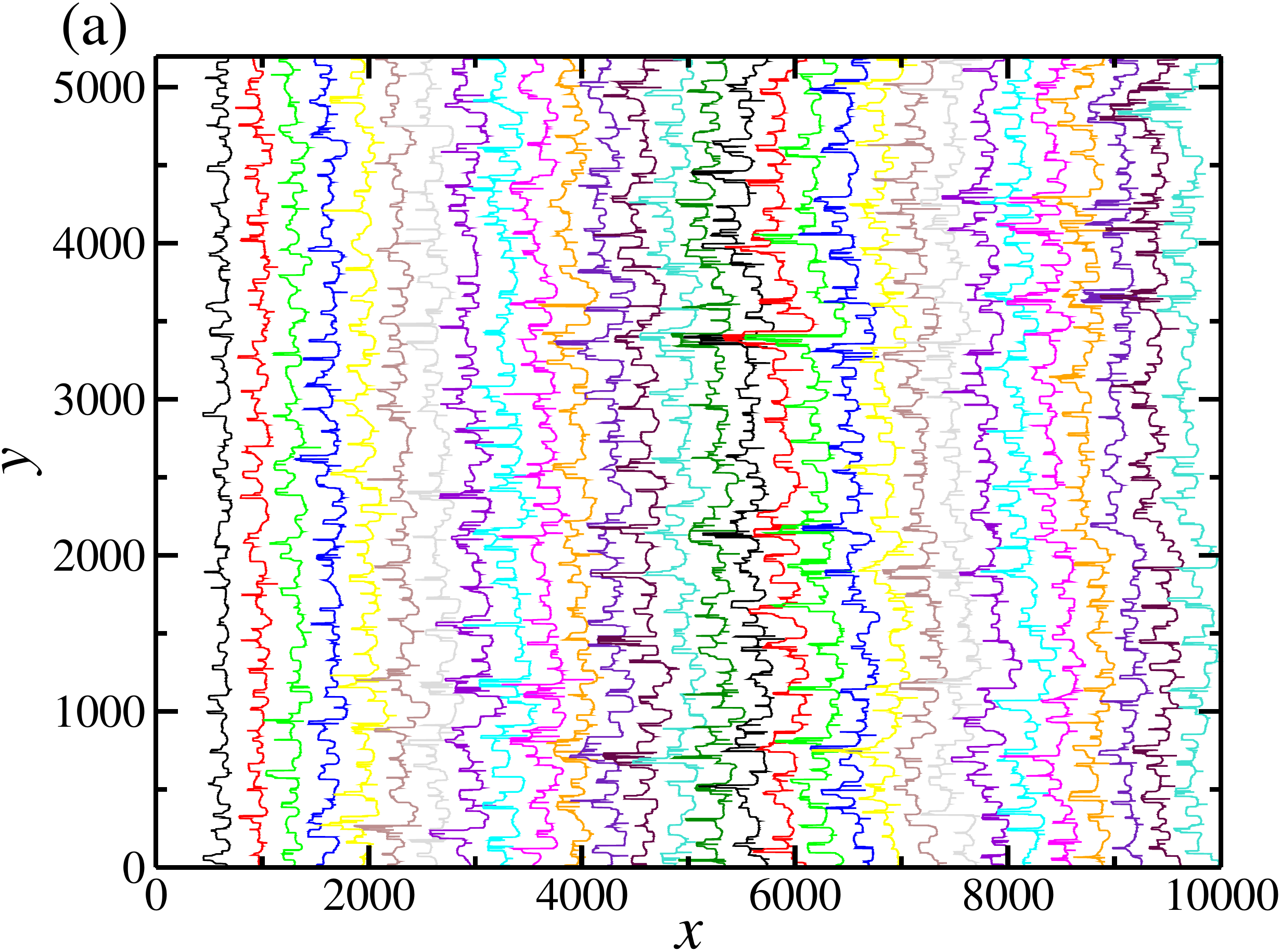}
\includegraphics*[width=0.4\textwidth]{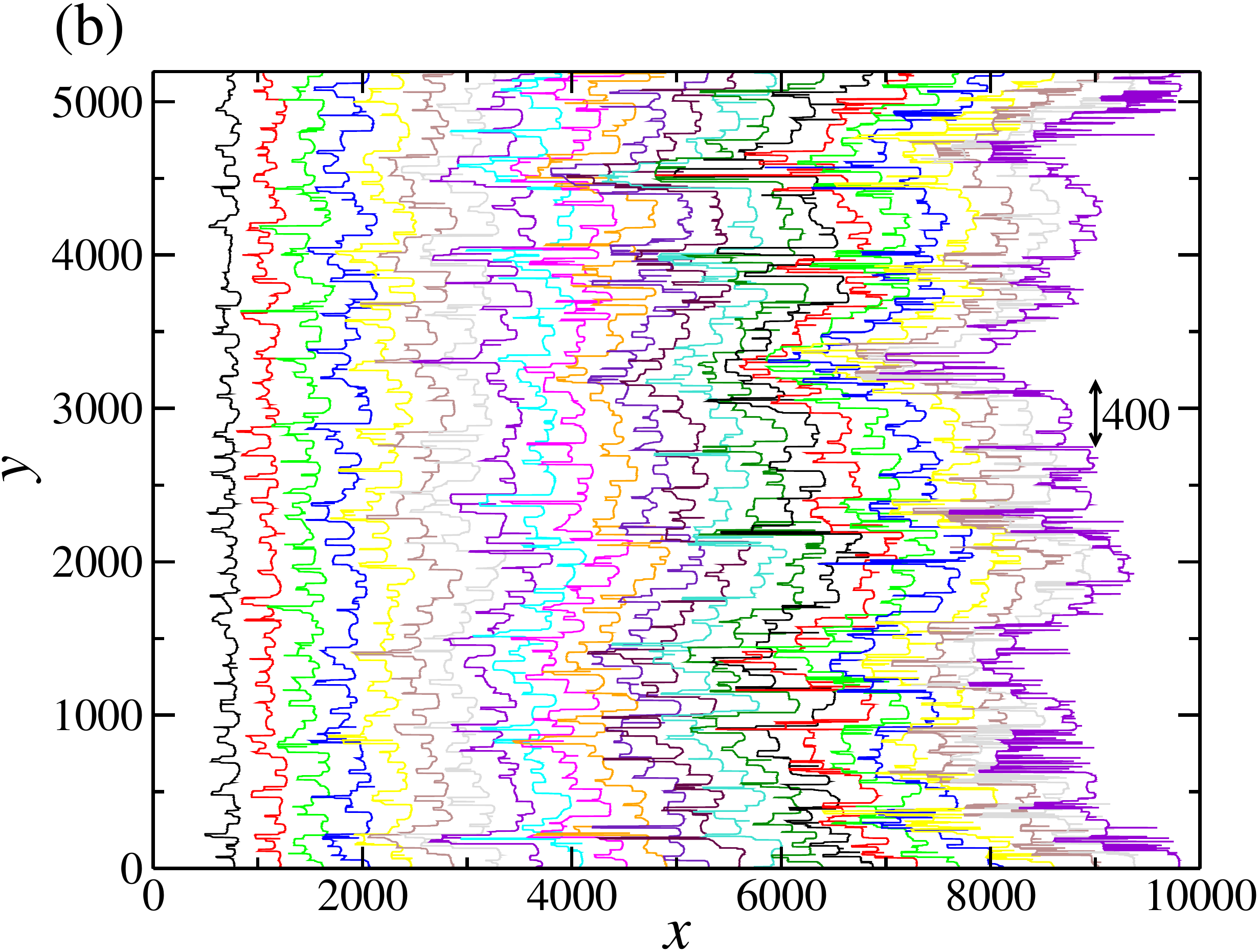}
\caption{(Color online) Height function $h(y,t)$ for systems with size 10000 x 6000 $\sqrt{3}/2$. In (a) the scattering rate of the red (light gray) and blue fluids are 0.001 (Case I), while in (b) the scattering rate of the red (light gray) fluid is 0.001 and the scattering rate of
the blue fluid is 0.002 (Case II). Time increases from left to right and each figure represents one realization of the
respective case.}
\label{ht}
\end{figure}

In Fig. \ref{simulacaoFoto} we present snapshots of the displacement of a blue fluid by the red fluid in a disordered medium in three regimes,
with the same forcing rate: 3.0 x 10$^{-4}$. In Fig. \ref{simulacaoFoto}(a) the scattering rate of the red fluid is higher
than the scattering rate of the blue fluid and the stable interface is flat in the void regions; in Fig. \ref{simulacaoFoto}(b) the scattering rate
of the two fluids are equal and the interface remains compact; while in Fig. \ref{simulacaoFoto}(c) the defending fluid has
a higher scattering rate compared to the invading, the interface displays a dendritic pattern, as expected
from Saffman-Taylor instability, and shows bubble production.

We shall now discuss in detail the interface dynamics in the following cases:
\begin{itemize}
\item\textbf{Case I}: blue and red fluids have the same scattering rate (0.001);
\item\textbf{Case II}: the scattering rate of the blue fluid is 0.002, while the scattering rate of the red fluid is 0.001.
\end{itemize}
A single-valued interface \textit{height function} $h(y,t)$ is defined as the horizontal position (with $h$ increasing from left to right)
of the rightmost red site in the horizontal line at $y$ for the time step $t$.
In Case I a constant forcing rate of
3 x 10$^{-4}$ is applied,
while in Case II a variable forcing rate ranging from 3 x 10$^{-4}$ to 5.341 x 10$^{-4}$ is used,
depending on the fraction of the system occupied by the red fluid.
The height functions so produced are presented in Fig. \ref{ht} for the two cases studied.

\begin{figure}[!htb]
\includegraphics*[width=0.4\textwidth]{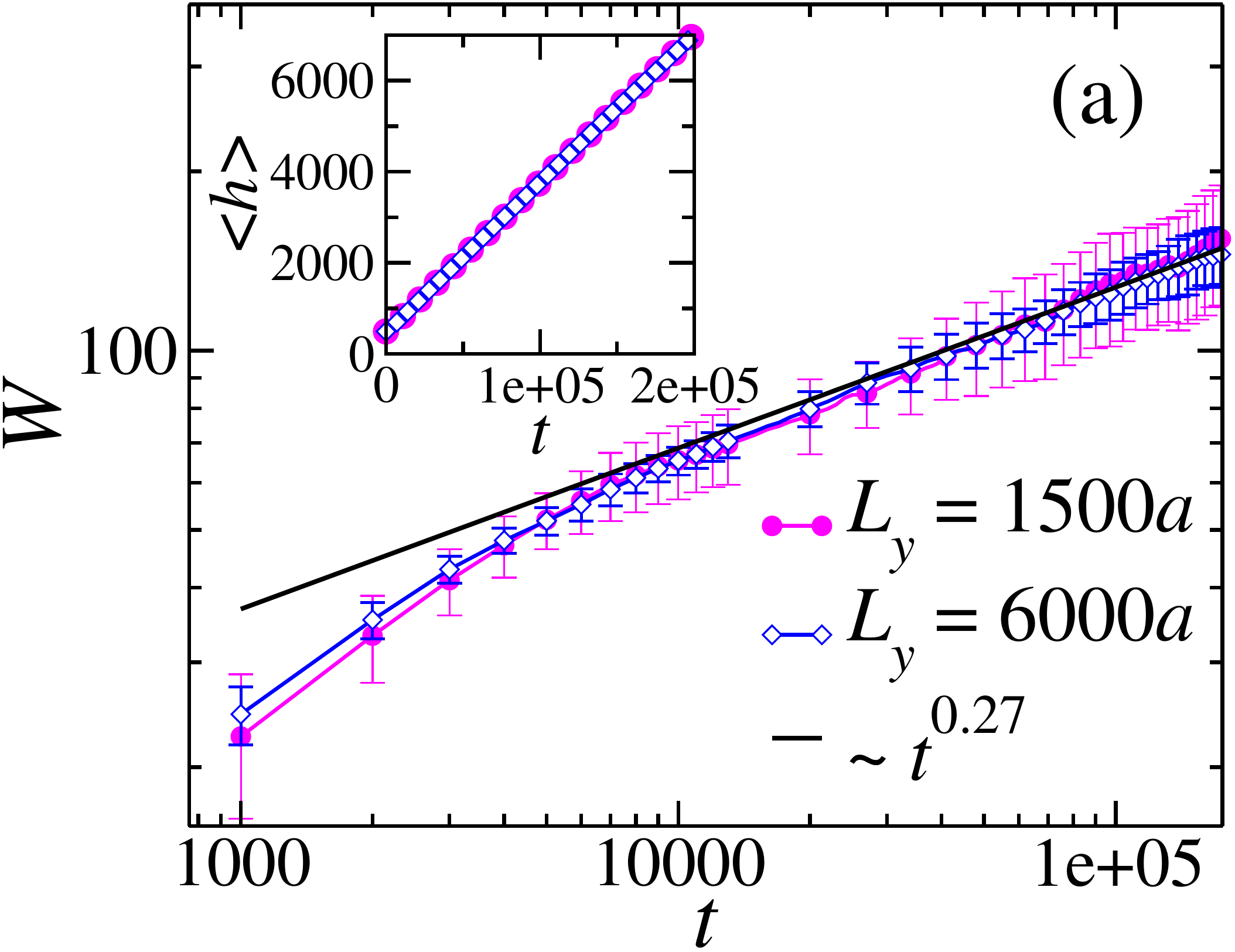}
\includegraphics*[width=0.4\textwidth]{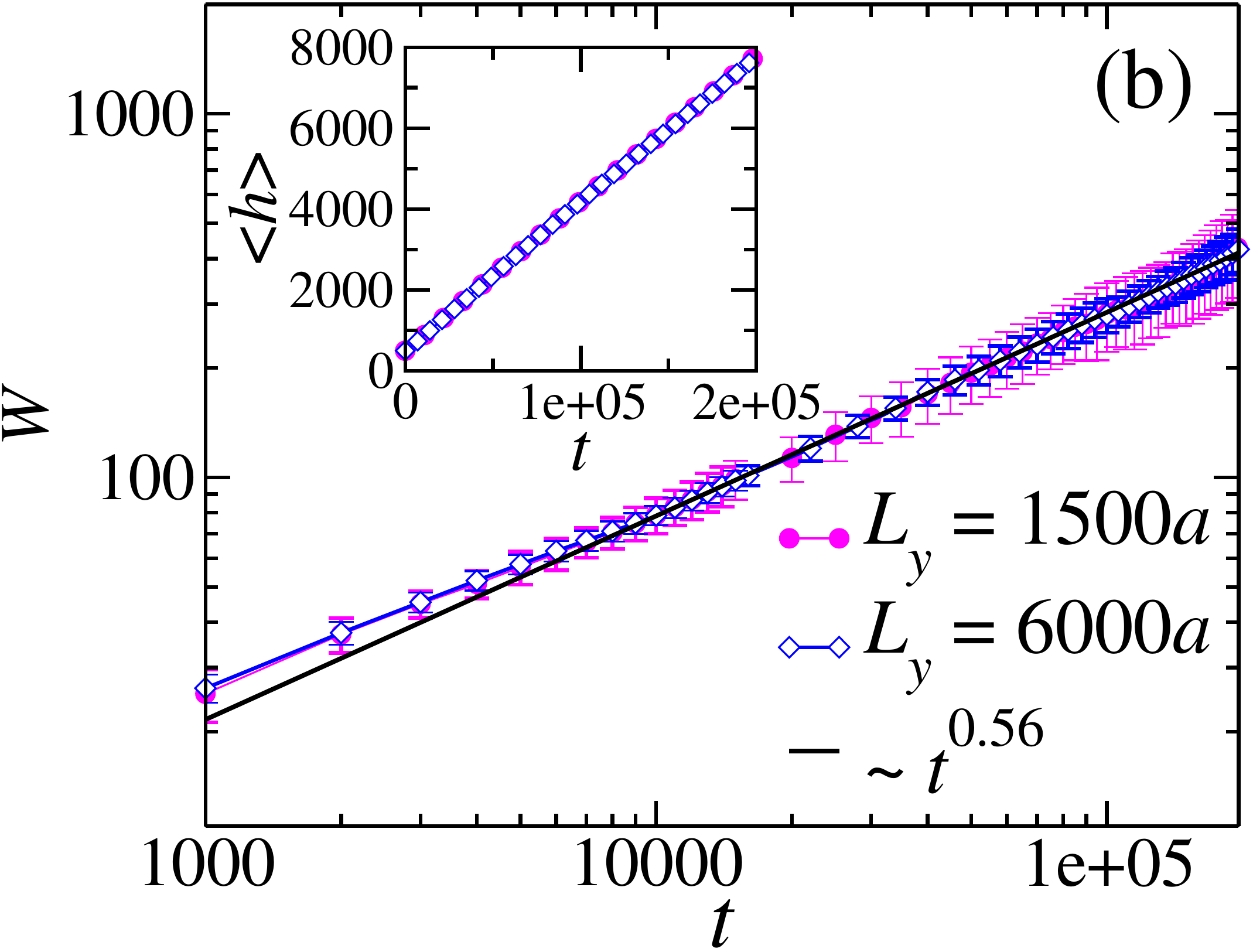}
\caption{(Color online) Interface width $W$ as a function of time for (a) Case I and (b) Case II for the indicated vertical dimensions, with $a=\sqrt{3}/2$, solid lines gives the best power-law function that best fits the last points of the curve for the largest system size.
Insets: average value of the interface height as a function of time.}
\label{W}
\end{figure}

The interface dynamics can be characterized by the time dependence of its width \cite{Barabasi}:
\begin{equation}
 W(L_y,t)=\sqrt{\langle [h(y,t)-\bar{h}(t)]^2\rangle},
\end{equation}
and the spatial dependence of the equal time correlation function:
\begin{equation}
 C(l,t)=\sqrt{\langle (h(y,t)-h(y^\prime,t))^2\rangle},\text{ for $l=|y-y^\prime|$},
\end{equation}
where angular brackets denote averages on $y$ and on different realizations of disorder (our data were obtained from simulations of a minimum of 50 and a
maximum of 100 disorder realizations), while $\bar{h}(t)$ denotes
the average value of $h$ along $y$ for a given disorder realization in time step $t$.
It is expected \cite{Barabasi} that $W$ and $C$ presents, respectively, a power-law behavior for short time and length scales:
\begin{equation}
 W(L_y,t)\sim t^\beta,\text{ for $t \ll t_\times$;}
\end{equation}
\begin{equation}
 C(l,t)\sim l^\alpha,\text{ for $l\ll \xi$,}
\end{equation}
where $\beta$ is the growth exponent, $\alpha$ is the roughness exponent, $t_\times\sim L^z$ defines the crossover
time above which the interface width saturates, where $z$ is the dynamic exponent, and $\xi$ is the correlation length parallel to the interface. 
The crossover time $t_\times$ is such that for $t\gtrsim t_\times$ the interface width saturates in a value
\begin{equation}
W_{sat}\sim {L_{y}}^\alpha.
\end{equation}
It is worth mentioning that the ILG model presents an intrinsic fluctuating interface with a very short width and for system sizes with $L_y\lesssim 64$,
the values $\alpha= 1/2$, $\beta= 1/3$, and $z=\alpha/\beta=3/2$ were estimated \cite{rothf,*PhysRevE.53.16}, which are the exponents
of the Kardar-Parisi-Zhang (KPZ) \cite{PhysRevLett.56} continuum model. However, for system sizes
with $L_y\gtrsim 64$, logarithmic corrections were observed \cite{rothf,*PhysRevE.53.16} in the scaling law for the saturation width.
On the other hand, under a Saffman-Taylor instability, the interface width is not expected to saturate.

In Fig. \ref{W} we present the growth of $W$ and the average height $<h>$ (see inset) as a function of time for the two cases studied.
First, we notice that $<h>$ is a linear function of time, as shown in the inset, and the velocity of the interface is constant, as mentioned above,
with a value of 3.3 x 10$^{-2}$ lattice units per time step in Case I and 3.6 x 10$^{-2}$
lattice units per time step in Case II. The interface width exhibits two distinct behaviors with a crossover at $t\sim 10000$, and $W \approx 50$, which is approximately the average
bead diameter: (i) for initial times, we find $\beta\approx 0.53$ and $\beta\approx 0.51$ for cases I and II, respectively; on the other hand, discarding 
the initial transient behavior, the exponent $\beta$ is found to be $\beta=0.27\pm0.07$ for Case I and $\beta=0.56\pm 0.07$ for Case II.
In Case I, for the large lattice sizes used ($L_y=1500\sqrt{3}/2$\text{ and }$6000\sqrt{3}/2$) and time interval of observation, i. e., 
($1\times10^3$ - $2\times 10^5$) time steps, we have not witness any sign of width saturation, which may occur in Hele-Shaw experiments with 
quenched disorder \cite{PhysRevE.55.2969}. 

\begin{figure}[!htb]
\includegraphics*[width=0.4\textwidth]{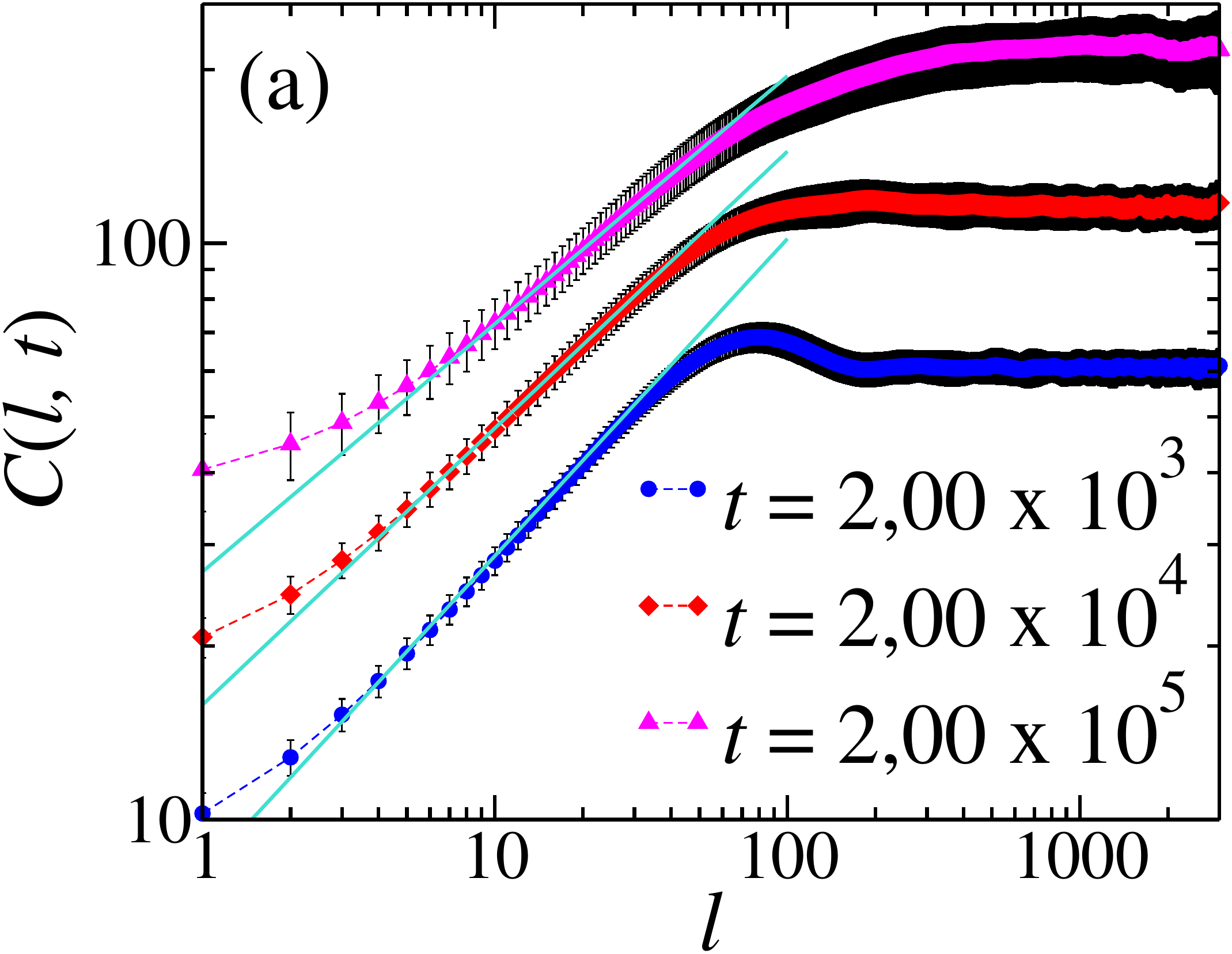}
\includegraphics*[width=0.4\textwidth]{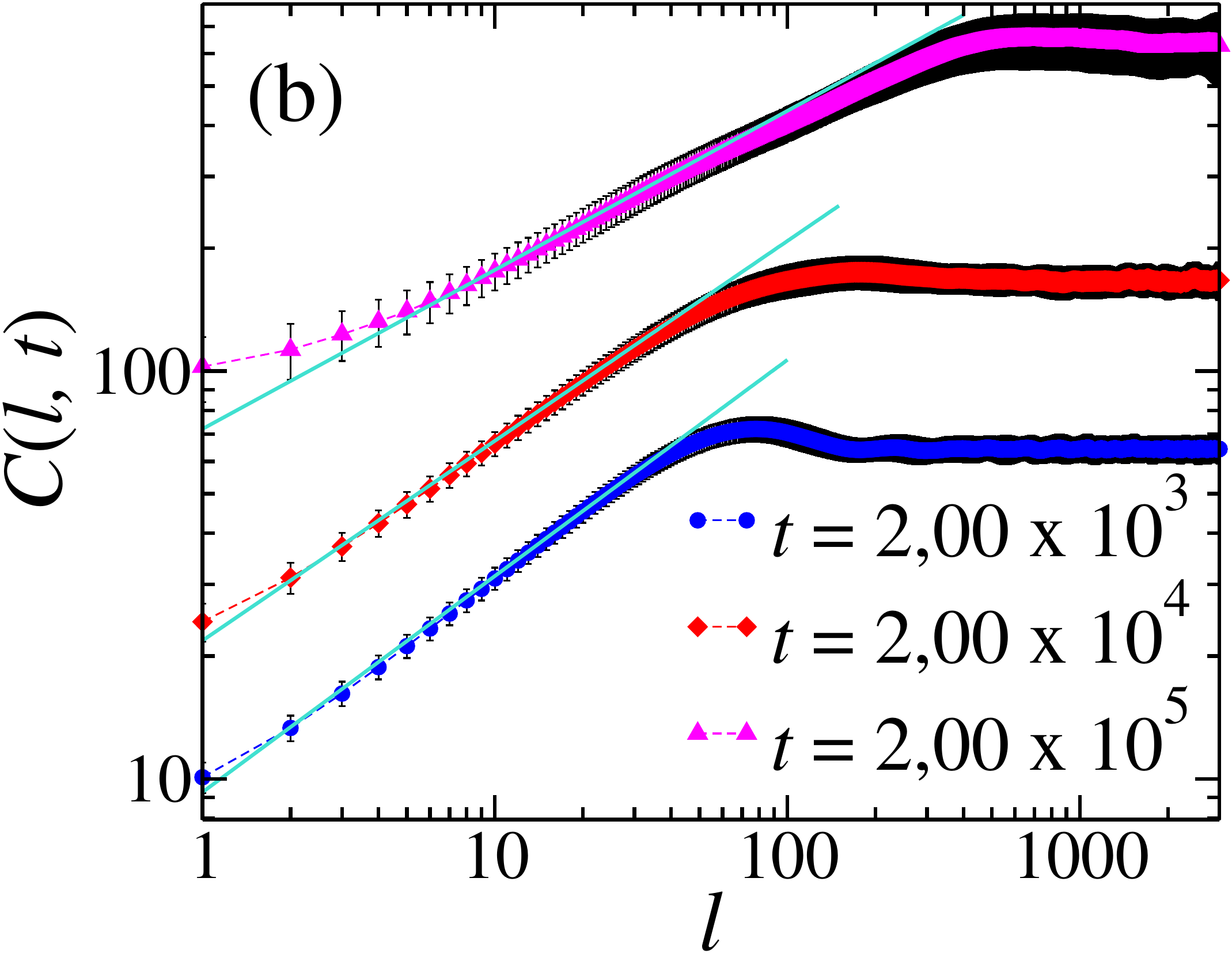}
\includegraphics*[width=0.4\textwidth]{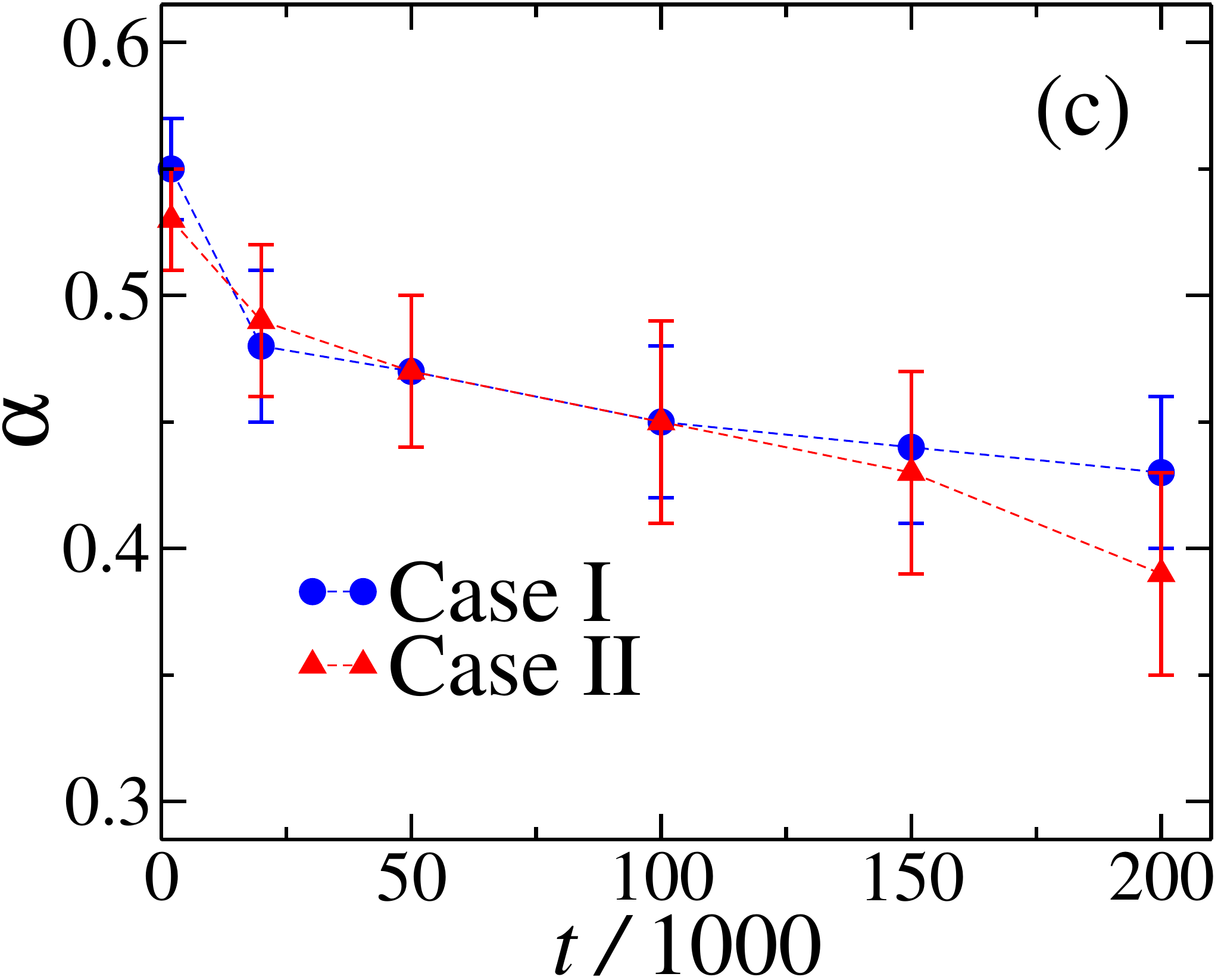}
\caption{(Color online) Equal time correlation function $C(l,t)$ as a function of distance $l$ for the indicated values of $t$ in (a) Case I and (b) Case II. (c) Roughness exponent $\alpha$ as a function of time for the two cases. Simulations are performed in a system of size 10000 x 6000 $\sqrt{3}/2$.}
\label{Corr}
\end{figure}

In Fig. \ref{Corr} we present the equal-time correlation functions for the two cases at three values of time: $t_1=2000, t_2=10t_1\text{ and }t_3=100t_1$.
The curves present an upturn for short values of $l$, which is possibly related to the short scales fluctuations of the ILG model and a power-law
behavior for intermediate values of $l$ before saturation for $l>100$; this length scale is higher for higher $t$ and
the value of $W$ also increases with time. The difference in the density of scatterers in Case II implies a higher value for the the correlation at saturation and
for the length scale associated with the crossover to this regime. We also calculate the time-dependent value of the roughness exponent
$\alpha$ for $10\leq l \leq 30$, which is shown in Fig. \ref{Corr} (c). We see that $\alpha$ decreases
slowly with time, starting from $\alpha\approx 0.5$, and reaches the lower average value in Case II in the longer time
studied: $\alpha=0.39\pm0.04$ at $t=200000$.

\subsection{Universality classes of the roughening process}
In Case I, the estimated values of the exponents: $\beta=0.27 \pm 0.07$ and $\alpha=(0.48\text{ - }0.43)\pm 0.03$ for $t\geq 2\times 10^4$, are very near the 
respective values for the discrete random deposition (RD) model with surface relaxation \cite{RDBwS} and the solution of the continuum
Edwards-Wilkinson (EW) equation \cite{Edwards08051982}, which are in
the same universality class and have the following exponents \cite{Barabasi}: $\beta=0.25$ and $\alpha=0.5$.
The reasoning for this coincidence can be understood by noticing that in Case I the scattering rate of the
blue and red fluids are equal, so that the interface dynamics is governed only by the random bead distribution and the surface tension. The
effect of the beads distribution is to increase the width in an uncorrelated fashion (like the RD process), while surface tension softens the
interface (surface relaxation effect), thereby inducing its correlated character.
\begin{figure}[!htb]
\includegraphics*[width=0.5\textwidth]{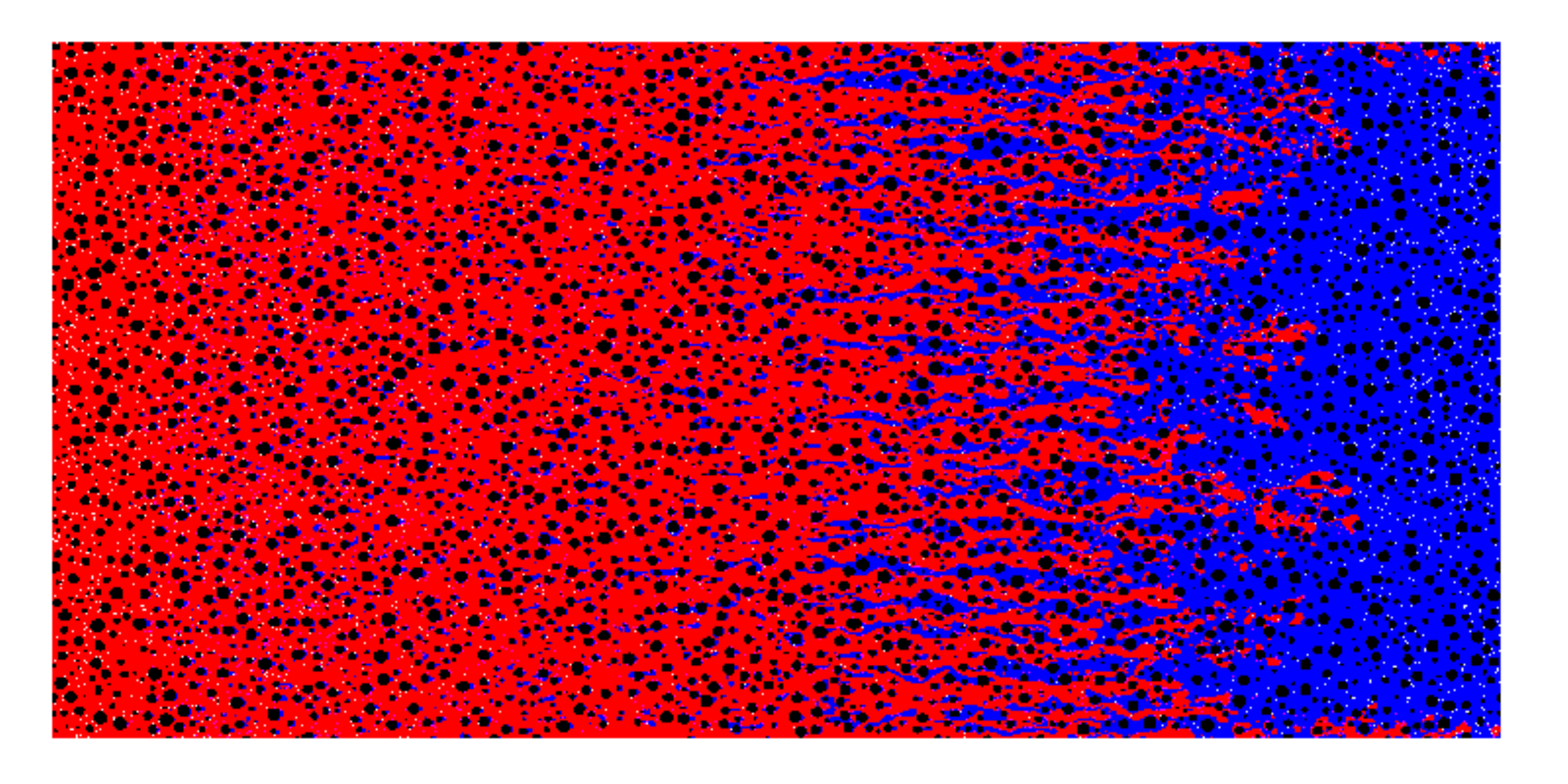}
\caption{(Color online) Same as in Fig. \ref{ht}(b): Snapshot of the system as the most advanced portion of the red (light gray) fluid meets the right boundary of the lattice.}
\label{lf}
\end{figure}

In Case II the interface is unstable to viscous finger formation, in which case the viscosity contrast in Eq. (\ref{Aparameter}) is $A=0.33$.
The simulation presents two important length scales, one of them
is the average diameter of the beads, $d \approx 50$, and the other is associated with the viscous finger dynamics, which is dominated by the perturbation
with wavelength $\lambda_m\approx 400$, i. e., $L_y/\lambda_m\approx 13$. Comparing Figs. \ref{ht}(a) and \ref{ht}(b), we notice a similar behavior for the
height functions of these two simulations for early times; while, for later times, the height function shown in Fig. \ref{ht}(b) exhibits an
oscillating pattern with wavelength $\sim \lambda_m$, superimposed by small spatial fluctuations with a characteristic length  $\sim d$ and caused
by the effective capillary pressure $\sim \sigma/d$ \cite{Barabasi}. As shown in Fig. \ref{lf}, for this time regime, the full integrity of the interface
is somewhat lost and the system displays a huge production of bubbles, thereby originating a large mixed zone and an interface height function associated
with the limits of this zone. In any event, the result of the quenched disorder manifests in a
time dependent roughness exponent, shown in Fig. \ref{Corr}(c); in fact, by mapping the interface fluctuations at a fixed time onto a fractional random
walk, with the $y$ direction playing the role of the random-walker time, the roughness exponent is identified, after a short transient time, with the Hurst exponent
of a subdiffusive fractional Brownian motion \cite{feder}, i. e., $\alpha(t) <0.5$. More precisely, 
we find for Case I: $\alpha(t)=$ (0.55 - 0.43), while for Case II: $\alpha(t)=(0.53\text{ - }0.39)$, with error bars indicated in Fig. \ref{Corr}(c). Therefore, the self-affine {\it local} fractal
dimension \cite{feder,mandelbrot}, $D_f(t)=2-\alpha(t)$, reads: $D_f(t)=$ (1.45 - 1.57) and $D_f(t)=$ (1.47 - 1.61) for cases I and II, respectively.  
In both cases, the early-time values of the fractal dimensions are consistent with the measured fractal dimension of 
immiscible displacement of viscosity-matched fluids in 2D porous media \cite{PhysRevE.55.2969}, which was suggested \cite{PhysRevE.55.2969}  
to be related to the fractal dimension of the external perimeter in invasion percolation \cite{PhysRevLett.67.584}. In addition, for Case II, 
it is quite satisfying to observe that our measured value of $D_f(t)$, at the breakthrough,  is consistent with the fractal dimension of the 
percolation cluster backbone \cite{PhysRevLett.53.1121,*law}, and also in agreement with fractal dimensions associated with: (i) fingering structure in a circular Hele-Shaw cell with quenched disorder and fluid invasion at the center 
of the cell for high Ca \cite{PhysRevLett.55.2688,*PhysRevA.36.318}; (ii) viscous fingering structure due to invasion of a less viscous fluid in a
more viscous one in a horizontal Hele-Shaw cell with quenched disorder for side-sizes $l$ (box counting) in the interval $L_y>l>a/\text{Ca}$ \cite{vadose}. 

\section{Summary and Conclusions}

In this work we presented a thorough study of a model based on lattice gas cellular automata techniques, including random dynamic scatterers. Our approach combines
the best features put forward in this class of models: microscopic kinetic rules, interface hydrodynamic fluctuations effects, and the presence of random dynamic
scattering sites. To the best of our knowledge, the combination of these techniques was not used in conjunction before the present work.
In fact, the main achievement of our work is to show that the dynamics of the proposed model is in very good agreement with the observed dynamics of immiscible
two-fluid flows in Hele-Shaw cells.

In order to fulfill this goal, we first showed that for a single fluid under a constant injection of momentum, the macroscopic Darcy's law equation is recovered, and the
ratio $k/\mu$ can be finely tuned through the model parameters. We then reported simulations of two distinct but related phenomena. (i) For two immiscible fluids with
the less viscous fluid displacing the more viscous one, the model exhibits
the Saffman-Taylor instability, in which case the number of fingers, shape, and time evolution is in very good agreement with the predictions from a
linear stability analysis and experimental observation of the interface dynamics.
We also mention that the finger dynamics is quite complex; indeed, as time increases, some fingers can give rise to both bubble formation and finger
amalgamation. (ii) Next, we studied the dynamics of the interface between immiscible fluids under quenched disorder, simulated by randomly distributed beads with
random diameters obeying the Poisson distribution. In the case of matched-viscosity fluids, the growth and roughness exponents
present values suggesting the same universality class of the
random deposition model with surface relaxation: $\beta=0.25$ and $\alpha=0.5$. In the case of the less viscous fluid displacing the more viscous one, we find $\beta\sim 0.5$,
and that the interface early-time dynamics is very similar to the former case; however, as time increases, viscous fingers develop with the subsequent production of
bubbles in the context of a complex dynamics, thereby giving rise to an increasing mixed zone. Finally, we emphasize that the time-dependent behavior of the roughness exponents are quite similar
in both cases, with a monotonic decreasing from their starting value: $\alpha\sim 0.5$; indeed,
after a short transient time, this exponent can be identified  with the Hurst exponent of a subdiffusive fractional Brownian motion. 
In particular, after the transient behavior, the early-time fractal dimension of the interfaces are consistent with
the measured fractal dimension of immiscible displacement of viscosity-matched fluids in porous media, and with the fractal dimension of the external
perimeter in invasion percolation.
Further, in the case of a less viscous fluid displacing a more viscous one, our measured value of the fractal dimension at the breakthrough is consistent with
the fractal dimension of the percolation cluster backbone, and in agreement with experimental results  
in Hele-Shaw cells with quenched disorder. 

The above summary clearly supports the conclusion that the two above-mentioned phenomena were successfully simulated by our model, 
with insights and interesting features unveiled. We thus expect that the reported results will stimulate further theoretical and experimental
studies of two-fluid flow in porous media. 

\begin{acknowledgments}
We acknowledge useful discussions with Leonardo P. Viana, G. L. Vasconcelos, J. S. Andrade and L. S. Lucena. We also acknowledge the financial support
of the Brazilian institutions:  FINEP, CNPq, FACEPE and CAPES, through the programs PRONEX and CT-PETRO.
\end{acknowledgments}

\bibliographystyle{apsrev4-1}
\bibliography{artigo}

%merlin.mbs apsrev4-1.bst 2010-07-25 4.21a (PWD, AO, DPC) hacked
%Control: key (0)
%Control: author (72) initials jnrlst
%Control: editor formatted (1) identically to author
%Control: production of article title (-1) disabled
%Control: page (0) single
%Control: year (1) truncated
%Control: production of eprint (0) enabled
\begin{thebibliography}{121}%
\makeatletter
\providecommand \@ifxundefined [1]{%
 \@ifx{#1\undefined}
}%
\providecommand \@ifnum [1]{%
 \ifnum #1\expandafter \@firstoftwo
 \else \expandafter \@secondoftwo
 \fi
}%
\providecommand \@ifx [1]{%
 \ifx #1\expandafter \@firstoftwo
 \else \expandafter \@secondoftwo
 \fi
}%
\providecommand \natexlab [1]{#1}%
\providecommand \enquote  [1]{``#1''}%
\providecommand \bibnamefont  [1]{#1}%
\providecommand \bibfnamefont [1]{#1}%
\providecommand \citenamefont [1]{#1}%
\providecommand \href@noop [0]{\@secondoftwo}%
\providecommand \href [0]{\begingroup \@sanitize@url \@href}%
\providecommand \@href[1]{\@@startlink{#1}\@@href}%
\providecommand \@@href[1]{\endgroup#1\@@endlink}%
\providecommand \@sanitize@url [0]{\catcode `\\12\catcode `\$12\catcode
  `\&12\catcode `\#12\catcode `\^12\catcode `\_12\catcode `\%12\relax}%
\providecommand \@@startlink[1]{}%
\providecommand \@@endlink[0]{}%
\providecommand \url  [0]{\begingroup\@sanitize@url \@url }%
\providecommand \@url [1]{\endgroup\@href {#1}{\urlprefix }}%
\providecommand \urlprefix  [0]{URL }%
\providecommand \Eprint [0]{\href }%
\providecommand \doibase [0]{http://dx.doi.org/}%
\providecommand \selectlanguage [0]{\@gobble}%
\providecommand \bibinfo  [0]{\@secondoftwo}%
\providecommand \bibfield  [0]{\@secondoftwo}%
\providecommand \translation [1]{[#1]}%
\providecommand \BibitemOpen [0]{}%
\providecommand \bibitemStop [0]{}%
\providecommand \bibitemNoStop [0]{.\EOS\space}%
\providecommand \EOS [0]{\spacefactor3000\relax}%
\providecommand \BibitemShut  [1]{\csname bibitem#1\endcsname}%
\let\auto@bib@innerbib\@empty
%</preamble>
\bibitem [{\citenamefont {Landau}\ and\ \citenamefont
  {Lifshitz}(1959)}]{landau}%
  \BibitemOpen
  \bibfield  {author} {\bibinfo {author} {\bibfnamefont {L.}~\bibnamefont
  {Landau}}\ and\ \bibinfo {author} {\bibfnamefont {E.}~\bibnamefont
  {Lifshitz}},\ }\href@noop {} {\emph {\bibinfo {title} {Fluid Mechanics}}}\
  (\bibinfo  {publisher} {Pergamon Press, New York},\ \bibinfo {year}
  {1959})\BibitemShut {NoStop}%
\bibitem [{\citenamefont {Spurk}(1997)}]{spurk}%
  \BibitemOpen
  \bibfield  {author} {\bibinfo {author} {\bibfnamefont {J.}~\bibnamefont
  {Spurk}},\ }\href@noop {} {\emph {\bibinfo {title} {Fluid Mechanics}}}\
  (\bibinfo  {publisher} {Springer, Berlin},\ \bibinfo {year}
  {1997})\BibitemShut {NoStop}%
\bibitem [{\citenamefont {Sahimi}(1993)}]{Sahimi_Review}%
  \BibitemOpen
  \bibfield  {author} {\bibinfo {author} {\bibfnamefont {M.}~\bibnamefont
  {Sahimi}},\ }\href@noop {} {\bibfield  {journal} {\bibinfo  {journal} {Rev.\
  Mod. Phys.}\ }\textbf {\bibinfo {volume} {65}},\ \bibinfo {pages} {1393}
  (\bibinfo {year} {1993})}\BibitemShut {NoStop}%
\bibitem [{\citenamefont {Sahimi}(2012)}]{sahimi2012flow}%
  \BibitemOpen
  \bibfield  {author} {\bibinfo {author} {\bibfnamefont {M.}~\bibnamefont
  {Sahimi}},\ }\href@noop {} {\emph {\bibinfo {title} {Flow and Transport in
  Porous Media and Fractured Rock}}}\ (\bibinfo  {publisher} {Wiley-VCH,
  Weinheim},\ \bibinfo {year} {2012})\BibitemShut {NoStop}%
\bibitem [{\citenamefont {Bear}(1988)}]{bear1972}%
  \BibitemOpen
  \bibfield  {author} {\bibinfo {author} {\bibfnamefont {J.}~\bibnamefont
  {Bear}},\ }\href@noop {} {\emph {\bibinfo {title} {Dynamics of Fluids in
  Porous Media}}}\ (\bibinfo  {publisher} {Dover Publications, New York},\
  \bibinfo {year} {1988})\BibitemShut {NoStop}%
\bibitem [{\citenamefont {Kang}\ \emph {et~al.}(2002)\citenamefont {Kang},
  \citenamefont {Zhang},\ and\ \citenamefont {Chen}}]{PhysRevE.66.056307}%
  \BibitemOpen
  \bibfield  {author} {\bibinfo {author} {\bibfnamefont {Q.}~\bibnamefont
  {Kang}}, \bibinfo {author} {\bibfnamefont {D.}~\bibnamefont {Zhang}}, \ and\
  \bibinfo {author} {\bibfnamefont {S.}~\bibnamefont {Chen}},\ }\href {\doibase
  10.1103/PhysRevE.66.056307} {\bibfield  {journal} {\bibinfo  {journal} {Phys.
  Rev. E}\ }\textbf {\bibinfo {volume} {66}},\ \bibinfo {pages} {056307}
  (\bibinfo {year} {2002})}\BibitemShut {NoStop}%
\bibitem [{tar()}]{tarta}%
  \BibitemOpen
  \href@noop {} {}\bibinfo {note} {A. M. Tartakovsky, S. P. Neuman and R. J.
  Lenhard, Phys. Fluids \textbf{15}, 3331 (2003); D. M. Tartakovsky and C. L.
  Winter, SIAM J. App. Math. \textbf{61}, 1857 (2001)}\BibitemShut {NoStop}%
\bibitem [{\citenamefont {Tartakovsky}\ \emph {et~al.}(2008)\citenamefont
  {Tartakovsky}, \citenamefont {Tartakovsky},\ and\ \citenamefont
  {Meakin}}]{tartaprl}%
  \BibitemOpen
  \bibfield  {author} {\bibinfo {author} {\bibfnamefont {A.~M.}\ \bibnamefont
  {Tartakovsky}}, \bibinfo {author} {\bibfnamefont {D.~M.}\ \bibnamefont
  {Tartakovsky}}, \ and\ \bibinfo {author} {\bibfnamefont {P.}~\bibnamefont
  {Meakin}},\ }\href {\doibase 10.1103/PhysRevLett.101.044502} {\bibfield
  {journal} {\bibinfo  {journal} {Phys. Rev. Lett.}\ }\textbf {\bibinfo
  {volume} {101}},\ \bibinfo {pages} {044502} (\bibinfo {year}
  {2008})}\BibitemShut {NoStop}%
\bibitem [{\citenamefont {Homsy}(1987)}]{Homsy}%
  \BibitemOpen
  \bibfield  {author} {\bibinfo {author} {\bibfnamefont {G.~M.}\ \bibnamefont
  {Homsy}},\ }\href {\doibase 10.1146/annurev.fl.19.010187.001415} {\bibfield
  {journal} {\bibinfo  {journal} {Annu. Rev. Fluid Mech.}\ }\textbf {\bibinfo
  {volume} {19}},\ \bibinfo {pages} {271} (\bibinfo {year} {1987})}\BibitemShut
  {NoStop}%
\bibitem [{mis()}]{mishracomsol}%
  \BibitemOpen
  \href@noop {} {}\bibinfo {note} {S. Pramanik, G. L. Kulukuru, and M. Mishra,
  \textit{Proceedings of the COMSOL Conference} (Bangalore, 2012)}\BibitemShut
  {NoStop}%
\bibitem [{\citenamefont {Maes}\ \emph {et~al.}(2010)\citenamefont {Maes},
  \citenamefont {Rousseaux}, \citenamefont {Scheid}, \citenamefont {Mishra},
  \citenamefont {Colinet},\ and\ \citenamefont {Wit}}]{mishrapf}%
  \BibitemOpen
  \bibfield  {author} {\bibinfo {author} {\bibfnamefont {R.}~\bibnamefont
  {Maes}}, \bibinfo {author} {\bibfnamefont {G.}~\bibnamefont {Rousseaux}},
  \bibinfo {author} {\bibfnamefont {B.}~\bibnamefont {Scheid}}, \bibinfo
  {author} {\bibfnamefont {M.}~\bibnamefont {Mishra}}, \bibinfo {author}
  {\bibfnamefont {P.}~\bibnamefont {Colinet}}, \ and\ \bibinfo {author}
  {\bibfnamefont {A.~D.}\ \bibnamefont {Wit}},\ }\href {\doibase
  10.1063/1.3528039} {\bibfield  {journal} {\bibinfo  {journal} {Phys. Fluids}\
  }\textbf {\bibinfo {volume} {22}},\ \bibinfo {eid} {123104} (\bibinfo {year}
  {2010})}\BibitemShut {NoStop}%
\bibitem [{\citenamefont {Mishra}\ \emph {et~al.}(2008)\citenamefont {Mishra},
  \citenamefont {Martin},\ and\ \citenamefont {De~Wit}}]{mishrapre}%
  \BibitemOpen
  \bibfield  {author} {\bibinfo {author} {\bibfnamefont {M.}~\bibnamefont
  {Mishra}}, \bibinfo {author} {\bibfnamefont {M.}~\bibnamefont {Martin}}, \
  and\ \bibinfo {author} {\bibfnamefont {A.}~\bibnamefont {De~Wit}},\ }\href
  {\doibase 10.1103/PhysRevE.78.066306} {\bibfield  {journal} {\bibinfo
  {journal} {Phys. Rev. E}\ }\textbf {\bibinfo {volume} {78}},\ \bibinfo
  {pages} {066306} (\bibinfo {year} {2008})}\BibitemShut {NoStop}%
\bibitem [{\citenamefont {De~Wit}(2001)}]{dewit}%
  \BibitemOpen
  \bibfield  {author} {\bibinfo {author} {\bibfnamefont {A.}~\bibnamefont
  {De~Wit}},\ }\href {\doibase 10.1103/PhysRevLett.87.054502} {\bibfield
  {journal} {\bibinfo  {journal} {Phys. Rev. Lett.}\ }\textbf {\bibinfo
  {volume} {87}},\ \bibinfo {pages} {054502} (\bibinfo {year}
  {2001})}\BibitemShut {NoStop}%
\bibitem [{\citenamefont {Chen}\ \emph {et~al.}(1991)\citenamefont {Chen},
  \citenamefont {Diemer}, \citenamefont {Doolen}, \citenamefont {Eggert},
  \citenamefont {Fu}, \citenamefont {Gutman},\ and\ \citenamefont
  {Travis}}]{Chen199172}%
  \BibitemOpen
  \bibfield  {author} {\bibinfo {author} {\bibfnamefont {S.}~\bibnamefont
  {Chen}}, \bibinfo {author} {\bibfnamefont {K.}~\bibnamefont {Diemer}},
  \bibinfo {author} {\bibfnamefont {G.~D.}\ \bibnamefont {Doolen}}, \bibinfo
  {author} {\bibfnamefont {K.}~\bibnamefont {Eggert}}, \bibinfo {author}
  {\bibfnamefont {C.}~\bibnamefont {Fu}}, \bibinfo {author} {\bibfnamefont
  {S.}~\bibnamefont {Gutman}}, \ and\ \bibinfo {author} {\bibfnamefont {B.~J.}\
  \bibnamefont {Travis}},\ }\href {\doibase 10.1016/0167-2789(91)90281-D}
  {\bibfield  {journal} {\bibinfo  {journal} {Physica D}\ }\textbf {\bibinfo
  {volume} {47}},\ \bibinfo {pages} {72 } (\bibinfo {year} {1991})}\BibitemShut
  {NoStop}%
\bibitem [{\citenamefont {Andrade}\ \emph {et~al.}(1999)\citenamefont
  {Andrade}, \citenamefont {Costa}, \citenamefont {Almeida}, \citenamefont
  {Makse},\ and\ \citenamefont {Stanley}}]{soares99}%
  \BibitemOpen
  \bibfield  {author} {\bibinfo {author} {\bibfnamefont {J.~S.}\ \bibnamefont
  {Andrade}}, \bibinfo {author} {\bibfnamefont {U.~M.~S.}\ \bibnamefont
  {Costa}}, \bibinfo {author} {\bibfnamefont {M.~P.}\ \bibnamefont {Almeida}},
  \bibinfo {author} {\bibfnamefont {H.~A.}\ \bibnamefont {Makse}}, \ and\
  \bibinfo {author} {\bibfnamefont {H.~E.}\ \bibnamefont {Stanley}},\ }\href
  {\doibase 10.1103/PhysRevLett.82.5249} {\bibfield  {journal} {\bibinfo
  {journal} {Phys. Rev. Lett.}\ }\textbf {\bibinfo {volume} {82}},\ \bibinfo
  {pages} {5249} (\bibinfo {year} {1999})}\BibitemShut {NoStop}%
\bibitem [{\citenamefont {Morais}\ \emph {et~al.}(2009)\citenamefont {Morais},
  \citenamefont {Seybold}, \citenamefont {Herrmann},\ and\ \citenamefont
  {Andrade}}]{soares2009}%
  \BibitemOpen
  \bibfield  {author} {\bibinfo {author} {\bibfnamefont {A.~F.}\ \bibnamefont
  {Morais}}, \bibinfo {author} {\bibfnamefont {H.}~\bibnamefont {Seybold}},
  \bibinfo {author} {\bibfnamefont {H.~J.}\ \bibnamefont {Herrmann}}, \ and\
  \bibinfo {author} {\bibfnamefont {J.~S.}\ \bibnamefont {Andrade}},\
  }\href@noop {} {\bibfield  {journal} {\bibinfo  {journal} {Phys. Rev. Lett.}\
  }\textbf {\bibinfo {volume} {103}},\ \bibinfo {pages} {194502} (\bibinfo
  {year} {2009})},\ \bibinfo {note} {and references therein}\BibitemShut
  {NoStop}%
\bibitem [{\citenamefont {Meybodi}\ and\ \citenamefont
  {Hassanzadeh}(2013)}]{PhysRevE.87.033009}%
  \BibitemOpen
  \bibfield  {author} {\bibinfo {author} {\bibfnamefont {H.~E.}\ \bibnamefont
  {Meybodi}}\ and\ \bibinfo {author} {\bibfnamefont {H.}~\bibnamefont
  {Hassanzadeh}},\ }\href {\doibase 10.1103/PhysRevE.87.033009} {\bibfield
  {journal} {\bibinfo  {journal} {Phys. Rev. E}\ }\textbf {\bibinfo {volume}
  {87}},\ \bibinfo {pages} {033009} (\bibinfo {year} {2013})}\BibitemShut
  {NoStop}%
\bibitem [{joe()}]{joekar}%
  \BibitemOpen
  \href@noop {} {}\bibinfo {note} {V. Joekar-Niasar, M. I. J. Dijke, and S. M.
  Hassanizadeh, Special issue on: \textit{Pore-Scale Modeling of Multiphase
  Flow and Transport: Achievements and Perspectives}, Transp. Porous Media
  \textbf{94}, 461 (2012)}\BibitemShut {NoStop}%
\bibitem [{\citenamefont {Holmes}\ \emph {et~al.}(2011)\citenamefont {Holmes},
  \citenamefont {Williams},\ and\ \citenamefont {Tilke}}]{NAG:NAG898}%
  \BibitemOpen
  \bibfield  {author} {\bibinfo {author} {\bibfnamefont {D.~W.}\ \bibnamefont
  {Holmes}}, \bibinfo {author} {\bibfnamefont {J.~R.}\ \bibnamefont
  {Williams}}, \ and\ \bibinfo {author} {\bibfnamefont {P.}~\bibnamefont
  {Tilke}},\ }\href {\doibase 10.1002/nag.898} {\bibfield  {journal} {\bibinfo
  {journal} {Int. J. Num. An. Meth. Geomech.}\ }\textbf {\bibinfo {volume}
  {35}},\ \bibinfo {pages} {419} (\bibinfo {year} {2011})},\ \bibinfo {note}
  {and references therein}\BibitemShut {NoStop}%
\bibitem [{\citenamefont {Feder}(1988)}]{feder}%
  \BibitemOpen
  \bibfield  {author} {\bibinfo {author} {\bibfnamefont {J.}~\bibnamefont
  {Feder}},\ }\href@noop {} {\emph {\bibinfo {title} {Fractals}}}\ (\bibinfo
  {publisher} {Plenum Press, New York},\ \bibinfo {year} {1988})\BibitemShut
  {NoStop}%
\bibitem [{\citenamefont {Barab\'{a}si}\ and\ \citenamefont
  {Stanley}(1995)}]{Barabasi}%
  \BibitemOpen
  \bibfield  {author} {\bibinfo {author} {\bibfnamefont {A.-L.}\ \bibnamefont
  {Barab\'{a}si}}\ and\ \bibinfo {author} {\bibfnamefont {H.}~\bibnamefont
  {Stanley}},\ }\href@noop {} {\emph {\bibinfo {title} {Fractal Concept in
  Surface Growth}}}\ (\bibinfo  {publisher} {Cambridge University Press,
  Cambridge},\ \bibinfo {year} {1995})\BibitemShut {NoStop}%
\bibitem [{\citenamefont {Frisch}\ \emph {et~al.}(1986)\citenamefont {Frisch},
  \citenamefont {Hasslacher},\ and\ \citenamefont
  {Pomeau}}]{PhysRevLett.56.1505}%
  \BibitemOpen
  \bibfield  {author} {\bibinfo {author} {\bibfnamefont {U.}~\bibnamefont
  {Frisch}}, \bibinfo {author} {\bibfnamefont {B.}~\bibnamefont {Hasslacher}},
  \ and\ \bibinfo {author} {\bibfnamefont {Y.}~\bibnamefont {Pomeau}},\ }\href
  {\doibase 10.1103/PhysRevLett.56.1505} {\bibfield  {journal} {\bibinfo
  {journal} {Phys. Rev. Lett.}\ }\textbf {\bibinfo {volume} {56}},\ \bibinfo
  {pages} {1505} (\bibinfo {year} {1986})}\BibitemShut {NoStop}%
\bibitem [{\citenamefont {Frisch}\ \emph {et~al.}(1987)\citenamefont {Frisch},
  \citenamefont {d'Humi\`{e}res}, \citenamefont {Hasslacher}, \citenamefont
  {Lallemand},\ and\ \citenamefont {Pomeau}}]{Frisch_et_al}%
  \BibitemOpen
  \bibfield  {author} {\bibinfo {author} {\bibfnamefont {U.}~\bibnamefont
  {Frisch}}, \bibinfo {author} {\bibfnamefont {D.}~\bibnamefont
  {d'Humi\`{e}res}}, \bibinfo {author} {\bibfnamefont {B.}~\bibnamefont
  {Hasslacher}}, \bibinfo {author} {\bibfnamefont {P.}~\bibnamefont
  {Lallemand}}, \ and\ \bibinfo {author} {\bibfnamefont {Y.}~\bibnamefont
  {Pomeau}},\ }\href@noop {} {\bibfield  {journal} {\bibinfo  {journal}
  {Complex Syst.}\ }\textbf {\bibinfo {volume} {1}},\ \bibinfo {pages} {169}
  (\bibinfo {year} {1987})}\BibitemShut {NoStop}%
\bibitem [{\citenamefont {Wolfram}(1986)}]{Wolfram_article}%
  \BibitemOpen
  \bibfield  {author} {\bibinfo {author} {\bibfnamefont {S.}~\bibnamefont
  {Wolfram}},\ }\href@noop {} {\bibfield  {journal} {\bibinfo  {journal} {J.\
  Stat. Phys.}\ }\textbf {\bibinfo {volume} {45}},\ \bibinfo {pages} {471}
  (\bibinfo {year} {1986})}\BibitemShut {NoStop}%
\bibitem [{\citenamefont {Rothman}\ and\ \citenamefont
  {Zaleski}(1997)}]{Rothman_book}%
  \BibitemOpen
  \bibfield  {author} {\bibinfo {author} {\bibfnamefont {D.~H.}\ \bibnamefont
  {Rothman}}\ and\ \bibinfo {author} {\bibfnamefont {S.}~\bibnamefont
  {Zaleski}},\ }\href@noop {} {\emph {\bibinfo {title} {Lattice-Gas Cellular
  Automata}}}\ (\bibinfo  {publisher} {Cambridge University Press, Cambridge},\
  \bibinfo {year} {1997})\BibitemShut {NoStop}%
\bibitem [{\citenamefont {d'Humi\`{e}res}\ \emph {et~al.}(1986)\citenamefont
  {d'Humi\`{e}res}, \citenamefont {Lallemand},\ and\ \citenamefont
  {Frisch}}]{0295-5075-2-4-006}%
  \BibitemOpen
  \bibfield  {author} {\bibinfo {author} {\bibfnamefont {D.}~\bibnamefont
  {d'Humi\`{e}res}}, \bibinfo {author} {\bibfnamefont {P.}~\bibnamefont
  {Lallemand}}, \ and\ \bibinfo {author} {\bibfnamefont {U.}~\bibnamefont
  {Frisch}},\ }\href@noop {} {\bibfield  {journal} {\bibinfo  {journal}
  {Europhys. Lett.}\ }\textbf {\bibinfo {volume} {2}},\ \bibinfo {pages} {291}
  (\bibinfo {year} {1986})}\BibitemShut {NoStop}%
\bibitem [{\citenamefont {Klales}\ \emph {et~al.}(2010)\citenamefont {Klales},
  \citenamefont {Cianci}, \citenamefont {Needell}, \citenamefont {Meyer},\ and\
  \citenamefont {Love}}]{PhysRevE.82.046705}%
  \BibitemOpen
  \bibfield  {author} {\bibinfo {author} {\bibfnamefont {A.}~\bibnamefont
  {Klales}}, \bibinfo {author} {\bibfnamefont {D.}~\bibnamefont {Cianci}},
  \bibinfo {author} {\bibfnamefont {Z.}~\bibnamefont {Needell}}, \bibinfo
  {author} {\bibfnamefont {D.~A.}\ \bibnamefont {Meyer}}, \ and\ \bibinfo
  {author} {\bibfnamefont {P.~J.}\ \bibnamefont {Love}},\ }\href {\doibase
  10.1103/PhysRevE.82.046705} {\bibfield  {journal} {\bibinfo  {journal} {Phys.
  Rev. E}\ }\textbf {\bibinfo {volume} {82}},\ \bibinfo {pages} {046705}
  (\bibinfo {year} {2010})}\BibitemShut {NoStop}%
\bibitem [{\citenamefont {Love}\ and\ \citenamefont
  {Cianci}(2011)}]{Love13062011}%
  \BibitemOpen
  \bibfield  {author} {\bibinfo {author} {\bibfnamefont {P.~J.}\ \bibnamefont
  {Love}}\ and\ \bibinfo {author} {\bibfnamefont {D.}~\bibnamefont {Cianci}},\
  }\href {\doibase 10.1098/rsta.2011.0097} {\bibfield  {journal} {\bibinfo
  {journal} {Philos. Trans. R. Soc. London A}\ }\textbf {\bibinfo {volume}
  {369}},\ \bibinfo {pages} {2362} (\bibinfo {year} {2011})}\BibitemShut
  {NoStop}%
\bibitem [{\citenamefont {Genabeek}\ and\ \citenamefont
  {Rothman}(1996)}]{rothman96}%
  \BibitemOpen
  \bibfield  {author} {\bibinfo {author} {\bibfnamefont {O.~v.}\ \bibnamefont
  {Genabeek}}\ and\ \bibinfo {author} {\bibfnamefont {D.~H.}\ \bibnamefont
  {Rothman}},\ }\href {\doibase 10.1146/annurev.earth.24.1.63} {\bibfield
  {journal} {\bibinfo  {journal} {Annu. Rev. Earth Planet. Sci.}\ }\textbf
  {\bibinfo {volume} {24}},\ \bibinfo {pages} {63} (\bibinfo {year}
  {1996})}\BibitemShut {NoStop}%
\bibitem [{\citenamefont {Rothman}(1988)}]{roth}%
  \BibitemOpen
  \bibfield  {author} {\bibinfo {author} {\bibfnamefont {D.~H.}\ \bibnamefont
  {Rothman}},\ }\href@noop {} {\bibfield  {journal} {\bibinfo  {journal}
  {Geophysics}\ }\textbf {\bibinfo {volume} {53}},\ \bibinfo {pages} {509}
  (\bibinfo {year} {1988})}\BibitemShut {NoStop}%
\bibitem [{\citenamefont {Humby}\ \emph {et~al.}(2002)\citenamefont {Humby},
  \citenamefont {Biggs},\ and\ \citenamefont {Tüzün}}]{Humby20021955}%
  \BibitemOpen
  \bibfield  {author} {\bibinfo {author} {\bibfnamefont {S.}~\bibnamefont
  {Humby}}, \bibinfo {author} {\bibfnamefont {M.}~\bibnamefont {Biggs}}, \ and\
  \bibinfo {author} {\bibfnamefont {U.}~\bibnamefont {Tüzün}},\ }\href
  {\doibase http://dx.doi.org/10.1016/S0009-2509(02)00103-3} {\bibfield
  {journal} {\bibinfo  {journal} {Chem. Eng. Sci.}\ }\textbf {\bibinfo {volume}
  {57}},\ \bibinfo {pages} {1955 } (\bibinfo {year} {2002})}\BibitemShut
  {NoStop}%
\bibitem [{Rot()}]{Rothman_Keller}%
  \BibitemOpen
  \href@noop {} {}\bibinfo {note} {D. H. Rothman and J. M. Keller, J. Stat.
  Phys. \textbf{52}, 1119 (1988). See also, C. Adler, D. d'Humi\`{e}res, and D.
  H. Rothman, J. Phys. I \textbf{4}, 29 (1994)}\BibitemShut {NoStop}%
\bibitem [{rot()}]{rothf}%
  \BibitemOpen
  \href@noop {} {}\bibinfo {note} {E. G. Flekk{\o}y and D. H. Rothman, Phys.
  Rev. Lett. \textbf{75}, 260 (1995)}\BibitemShut {NoStop}%
\bibitem [{\citenamefont {Flekk{\o}y}\ and\ \citenamefont
  {Rothman}(1996)}]{PhysRevE.53.16}%
  \BibitemOpen
  \bibfield  {author} {\bibinfo {author} {\bibfnamefont {E.~G.}\ \bibnamefont
  {Flekk{\o}y}}\ and\ \bibinfo {author} {\bibfnamefont {D.~H.}\ \bibnamefont
  {Rothman}},\ }\href {\doibase 10.1103/PhysRevE.53.1622} {\bibfield  {journal}
  {\bibinfo  {journal} {Phys. Rev. E}\ }\textbf {\bibinfo {volume} {53}},\
  \bibinfo {pages} {1622} (\bibinfo {year} {1996})}\BibitemShut {NoStop}%
\bibitem [{\citenamefont {Kardar}\ \emph {et~al.}(1986)\citenamefont {Kardar},
  \citenamefont {Parisi},\ and\ \citenamefont {Zhang}}]{PhysRevLett.56}%
  \BibitemOpen
  \bibfield  {author} {\bibinfo {author} {\bibfnamefont {M.}~\bibnamefont
  {Kardar}}, \bibinfo {author} {\bibfnamefont {G.}~\bibnamefont {Parisi}}, \
  and\ \bibinfo {author} {\bibfnamefont {Y.-C.}\ \bibnamefont {Zhang}},\ }\href
  {\doibase 10.1103/PhysRevLett.56.889} {\bibfield  {journal} {\bibinfo
  {journal} {Phys. Rev. Lett.}\ }\textbf {\bibinfo {volume} {56}},\ \bibinfo
  {pages} {889} (\bibinfo {year} {1986})}\BibitemShut {NoStop}%
\bibitem [{\citenamefont {Oliveira}\ \emph {et~al.}(2012)\citenamefont
  {Oliveira}, \citenamefont {Ferreira},\ and\ \citenamefont
  {Alves}}]{PhysRevE.85.010601}%
  \BibitemOpen
  \bibfield  {author} {\bibinfo {author} {\bibfnamefont {T.~J.}\ \bibnamefont
  {Oliveira}}, \bibinfo {author} {\bibfnamefont {S.~C.}\ \bibnamefont
  {Ferreira}}, \ and\ \bibinfo {author} {\bibfnamefont {S.~G.}\ \bibnamefont
  {Alves}},\ }\href {\doibase 10.1103/PhysRevE.85.010601} {\bibfield  {journal}
  {\bibinfo  {journal} {Phys. Rev. E}\ }\textbf {\bibinfo {volume} {85}},\
  \bibinfo {pages} {010601} (\bibinfo {year} {2012})}\BibitemShut {NoStop}%
\bibitem [{\citenamefont {Olson}\ and\ \citenamefont
  {Rothman}(1997)}]{FLM:13517}%
  \BibitemOpen
  \bibfield  {author} {\bibinfo {author} {\bibfnamefont {J.~F.}\ \bibnamefont
  {Olson}}\ and\ \bibinfo {author} {\bibfnamefont {D.~H.}\ \bibnamefont
  {Rothman}},\ }\href {\doibase 10.1017/S0022112097005533} {\bibfield
  {journal} {\bibinfo  {journal} {J. Fluid Mech.}\ }\textbf {\bibinfo {volume}
  {341}},\ \bibinfo {pages} {343} (\bibinfo {year} {1997})}\BibitemShut
  {NoStop}%
\bibitem [{\citenamefont {Maillet}\ and\ \citenamefont
  {Coveney}(2000)}]{PhysRevE.62.2898}%
  \BibitemOpen
  \bibfield  {author} {\bibinfo {author} {\bibfnamefont {J.-B.}\ \bibnamefont
  {Maillet}}\ and\ \bibinfo {author} {\bibfnamefont {P.~V.}\ \bibnamefont
  {Coveney}},\ }\href {\doibase 10.1103/PhysRevE.62.2898} {\bibfield  {journal}
  {\bibinfo  {journal} {Phys. Rev. E}\ }\textbf {\bibinfo {volume} {62}},\
  \bibinfo {pages} {2898} (\bibinfo {year} {2000})}\BibitemShut {NoStop}%
\bibitem [{\citenamefont {Love}\ \emph {et~al.}(2001)\citenamefont {Love},
  \citenamefont {Maillet},\ and\ \citenamefont {Coveney}}]{PhysRevE.64.061302}%
  \BibitemOpen
  \bibfield  {author} {\bibinfo {author} {\bibfnamefont {P.~J.}\ \bibnamefont
  {Love}}, \bibinfo {author} {\bibfnamefont {J.-B.}\ \bibnamefont {Maillet}}, \
  and\ \bibinfo {author} {\bibfnamefont {P.~V.}\ \bibnamefont {Coveney}},\
  }\href {\doibase 10.1103/PhysRevE.64.061302} {\bibfield  {journal} {\bibinfo
  {journal} {Phys. Rev. E}\ }\textbf {\bibinfo {volume} {64}},\ \bibinfo
  {pages} {061302} (\bibinfo {year} {2001})}\BibitemShut {NoStop}%
\bibitem [{\citenamefont {Love}\ \emph {et~al.}(2003)\citenamefont {Love},
  \citenamefont {Nekovee}, \citenamefont {Coveney}, \citenamefont {Chin},
  \citenamefont {Gonz\'alez-Segredo},\ and\ \citenamefont
  {Martin}}]{Love2003340}%
  \BibitemOpen
  \bibfield  {author} {\bibinfo {author} {\bibfnamefont {P.~J.}\ \bibnamefont
  {Love}}, \bibinfo {author} {\bibfnamefont {M.}~\bibnamefont {Nekovee}},
  \bibinfo {author} {\bibfnamefont {P.~V.}\ \bibnamefont {Coveney}}, \bibinfo
  {author} {\bibfnamefont {J.}~\bibnamefont {Chin}}, \bibinfo {author}
  {\bibfnamefont {N.}~\bibnamefont {Gonz\'alez-Segredo}}, \ and\ \bibinfo
  {author} {\bibfnamefont {J.~M.~R.}\ \bibnamefont {Martin}},\ }\href {\doibase
  http://dx.doi.org/10.1016/S0010-4655(03)00200-5} {\bibfield  {journal}
  {\bibinfo  {journal} {Comput. Phys. Commun.}\ }\textbf {\bibinfo {volume}
  {153}},\ \bibinfo {pages} {340 } (\bibinfo {year} {2003})}\BibitemShut
  {NoStop}%
\bibitem [{\citenamefont {McNamara}\ and\ \citenamefont
  {Zanetti}(1988)}]{PhysRevLett.61.2332}%
  \BibitemOpen
  \bibfield  {author} {\bibinfo {author} {\bibfnamefont {G.~R.}\ \bibnamefont
  {McNamara}}\ and\ \bibinfo {author} {\bibfnamefont {G.}~\bibnamefont
  {Zanetti}},\ }\href {\doibase 10.1103/PhysRevLett.61.2332} {\bibfield
  {journal} {\bibinfo  {journal} {Phys. Rev. Lett.}\ }\textbf {\bibinfo
  {volume} {61}},\ \bibinfo {pages} {2332} (\bibinfo {year}
  {1988})}\BibitemShut {NoStop}%
\bibitem [{\citenamefont {Chen}\ and\ \citenamefont {Doolen}(1998)}]{revlbm}%
  \BibitemOpen
  \bibfield  {author} {\bibinfo {author} {\bibfnamefont {S.}~\bibnamefont
  {Chen}}\ and\ \bibinfo {author} {\bibfnamefont {G.~D.}\ \bibnamefont
  {Doolen}},\ }\href {\doibase 10.1146/annurev.fluid.30.1.329} {\bibfield
  {journal} {\bibinfo  {journal} {Annu. Rev. Fluid Mech.}\ }\textbf {\bibinfo
  {volume} {30}},\ \bibinfo {pages} {329} (\bibinfo {year} {1998})}\BibitemShut
  {NoStop}%
\bibitem [{hig()}]{higuera}%
  \BibitemOpen
  \href@noop {} {}\bibinfo {note} {F. Higuera and J. Jimenez, Europhys. Lett.
  \textbf{9}, 663 (1989)}\BibitemShut {NoStop}%
\bibitem [{ext()}]{extralbm}%
  \BibitemOpen
  \href@noop {} {}\bibinfo {note} {Y. Qian, D. d'Humi\`eres, and P. Lallemand,
  Europhys. Lett. \textbf{17}, 479 (1992); H. Chen, S. Chen, and W. H.
  Matthaeus, Phys. Rev. A \textbf{45}, R5339 (1992)}\BibitemShut {NoStop}%
\bibitem [{\citenamefont {Succi}\ \emph {et~al.}(1989)\citenamefont {Succi},
  \citenamefont {Foti},\ and\ \citenamefont {Higuera}}]{0295-5075-10-5-008}%
  \BibitemOpen
  \bibfield  {author} {\bibinfo {author} {\bibfnamefont {S.}~\bibnamefont
  {Succi}}, \bibinfo {author} {\bibfnamefont {E.}~\bibnamefont {Foti}}, \ and\
  \bibinfo {author} {\bibfnamefont {F.}~\bibnamefont {Higuera}},\ }\href@noop
  {} {\bibfield  {journal} {\bibinfo  {journal} {Europhys. Lett.}\ }\textbf
  {\bibinfo {volume} {10}},\ \bibinfo {pages} {433} (\bibinfo {year}
  {1989})}\BibitemShut {NoStop}%
\bibitem [{\citenamefont {Cancelliere}\ \emph {et~al.}(1990)\citenamefont
  {Cancelliere}, \citenamefont {Chang}, \citenamefont {Foti}, \citenamefont
  {Rothman},\ and\ \citenamefont {Succi}}]{cancelliere:2085}%
  \BibitemOpen
  \bibfield  {author} {\bibinfo {author} {\bibfnamefont {A.}~\bibnamefont
  {Cancelliere}}, \bibinfo {author} {\bibfnamefont {C.}~\bibnamefont {Chang}},
  \bibinfo {author} {\bibfnamefont {E.}~\bibnamefont {Foti}}, \bibinfo {author}
  {\bibfnamefont {D.~H.}\ \bibnamefont {Rothman}}, \ and\ \bibinfo {author}
  {\bibfnamefont {S.}~\bibnamefont {Succi}},\ }\href {\doibase
  10.1063/1.857793} {\bibfield  {journal} {\bibinfo  {journal} {Phys. Fluids
  A}\ }\textbf {\bibinfo {volume} {2}},\ \bibinfo {pages} {2085} (\bibinfo
  {year} {1990})}\BibitemShut {NoStop}%
\bibitem [{\citenamefont {Ferr\'eol}\ and\ \citenamefont
  {Rothman}(1995)}]{rothman95}%
  \BibitemOpen
  \bibfield  {author} {\bibinfo {author} {\bibfnamefont {B.}~\bibnamefont
  {Ferr\'eol}}\ and\ \bibinfo {author} {\bibfnamefont {D.~H.}\ \bibnamefont
  {Rothman}},\ }\href {\doibase 10.1007/BF00616923} {\bibfield  {journal}
  {\bibinfo  {journal} {Transp. Porous Media}\ }\textbf {\bibinfo {volume}
  {20}},\ \bibinfo {pages} {3} (\bibinfo {year} {1995})}\BibitemShut {NoStop}%
\bibitem [{\citenamefont {Martys}\ and\ \citenamefont
  {Chen}(1996)}]{PhysRevE.53.743}%
  \BibitemOpen
  \bibfield  {author} {\bibinfo {author} {\bibfnamefont {N.~S.}\ \bibnamefont
  {Martys}}\ and\ \bibinfo {author} {\bibfnamefont {H.}~\bibnamefont {Chen}},\
  }\href {\doibase 10.1103/PhysRevE.53.743} {\bibfield  {journal} {\bibinfo
  {journal} {Phys. Rev. E}\ }\textbf {\bibinfo {volume} {53}},\ \bibinfo
  {pages} {743} (\bibinfo {year} {1996})}\BibitemShut {NoStop}%
\bibitem [{\citenamefont {Gonz\'alez-Segredo}\ \emph
  {et~al.}(2003)\citenamefont {Gonz\'alez-Segredo}, \citenamefont {Nekovee},\
  and\ \citenamefont {Coveney}}]{PhysRevE.67.046304}%
  \BibitemOpen
  \bibfield  {author} {\bibinfo {author} {\bibfnamefont {N.}~\bibnamefont
  {Gonz\'alez-Segredo}}, \bibinfo {author} {\bibfnamefont {M.}~\bibnamefont
  {Nekovee}}, \ and\ \bibinfo {author} {\bibfnamefont {P.~V.}\ \bibnamefont
  {Coveney}},\ }\href {\doibase 10.1103/PhysRevE.67.046304} {\bibfield
  {journal} {\bibinfo  {journal} {Phys. Rev. E}\ }\textbf {\bibinfo {volume}
  {67}},\ \bibinfo {pages} {046304} (\bibinfo {year} {2003})}\BibitemShut
  {NoStop}%
\bibitem [{\citenamefont {Li}\ \emph {et~al.}(2005)\citenamefont {Li},
  \citenamefont {Pan},\ and\ \citenamefont {Miller}}]{PhysRevE.72.026705}%
  \BibitemOpen
  \bibfield  {author} {\bibinfo {author} {\bibfnamefont {H.}~\bibnamefont
  {Li}}, \bibinfo {author} {\bibfnamefont {C.}~\bibnamefont {Pan}}, \ and\
  \bibinfo {author} {\bibfnamefont {C.~T.}\ \bibnamefont {Miller}},\ }\href
  {\doibase 10.1103/PhysRevE.72.026705} {\bibfield  {journal} {\bibinfo
  {journal} {Phys. Rev. E}\ }\textbf {\bibinfo {volume} {72}},\ \bibinfo
  {pages} {026705} (\bibinfo {year} {2005})}\BibitemShut {NoStop}%
\bibitem [{\citenamefont {Boek}\ and\ \citenamefont
  {Venturoli}(2010)}]{Boek20102305}%
  \BibitemOpen
  \bibfield  {author} {\bibinfo {author} {\bibfnamefont {E.~S.}\ \bibnamefont
  {Boek}}\ and\ \bibinfo {author} {\bibfnamefont {M.}~\bibnamefont
  {Venturoli}},\ }\href {\doibase 10.1016/j.camwa.2009.08.063} {\bibfield
  {journal} {\bibinfo  {journal} {Comput. Math. Appl.}\ }\textbf {\bibinfo
  {volume} {59}},\ \bibinfo {pages} {2305 } (\bibinfo {year}
  {2010})}\BibitemShut {NoStop}%
\bibitem [{\citenamefont {Guo}\ and\ \citenamefont
  {Zhao}(2002)}]{PhysRevE.66.036304}%
  \BibitemOpen
  \bibfield  {author} {\bibinfo {author} {\bibfnamefont {Z.}~\bibnamefont
  {Guo}}\ and\ \bibinfo {author} {\bibfnamefont {T.~S.}\ \bibnamefont {Zhao}},\
  }\href {\doibase 10.1103/PhysRevE.66.036304} {\bibfield  {journal} {\bibinfo
  {journal} {Phys. Rev. E}\ }\textbf {\bibinfo {volume} {66}},\ \bibinfo
  {pages} {036304} (\bibinfo {year} {2002})}\BibitemShut {NoStop}%
\bibitem [{\citenamefont {Nithiarasu}\ \emph {et~al.}(1997)\citenamefont
  {Nithiarasu}, \citenamefont {Seetharamu},\ and\ \citenamefont
  {Sundararajan}}]{Nithiarasu19973955}%
  \BibitemOpen
  \bibfield  {author} {\bibinfo {author} {\bibfnamefont {P.}~\bibnamefont
  {Nithiarasu}}, \bibinfo {author} {\bibfnamefont {K.}~\bibnamefont
  {Seetharamu}}, \ and\ \bibinfo {author} {\bibfnamefont {T.}~\bibnamefont
  {Sundararajan}},\ }\href@noop {} {\bibfield  {journal} {\bibinfo  {journal}
  {Int. J. Heat Mass Transfer}\ }\textbf {\bibinfo {volume} {40}},\ \bibinfo
  {pages} {3955 } (\bibinfo {year} {1997})}\BibitemShut {NoStop}%
\bibitem [{\citenamefont {Rakotomalala}\ \emph {et~al.}(1997)\citenamefont
  {Rakotomalala}, \citenamefont {Salin},\ and\ \citenamefont {Watzky}}]{rako}%
  \BibitemOpen
  \bibfield  {author} {\bibinfo {author} {\bibfnamefont {N.}~\bibnamefont
  {Rakotomalala}}, \bibinfo {author} {\bibfnamefont {D.}~\bibnamefont {Salin}},
  \ and\ \bibinfo {author} {\bibfnamefont {P.}~\bibnamefont {Watzky}},\
  }\href@noop {} {\bibfield  {journal} {\bibinfo  {journal} {J. Fluid Mech.}\
  }\textbf {\bibinfo {volume} {338}},\ \bibinfo {pages} {277} (\bibinfo {year}
  {1997})}\BibitemShut {NoStop}%
\bibitem [{\citenamefont {Langaas}\ and\ \citenamefont
  {Yeomans}(2000)}]{langaas}%
  \BibitemOpen
  \bibfield  {author} {\bibinfo {author} {\bibfnamefont {K.}~\bibnamefont
  {Langaas}}\ and\ \bibinfo {author} {\bibfnamefont {J.}~\bibnamefont
  {Yeomans}},\ }\href {\doibase 10.1007/s100510051107} {\bibfield  {journal}
  {\bibinfo  {journal} {Eur. Phys. J. B}\ }\textbf {\bibinfo {volume} {15}},\
  \bibinfo {pages} {133} (\bibinfo {year} {2000})}\BibitemShut {NoStop}%
\bibitem [{\citenamefont {Denniston}\ \emph {et~al.}(2004)\citenamefont
  {Denniston}, \citenamefont {Marenduzzo}, \citenamefont {Orlandini},\ and\
  \citenamefont {Yeomans}}]{15082004}%
  \BibitemOpen
  \bibfield  {author} {\bibinfo {author} {\bibfnamefont {C.}~\bibnamefont
  {Denniston}}, \bibinfo {author} {\bibfnamefont {D.}~\bibnamefont
  {Marenduzzo}}, \bibinfo {author} {\bibfnamefont {E.}~\bibnamefont
  {Orlandini}}, \ and\ \bibinfo {author} {\bibfnamefont {J.~M.}\ \bibnamefont
  {Yeomans}},\ }\href {\doibase 10.1098/rsta.2004.1416} {\bibfield  {journal}
  {\bibinfo  {journal} {Philos. Trans. R. Soc. London A}\ }\textbf {\bibinfo
  {volume} {362}},\ \bibinfo {pages} {1745} (\bibinfo {year}
  {2004})}\BibitemShut {NoStop}%
\bibitem [{\citenamefont {Ravnik}\ and\ \citenamefont
  {Yeomans}(2013)}]{PhysRevLett.110.026001}%
  \BibitemOpen
  \bibfield  {author} {\bibinfo {author} {\bibfnamefont {M.}~\bibnamefont
  {Ravnik}}\ and\ \bibinfo {author} {\bibfnamefont {J.~M.}\ \bibnamefont
  {Yeomans}},\ }\href {\doibase 10.1103/PhysRevLett.110.026001} {\bibfield
  {journal} {\bibinfo  {journal} {Phys. Rev. Lett.}\ }\textbf {\bibinfo
  {volume} {110}},\ \bibinfo {pages} {026001} (\bibinfo {year}
  {2013})}\BibitemShut {NoStop}%
\bibitem [{\citenamefont {Sengupta}\ \emph {et~al.}(2013)\citenamefont
  {Sengupta}, \citenamefont {Tkalec}, \citenamefont {Ravnik}, \citenamefont
  {Yeomans}, \citenamefont {Bahr},\ and\ \citenamefont
  {Herminghaus}}]{PhysRevLett.110.048303}%
  \BibitemOpen
  \bibfield  {author} {\bibinfo {author} {\bibfnamefont {A.}~\bibnamefont
  {Sengupta}}, \bibinfo {author} {\bibfnamefont {U.}~\bibnamefont {Tkalec}},
  \bibinfo {author} {\bibfnamefont {M.}~\bibnamefont {Ravnik}}, \bibinfo
  {author} {\bibfnamefont {J.~M.}\ \bibnamefont {Yeomans}}, \bibinfo {author}
  {\bibfnamefont {C.}~\bibnamefont {Bahr}}, \ and\ \bibinfo {author}
  {\bibfnamefont {S.}~\bibnamefont {Herminghaus}},\ }\href {\doibase
  10.1103/PhysRevLett.110.048303} {\bibfield  {journal} {\bibinfo  {journal}
  {Phys. Rev. Lett.}\ }\textbf {\bibinfo {volume} {110}},\ \bibinfo {pages}
  {048303} (\bibinfo {year} {2013})}\BibitemShut {NoStop}%
\bibitem [{Pro()}]{Proc.R.Soc.}%
  \BibitemOpen
  \href@noop {} {}\bibinfo {note} {P. Saffman and G. Taylor, Proc. R. Soc.
  London, Sect. A \textbf{245}, 312 (1958)}\BibitemShut {NoStop}%
\bibitem [{chu()}]{chuoke}%
  \BibitemOpen
  \href@noop {} {}\bibinfo {note} {R. L. Chuoke, P. van Meurs and C. van der
  Poel, J. Pet. Technol. \textbf{11}, 64 (1959)}\BibitemShut {NoStop}%
\bibitem [{mah()}]{maher}%
  \BibitemOpen
  \href@noop {} {}\bibinfo {note} {J. V. Maher, Phys. Rev. Lett. \textbf{54},
  1498 (1985)}\BibitemShut {NoStop}%
\bibitem [{\citenamefont {Tabeling}\ and\ \citenamefont
  {Libchaber}(1986)}]{PhysRevA.33.794}%
  \BibitemOpen
  \bibfield  {author} {\bibinfo {author} {\bibfnamefont {P.}~\bibnamefont
  {Tabeling}}\ and\ \bibinfo {author} {\bibfnamefont {A.}~\bibnamefont
  {Libchaber}},\ }\href {\doibase 10.1103/PhysRevA.33.794} {\bibfield
  {journal} {\bibinfo  {journal} {Phys. Rev. A}\ }\textbf {\bibinfo {volume}
  {33}},\ \bibinfo {pages} {794} (\bibinfo {year} {1986})}\BibitemShut
  {NoStop}%
\bibitem [{jfm()}]{jfmlib}%
  \BibitemOpen
  \href@noop {} {}\bibinfo {note} {P. Tabeling, G. Zoochi, and A. Libchaber, J.
  Fluid Mech. \textbf{177}, 67 (1987)}\BibitemShut {NoStop}%
\bibitem [{\citenamefont {Zocchi}\ \emph {et~al.}(1987)\citenamefont {Zocchi},
  \citenamefont {Shaw}, \citenamefont {Libchaber},\ and\ \citenamefont
  {Kadanoff}}]{PhysRevA.36.1894}%
  \BibitemOpen
  \bibfield  {author} {\bibinfo {author} {\bibfnamefont {G.}~\bibnamefont
  {Zocchi}}, \bibinfo {author} {\bibfnamefont {B.~E.}\ \bibnamefont {Shaw}},
  \bibinfo {author} {\bibfnamefont {A.}~\bibnamefont {Libchaber}}, \ and\
  \bibinfo {author} {\bibfnamefont {L.~P.}\ \bibnamefont {Kadanoff}},\ }\href
  {\doibase 10.1103/PhysRevA.36.1894} {\bibfield  {journal} {\bibinfo
  {journal} {Phys. Rev. A}\ }\textbf {\bibinfo {volume} {36}},\ \bibinfo
  {pages} {1894} (\bibinfo {year} {1987})}\BibitemShut {NoStop}%
\bibitem [{\citenamefont {Tabeling}\ \emph {et~al.}(1989)\citenamefont
  {Tabeling}, \citenamefont {Zocchi},\ and\ \citenamefont
  {Libchaber}}]{tabeling1989experimental}%
  \BibitemOpen
  \bibfield  {author} {\bibinfo {author} {\bibfnamefont {P.}~\bibnamefont
  {Tabeling}}, \bibinfo {author} {\bibfnamefont {G.}~\bibnamefont {Zocchi}}, \
  and\ \bibinfo {author} {\bibfnamefont {A.}~\bibnamefont {Libchaber}},\ }in\
  \href@noop {} {\emph {\bibinfo {booktitle} {Physicochemical Hydrodynamics}}}\
  (\bibinfo  {publisher} {Springer},\ \bibinfo {year} {1989})\ pp.\ \bibinfo
  {pages} {515--525}\BibitemShut {NoStop}%
\bibitem [{\citenamefont {Bensimon}\ \emph {et~al.}(1986)\citenamefont
  {Bensimon}, \citenamefont {Kadanoff}, \citenamefont {Liang}, \citenamefont
  {Shraiman},\ and\ \citenamefont {Tang}}]{RevModPhys.58}%
  \BibitemOpen
  \bibfield  {author} {\bibinfo {author} {\bibfnamefont {D.}~\bibnamefont
  {Bensimon}}, \bibinfo {author} {\bibfnamefont {L.~P.}\ \bibnamefont
  {Kadanoff}}, \bibinfo {author} {\bibfnamefont {S.}~\bibnamefont {Liang}},
  \bibinfo {author} {\bibfnamefont {B.~I.}\ \bibnamefont {Shraiman}}, \ and\
  \bibinfo {author} {\bibfnamefont {C.}~\bibnamefont {Tang}},\ }\href {\doibase
  10.1103/RevModPhys.58.977} {\bibfield  {journal} {\bibinfo  {journal} {Rev.
  Mod. Phys.}\ }\textbf {\bibinfo {volume} {58}},\ \bibinfo {pages} {977}
  (\bibinfo {year} {1986})}\BibitemShut {NoStop}%
\bibitem [{are()}]{aref}%
  \BibitemOpen
  \href@noop {} {}\bibinfo {note} {G. Tryggvason and H. Aref, J. Fluid Mech.
  \textbf{136}, 1 (1983); \textit{ibid.} \textbf{154}, 287 (1985)}\BibitemShut
  {NoStop}%
\bibitem [{\citenamefont {Folch}\ \emph {et~al.}(1999)\citenamefont {Folch},
  \citenamefont {Casademunt}, \citenamefont {Hern\'andez-Machado},\ and\
  \citenamefont {Ram\'{\i}rez-Piscina}}]{PhysRevE.60.1734}%
  \BibitemOpen
  \bibfield  {author} {\bibinfo {author} {\bibfnamefont {R.}~\bibnamefont
  {Folch}}, \bibinfo {author} {\bibfnamefont {J.}~\bibnamefont {Casademunt}},
  \bibinfo {author} {\bibfnamefont {A.}~\bibnamefont {Hern\'andez-Machado}}, \
  and\ \bibinfo {author} {\bibfnamefont {L.}~\bibnamefont
  {Ram\'{\i}rez-Piscina}},\ }\href {\doibase 10.1103/PhysRevE.60.1734}
  {\bibfield  {journal} {\bibinfo  {journal} {Phys. Rev. E}\ }\textbf {\bibinfo
  {volume} {60}},\ \bibinfo {pages} {1734} (\bibinfo {year}
  {1999})}\BibitemShut {NoStop}%
\bibitem [{\citenamefont {Paterson}(1981)}]{FLM:389437}%
  \BibitemOpen
  \bibfield  {author} {\bibinfo {author} {\bibfnamefont {L.}~\bibnamefont
  {Paterson}},\ }\href {\doibase 10.1017/S0022112081003613} {\bibfield
  {journal} {\bibinfo  {journal} {J. Fluid Mech.}\ }\textbf {\bibinfo {volume}
  {113}},\ \bibinfo {pages} {513} (\bibinfo {year} {1981})}\BibitemShut
  {NoStop}%
\bibitem [{\citenamefont {Maxworthy}(1989)}]{PhysRevA.39.5863}%
  \BibitemOpen
  \bibfield  {author} {\bibinfo {author} {\bibfnamefont {T.}~\bibnamefont
  {Maxworthy}},\ }\href {\doibase 10.1103/PhysRevA.39.5863} {\bibfield
  {journal} {\bibinfo  {journal} {Phys. Rev. A}\ }\textbf {\bibinfo {volume}
  {39}},\ \bibinfo {pages} {5863} (\bibinfo {year} {1989})}\BibitemShut
  {NoStop}%
\bibitem [{\citenamefont {Kim}\ \emph {et~al.}(2009)\citenamefont {Kim},
  \citenamefont {Funada}, \citenamefont {Joseph},\ and\ \citenamefont
  {Homsy}}]{kim:074106}%
  \BibitemOpen
  \bibfield  {author} {\bibinfo {author} {\bibfnamefont {H.}~\bibnamefont
  {Kim}}, \bibinfo {author} {\bibfnamefont {T.}~\bibnamefont {Funada}},
  \bibinfo {author} {\bibfnamefont {D.~D.}\ \bibnamefont {Joseph}}, \ and\
  \bibinfo {author} {\bibfnamefont {G.~M.}\ \bibnamefont {Homsy}},\ }\href
  {\doibase 10.1063/1.3184574} {\bibfield  {journal} {\bibinfo  {journal}
  {Phys. Fluids}\ }\textbf {\bibinfo {volume} {21}},\ \bibinfo {eid} {074106}
  (\bibinfo {year} {2009})}\BibitemShut {NoStop}%
\bibitem [{\citenamefont {Vasconcelos}\ and\ \citenamefont
  {Kadanoff}(1991)}]{PhysRevA.44.6490}%
  \BibitemOpen
  \bibfield  {author} {\bibinfo {author} {\bibfnamefont {G.~L.}\ \bibnamefont
  {Vasconcelos}}\ and\ \bibinfo {author} {\bibfnamefont {L.~P.}\ \bibnamefont
  {Kadanoff}},\ }\href {\doibase 10.1103/PhysRevA.44.6490} {\bibfield
  {journal} {\bibinfo  {journal} {Phys. Rev. A}\ }\textbf {\bibinfo {volume}
  {44}},\ \bibinfo {pages} {6490} (\bibinfo {year} {1991})}\BibitemShut
  {NoStop}%
\bibitem [{\citenamefont {Mineev-Weinstein}(1998)}]{PhysRevLett.80.2113}%
  \BibitemOpen
  \bibfield  {author} {\bibinfo {author} {\bibfnamefont {M.}~\bibnamefont
  {Mineev-Weinstein}},\ }\href {\doibase 10.1103/PhysRevLett.80.2113}
  {\bibfield  {journal} {\bibinfo  {journal} {Phys. Rev. Lett.}\ }\textbf
  {\bibinfo {volume} {80}},\ \bibinfo {pages} {2113} (\bibinfo {year}
  {1998})}\BibitemShut {NoStop}%
\bibitem [{\citenamefont {Mineev-Weinstein}\ \emph {et~al.}(2000)\citenamefont
  {Mineev-Weinstein}, \citenamefont {Wiegmann},\ and\ \citenamefont
  {Zabrodin}}]{PhysRevLett.84.5106}%
  \BibitemOpen
  \bibfield  {author} {\bibinfo {author} {\bibfnamefont {M.}~\bibnamefont
  {Mineev-Weinstein}}, \bibinfo {author} {\bibfnamefont {P.~B.}\ \bibnamefont
  {Wiegmann}}, \ and\ \bibinfo {author} {\bibfnamefont {A.}~\bibnamefont
  {Zabrodin}},\ }\href {\doibase 10.1103/PhysRevLett.84.5106} {\bibfield
  {journal} {\bibinfo  {journal} {Phys. Rev. Lett.}\ }\textbf {\bibinfo
  {volume} {84}},\ \bibinfo {pages} {5106} (\bibinfo {year}
  {2000})}\BibitemShut {NoStop}%
\bibitem [{\citenamefont {Vasconcelos}(2001)}]{FLM:85633}%
  \BibitemOpen
  \bibfield  {author} {\bibinfo {author} {\bibfnamefont {G.~L.}\ \bibnamefont
  {Vasconcelos}},\ }\href {\doibase 10.1017/S0022112001005365} {\bibfield
  {journal} {\bibinfo  {journal} {J. Fluid Mech.}\ }\textbf {\bibinfo {volume}
  {444}},\ \bibinfo {pages} {175} (\bibinfo {year} {2001})}\BibitemShut
  {NoStop}%
\bibitem [{Cro()}]{Crowdy01112006}%
  \BibitemOpen
  \href@noop {} {\ }\bibinfo {note} {D. G. Crowdy, Quart. J. Mech. Appl. Math.
  \textbf{59}, 475 (2006)}\BibitemShut {NoStop}%
\bibitem [{\citenamefont {Balasubramanian}\ \emph {et~al.}(1987)\citenamefont
  {Balasubramanian}, \citenamefont {Hayot},\ and\ \citenamefont
  {Saam}}]{Hayot1}%
  \BibitemOpen
  \bibfield  {author} {\bibinfo {author} {\bibfnamefont {K.}~\bibnamefont
  {Balasubramanian}}, \bibinfo {author} {\bibfnamefont {F.}~\bibnamefont
  {Hayot}}, \ and\ \bibinfo {author} {\bibfnamefont {W.}~\bibnamefont {Saam}},\
  }\href@noop {} {\bibfield  {journal} {\bibinfo  {journal} {Phys.\ Rev. A}\
  }\textbf {\bibinfo {volume} {36}},\ \bibinfo {pages} {2248} (\bibinfo {year}
  {1987})}\BibitemShut {NoStop}%
\bibitem [{not()}]{notagrosfils}%
  \BibitemOpen
  \href@noop {} {}\bibinfo {note} {Models based on LBM with effective
  dissipation term in the Navier-Stokes equations (plus a reaction-diffusion
  equation) were proposed to study: (i) single-phase flow around a circular
  obstacle: E. G. Flekk\o{}y, U. Oxaal, J. Feder, and T. J\o{}ssang, Phys. Rev.
  E \textbf{52}, 4952 (1995); (ii) and two-phase miscible fingers: P. Grosfils,
  J. P. Boon, J. Chin, and E. S. Boek, Philos. Trans. R. Soc. London A
  \textbf{362}, 1723 (2004), in Hele-Shaw cells}\BibitemShut {NoStop}%
\bibitem [{\citenamefont {Burgess}\ \emph {et~al.}(1989)\citenamefont
  {Burgess}, \citenamefont {Hayot},\ and\ \citenamefont {Saam}}]{Hayot2}%
  \BibitemOpen
  \bibfield  {author} {\bibinfo {author} {\bibfnamefont {D.}~\bibnamefont
  {Burgess}}, \bibinfo {author} {\bibfnamefont {F.}~\bibnamefont {Hayot}}, \
  and\ \bibinfo {author} {\bibfnamefont {W.~F.}\ \bibnamefont {Saam}},\ }\href
  {\doibase 10.1103/PhysRevA.39.4695} {\bibfield  {journal} {\bibinfo
  {journal} {Phys. Rev. A}\ }\textbf {\bibinfo {volume} {39}},\ \bibinfo
  {pages} {4695} (\bibinfo {year} {1989})}\BibitemShut {NoStop}%
\bibitem [{\citenamefont {Hayot}\ \emph {et~al.}(1989)\citenamefont {Hayot},
  \citenamefont {Mandal},\ and\ \citenamefont {Sadayappan}}]{Hayot1989277}%
  \BibitemOpen
  \bibfield  {author} {\bibinfo {author} {\bibfnamefont {F.}~\bibnamefont
  {Hayot}}, \bibinfo {author} {\bibfnamefont {M.}~\bibnamefont {Mandal}}, \
  and\ \bibinfo {author} {\bibfnamefont {P.}~\bibnamefont {Sadayappan}},\
  }\href {\doibase http://dx.doi.org/10.1016/0021-9991(89)90100-9} {\bibfield
  {journal} {\bibinfo  {journal} {J. Comput. Phys.}\ }\textbf {\bibinfo
  {volume} {80}},\ \bibinfo {pages} {277 } (\bibinfo {year}
  {1989})}\BibitemShut {NoStop}%
\bibitem [{\citenamefont {Burgess}\ and\ \citenamefont
  {Hayot}(1989)}]{Hayot89}%
  \BibitemOpen
  \bibfield  {author} {\bibinfo {author} {\bibfnamefont {D.}~\bibnamefont
  {Burgess}}\ and\ \bibinfo {author} {\bibfnamefont {F.}~\bibnamefont
  {Hayot}},\ }\href {\doibase 10.1103/PhysRevA.40.5187} {\bibfield  {journal}
  {\bibinfo  {journal} {Phys. Rev. A}\ }\textbf {\bibinfo {volume} {40}},\
  \bibinfo {pages} {5187} (\bibinfo {year} {1989})}\BibitemShut {NoStop}%
\bibitem [{\citenamefont {Hayot}(1991)}]{Hayot199164}%
  \BibitemOpen
  \bibfield  {author} {\bibinfo {author} {\bibfnamefont {F.}~\bibnamefont
  {Hayot}},\ }\href {\doibase http://dx.doi.org/10.1016/0167-2789(91)90280-M}
  {\bibfield  {journal} {\bibinfo  {journal} {Physica D}\ }\textbf {\bibinfo
  {volume} {47}},\ \bibinfo {pages} {64 } (\bibinfo {year} {1991})}\BibitemShut
  {NoStop}%
\bibitem [{\citenamefont {Somers}\ and\ \citenamefont
  {Rem}(1991)}]{Somers199139}%
  \BibitemOpen
  \bibfield  {author} {\bibinfo {author} {\bibfnamefont {J.}~\bibnamefont
  {Somers}}\ and\ \bibinfo {author} {\bibfnamefont {P.}~\bibnamefont {Rem}},\
  }\href@noop {} {\bibfield  {journal} {\bibinfo  {journal} {Physica D}\
  }\textbf {\bibinfo {volume} {47}},\ \bibinfo {pages} {39 } (\bibinfo {year}
  {1991})}\BibitemShut {NoStop}%
\bibitem [{lut()}]{lutsko}%
  \BibitemOpen
  \href@noop {} {}\bibinfo {note} {J. Lutsko, J. Boon, and J. Somers,
  \textit{Numerical Methods for the Simulation of Multi-Phase and Complex
  Flows}, Lecture Notes in Physics \textbf{398}, 124 (Springer, Berlin,
  1992)}\BibitemShut {NoStop}%
\bibitem [{\citenamefont {M\aa{}l\o{}y}\ \emph {et~al.}(1985)\citenamefont
  {M\aa{}l\o{}y}, \citenamefont {Feder},\ and\ \citenamefont
  {J\o{}ssang}}]{PhysRevLett.55.2688}%
  \BibitemOpen
  \bibfield  {author} {\bibinfo {author} {\bibfnamefont {K.~J.}\ \bibnamefont
  {M\aa{}l\o{}y}}, \bibinfo {author} {\bibfnamefont {J.}~\bibnamefont {Feder}},
  \ and\ \bibinfo {author} {\bibfnamefont {T.}~\bibnamefont {J\o{}ssang}},\
  }\href {\doibase 10.1103/PhysRevLett.55.2688} {\bibfield  {journal} {\bibinfo
   {journal} {Phys. Rev. Lett.}\ }\textbf {\bibinfo {volume} {55}},\ \bibinfo
  {pages} {2688} (\bibinfo {year} {1985})}\BibitemShut {NoStop}%
\bibitem [{\citenamefont {M{\aa}l{\o}y}\ \emph {et~al.}(1987)\citenamefont
  {M{\aa}l{\o}y}, \citenamefont {Boger}, \citenamefont {Feder}, \citenamefont
  {J\o{}ssang},\ and\ \citenamefont {Meakin}}]{PhysRevA.36.318}%
  \BibitemOpen
  \bibfield  {author} {\bibinfo {author} {\bibfnamefont {K.~J.}\ \bibnamefont
  {M{\aa}l{\o}y}}, \bibinfo {author} {\bibfnamefont {F.}~\bibnamefont {Boger}},
  \bibinfo {author} {\bibfnamefont {J.}~\bibnamefont {Feder}}, \bibinfo
  {author} {\bibfnamefont {T.}~\bibnamefont {J\o{}ssang}}, \ and\ \bibinfo
  {author} {\bibfnamefont {P.}~\bibnamefont {Meakin}},\ }\href {\doibase
  10.1103/PhysRevA.36.318} {\bibfield  {journal} {\bibinfo  {journal} {Phys.
  Rev. A}\ }\textbf {\bibinfo {volume} {36}},\ \bibinfo {pages} {318} (\bibinfo
  {year} {1987})}\BibitemShut {NoStop}%
\bibitem [{\citenamefont {Paterson}(1984)}]{PhysRevLett.52.1621}%
  \BibitemOpen
  \bibfield  {author} {\bibinfo {author} {\bibfnamefont {L.}~\bibnamefont
  {Paterson}},\ }\href {\doibase 10.1103/PhysRevLett.52.1621} {\bibfield
  {journal} {\bibinfo  {journal} {Phys. Rev. Lett.}\ }\textbf {\bibinfo
  {volume} {52}},\ \bibinfo {pages} {1621} (\bibinfo {year}
  {1984})}\BibitemShut {NoStop}%
\bibitem [{\citenamefont {Chen}\ and\ \citenamefont
  {Wilkinson}(1985)}]{PhysRevLett.55.1892}%
  \BibitemOpen
  \bibfield  {author} {\bibinfo {author} {\bibfnamefont {J.-D.}\ \bibnamefont
  {Chen}}\ and\ \bibinfo {author} {\bibfnamefont {D.}~\bibnamefont
  {Wilkinson}},\ }\href {\doibase 10.1103/PhysRevLett.55.1892} {\bibfield
  {journal} {\bibinfo  {journal} {Phys. Rev. Lett.}\ }\textbf {\bibinfo
  {volume} {55}},\ \bibinfo {pages} {1892} (\bibinfo {year}
  {1985})}\BibitemShut {NoStop}%
\bibitem [{\citenamefont {Witten}\ and\ \citenamefont
  {Sander}(1981)}]{witten1}%
  \BibitemOpen
  \bibfield  {author} {\bibinfo {author} {\bibfnamefont {T.~A.}\ \bibnamefont
  {Witten}}\ and\ \bibinfo {author} {\bibfnamefont {L.~M.}\ \bibnamefont
  {Sander}},\ }\href {\doibase 10.1103/PhysRevLett.47.1400} {\bibfield
  {journal} {\bibinfo  {journal} {Phys. Rev. Lett.}\ }\textbf {\bibinfo
  {volume} {47}},\ \bibinfo {pages} {1400} (\bibinfo {year}
  {1981})}\BibitemShut {NoStop}%
\bibitem [{\citenamefont {Witten}\ and\ \citenamefont
  {Sander}(1983)}]{witten2}%
  \BibitemOpen
  \bibfield  {author} {\bibinfo {author} {\bibfnamefont {T.~A.}\ \bibnamefont
  {Witten}}\ and\ \bibinfo {author} {\bibfnamefont {L.~M.}\ \bibnamefont
  {Sander}},\ }\href {\doibase 10.1103/PhysRevB.27.5686} {\bibfield  {journal}
  {\bibinfo  {journal} {Phys. Rev. B}\ }\textbf {\bibinfo {volume} {27}},\
  \bibinfo {pages} {5686} (\bibinfo {year} {1983})}\BibitemShut {NoStop}%
\bibitem [{\citenamefont {Herrmann}\ and\ \citenamefont
  {Stanley}(1984)}]{PhysRevLett.53.1121}%
  \BibitemOpen
  \bibfield  {author} {\bibinfo {author} {\bibfnamefont {H.~J.}\ \bibnamefont
  {Herrmann}}\ and\ \bibinfo {author} {\bibfnamefont {H.~E.}\ \bibnamefont
  {Stanley}},\ }\href {\doibase 10.1103/PhysRevLett.53.1121} {\bibfield
  {journal} {\bibinfo  {journal} {Phys. Rev. Lett.}\ }\textbf {\bibinfo
  {volume} {53}},\ \bibinfo {pages} {1121} (\bibinfo {year}
  {1984})}\BibitemShut {NoStop}%
\bibitem [{\citenamefont {Laidlaw}\ \emph {et~al.}(1987)\citenamefont
  {Laidlaw}, \citenamefont {MacKay},\ and\ \citenamefont {Jan}}]{law}%
  \BibitemOpen
  \bibfield  {author} {\bibinfo {author} {\bibfnamefont {D.}~\bibnamefont
  {Laidlaw}}, \bibinfo {author} {\bibfnamefont {G.}~\bibnamefont {MacKay}}, \
  and\ \bibinfo {author} {\bibfnamefont {N.}~\bibnamefont {Jan}},\ }\href
  {\doibase 10.1007/BF01013371} {\bibfield  {journal} {\bibinfo  {journal} {J.
  Stat. Phys.}\ }\textbf {\bibinfo {volume} {46}},\ \bibinfo {pages} {507}
  (\bibinfo {year} {1987})}\BibitemShut {NoStop}%
\bibitem [{\citenamefont {Murat}\ and\ \citenamefont
  {Aharony}(1986)}]{PhysRevLett.57.1875}%
  \BibitemOpen
  \bibfield  {author} {\bibinfo {author} {\bibfnamefont {M.}~\bibnamefont
  {Murat}}\ and\ \bibinfo {author} {\bibfnamefont {A.}~\bibnamefont
  {Aharony}},\ }\href {\doibase 10.1103/PhysRevLett.57.1875} {\bibfield
  {journal} {\bibinfo  {journal} {Phys. Rev. Lett.}\ }\textbf {\bibinfo
  {volume} {57}},\ \bibinfo {pages} {1875} (\bibinfo {year}
  {1986})}\BibitemShut {NoStop}%
\bibitem [{\citenamefont {Wilkinson}\ and\ \citenamefont
  {Willemsen}(1983)}]{wilkinson}%
  \BibitemOpen
  \bibfield  {author} {\bibinfo {author} {\bibfnamefont {D.}~\bibnamefont
  {Wilkinson}}\ and\ \bibinfo {author} {\bibfnamefont {J.~F.}\ \bibnamefont
  {Willemsen}},\ }\href {http://stacks.iop.org/0305-4470/16/i=14/a=028}
  {\bibfield  {journal} {\bibinfo  {journal} {J. Phys. A}\ }\textbf {\bibinfo
  {volume} {16}},\ \bibinfo {pages} {3365} (\bibinfo {year}
  {1983})}\BibitemShut {NoStop}%
\bibitem [{\citenamefont {Lenormand}\ and\ \citenamefont
  {Zarcone}(1985)}]{PhysRevLett.54.2226}%
  \BibitemOpen
  \bibfield  {author} {\bibinfo {author} {\bibfnamefont {R.}~\bibnamefont
  {Lenormand}}\ and\ \bibinfo {author} {\bibfnamefont {C.}~\bibnamefont
  {Zarcone}},\ }\href {\doibase 10.1103/PhysRevLett.54.2226} {\bibfield
  {journal} {\bibinfo  {journal} {Phys. Rev. Lett.}\ }\textbf {\bibinfo
  {volume} {54}},\ \bibinfo {pages} {2226} (\bibinfo {year}
  {1985})}\BibitemShut {NoStop}%
\bibitem [{\citenamefont {Stauffer}(1985)}]{stauffer}%
  \BibitemOpen
  \bibfield  {author} {\bibinfo {author} {\bibfnamefont {D.}~\bibnamefont
  {Stauffer}},\ }\href@noop {} {\emph {\bibinfo {title} {Introduction to
  Percolation Theory}}}\ (\bibinfo  {publisher} {Taylor \& Francis, London},\
  \bibinfo {year} {1985})\BibitemShut {NoStop}%
\bibitem [{\citenamefont {Dias}\ and\ \citenamefont {Wilkinson}(1986)}]{dias}%
  \BibitemOpen
  \bibfield  {author} {\bibinfo {author} {\bibfnamefont {M.~M.}\ \bibnamefont
  {Dias}}\ and\ \bibinfo {author} {\bibfnamefont {D.}~\bibnamefont
  {Wilkinson}},\ }\href {http://stacks.iop.org/0305-4470/19/i=15/a=034}
  {\bibfield  {journal} {\bibinfo  {journal} {J. Phys. A}\ }\textbf {\bibinfo
  {volume} {19}},\ \bibinfo {pages} {3131} (\bibinfo {year}
  {1986})}\BibitemShut {NoStop}%
\bibitem [{\citenamefont {Ara\'ujo}\ \emph {et~al.}(2005)\citenamefont
  {Ara\'ujo}, \citenamefont {Vasconcelos}, \citenamefont {Moreira},
  \citenamefont {Lucena},\ and\ \citenamefont {Andrade}}]{liacir}%
  \BibitemOpen
  \bibfield  {author} {\bibinfo {author} {\bibfnamefont {A.~D.}\ \bibnamefont
  {Ara\'ujo}}, \bibinfo {author} {\bibfnamefont {T.~F.}\ \bibnamefont
  {Vasconcelos}}, \bibinfo {author} {\bibfnamefont {A.~A.}\ \bibnamefont
  {Moreira}}, \bibinfo {author} {\bibfnamefont {L.~S.}\ \bibnamefont {Lucena}},
  \ and\ \bibinfo {author} {\bibfnamefont {J.~S.}\ \bibnamefont {Andrade}},\
  }\href {\doibase 10.1103/PhysRevE.72.041404} {\bibfield  {journal} {\bibinfo
  {journal} {Phys. Rev. E}\ }\textbf {\bibinfo {volume} {72}},\ \bibinfo
  {pages} {041404} (\bibinfo {year} {2005})}\BibitemShut {NoStop}%
\bibitem [{sap()}]{sapoval}%
  \BibitemOpen
  \href@noop {} {}\bibinfo {note} {B. Sapoval, M. Rosso, and J. F. Gouyet, J.
  Phys., Lett. \textbf{46}, L149 (1985)}\BibitemShut {NoStop}%
\bibitem [{\citenamefont {Rosso}\ \emph {et~al.}(1985)\citenamefont {Rosso},
  \citenamefont {Gouyet},\ and\ \citenamefont {Sapoval}}]{PhysRevB.32.6053}%
  \BibitemOpen
  \bibfield  {author} {\bibinfo {author} {\bibfnamefont {M.}~\bibnamefont
  {Rosso}}, \bibinfo {author} {\bibfnamefont {J.~F.}\ \bibnamefont {Gouyet}}, \
  and\ \bibinfo {author} {\bibfnamefont {B.}~\bibnamefont {Sapoval}},\ }\href
  {\doibase 10.1103/PhysRevB.32.6053} {\bibfield  {journal} {\bibinfo
  {journal} {Phys. Rev. B}\ }\textbf {\bibinfo {volume} {32}},\ \bibinfo
  {pages} {6053} (\bibinfo {year} {1985})}\BibitemShut {NoStop}%
\bibitem [{\citenamefont {Rosso}\ \emph {et~al.}(1986)\citenamefont {Rosso},
  \citenamefont {Gouyet},\ and\ \citenamefont {Sapoval}}]{PhysRevLett.57.3195}%
  \BibitemOpen
  \bibfield  {author} {\bibinfo {author} {\bibfnamefont {M.}~\bibnamefont
  {Rosso}}, \bibinfo {author} {\bibfnamefont {J.~F.}\ \bibnamefont {Gouyet}}, \
  and\ \bibinfo {author} {\bibfnamefont {B.}~\bibnamefont {Sapoval}},\ }\href
  {\doibase 10.1103/PhysRevLett.57.3195} {\bibfield  {journal} {\bibinfo
  {journal} {Phys. Rev. Lett.}\ }\textbf {\bibinfo {volume} {57}},\ \bibinfo
  {pages} {3195} (\bibinfo {year} {1986})}\BibitemShut {NoStop}%
\bibitem [{sto()}]{stokes}%
  \BibitemOpen
  \href@noop {} {}\bibinfo {note} {J. P. Stokes, D. A. Weitz, J. P. Gollub, A.
  Dougherty, M. O. Robbins, P. M. Chaikin, and H. M. Lindsay, Phys. Rev. Lett.
  \textbf{57}, 1718 (1986); see also: D. A. Weitz, J. P. Stokes, R. C. Ball,
  and A. P. Kushnick, Phys. Rev. Lett. \textbf{59}, 2967 (1987)}\BibitemShut
  {NoStop}%
\bibitem [{\citenamefont {Cieplak}\ and\ \citenamefont
  {Robbins}(1988)}]{PhysRevLett.60.2042}%
  \BibitemOpen
  \bibfield  {author} {\bibinfo {author} {\bibfnamefont {M.}~\bibnamefont
  {Cieplak}}\ and\ \bibinfo {author} {\bibfnamefont {M.~O.}\ \bibnamefont
  {Robbins}},\ }\href {\doibase 10.1103/PhysRevLett.60.2042} {\bibfield
  {journal} {\bibinfo  {journal} {Phys. Rev. Lett.}\ }\textbf {\bibinfo
  {volume} {60}},\ \bibinfo {pages} {2042} (\bibinfo {year}
  {1988})}\BibitemShut {NoStop}%
\bibitem [{man()}]{mandelbrot}%
  \BibitemOpen
  \href@noop {} {}\bibinfo {note} {See, e. g., B. B. Mandelbrot, Physica
  Scripta \textbf{32}, 257 (1985); K. J. Falconer, Math. Proc. Cambridge
  Philos. Soc. \textbf{103}, 339 (1988); \textbf{111}, 164 (1992); A.-L.
  Barab\'asi and T. Vicsek, Phys. Rev. A \textbf{44}, 2730 (1991)}\BibitemShut
  {NoStop}%
\bibitem [{rub()}]{rubio}%
  \BibitemOpen
  \href@noop {} {}\bibinfo {note} {M. A. Rubio, C. A. Edwards, A. Dougherty,
  and J. P. Gollub, Phys. Rev. Lett. \textbf{63}, 1685 (1989); V. K. Horv\'ath,
  F. Family, and T. Vicsek, Phys. Rev. Lett. \textbf{65}, 1388 (1990); M. A.
  Rubio, A. Dougherty, and J. P. Gollub, Phys. Rev. Lett. \textbf{65}, 1389
  (1990)}\BibitemShut {NoStop}%
\bibitem [{\citenamefont {He}\ \emph {et~al.}(1992)\citenamefont {He},
  \citenamefont {Kahanda},\ and\ \citenamefont {Wong}}]{PhysRevLett.69}%
  \BibitemOpen
  \bibfield  {author} {\bibinfo {author} {\bibfnamefont {S.}~\bibnamefont
  {He}}, \bibinfo {author} {\bibfnamefont {G.~L. M. K.~S.}\ \bibnamefont
  {Kahanda}}, \ and\ \bibinfo {author} {\bibfnamefont {P.-z.}\ \bibnamefont
  {Wong}},\ }\href {\doibase 10.1103/PhysRevLett.69.3731} {\bibfield  {journal}
  {\bibinfo  {journal} {Phys. Rev. Lett.}\ }\textbf {\bibinfo {volume} {69}},\
  \bibinfo {pages} {3731} (\bibinfo {year} {1992})}\BibitemShut {NoStop}%
\bibitem [{\citenamefont {Nolle}\ \emph {et~al.}(1993)\citenamefont {Nolle},
  \citenamefont {Koiller}, \citenamefont {Martys},\ and\ \citenamefont
  {Robbins}}]{PhysRevLett.71.2074}%
  \BibitemOpen
  \bibfield  {author} {\bibinfo {author} {\bibfnamefont {C.~S.}\ \bibnamefont
  {Nolle}}, \bibinfo {author} {\bibfnamefont {B.}~\bibnamefont {Koiller}},
  \bibinfo {author} {\bibfnamefont {N.}~\bibnamefont {Martys}}, \ and\ \bibinfo
  {author} {\bibfnamefont {M.~O.}\ \bibnamefont {Robbins}},\ }\href {\doibase
  10.1103/PhysRevLett.71.2074} {\bibfield  {journal} {\bibinfo  {journal}
  {Phys. Rev. Lett.}\ }\textbf {\bibinfo {volume} {71}},\ \bibinfo {pages}
  {2074} (\bibinfo {year} {1993})}\BibitemShut {NoStop}%
\bibitem [{\citenamefont {Horvath}\ \emph {et~al.}(1991)\citenamefont
  {Horvath}, \citenamefont {Family},\ and\ \citenamefont
  {Vicsek}}]{0305-4470241}%
  \BibitemOpen
  \bibfield  {author} {\bibinfo {author} {\bibfnamefont {V.~K.}\ \bibnamefont
  {Horvath}}, \bibinfo {author} {\bibfnamefont {F.}~\bibnamefont {Family}}, \
  and\ \bibinfo {author} {\bibfnamefont {T.}~\bibnamefont {Vicsek}},\
  }\href@noop {} {\bibfield  {journal} {\bibinfo  {journal} {J. Phys. A}\
  }\textbf {\bibinfo {volume} {24}},\ \bibinfo {pages} {L25} (\bibinfo {year}
  {1991})}\BibitemShut {NoStop}%
\bibitem [{aur()}]{aurora}%
  \BibitemOpen
  \href@noop {} {}\bibinfo {note} {We mention a series of studies of forced
  fluid invasion of Hele-Shaw cells under quenched disorder filled with a less
  viscous fluid: A. Hern\'andez-Machado, J. Soriano, A. M. Lacasta, M. A.
  Rodr\'{\i}guez, L. Ram\'{\i}rez-Piscina, and J. Ort\'{\i}n, Europhys. Lett.
  \textbf{55}, 194 (2001); J. Soriano, J. J. Ramasco, M. A. Rodr\'{\i}guez, A.
  Hern\'andez-Machado, and J. Ort\'{\i}n, Phys. Rev. Lett. \textbf{89}, 026102
  (2002); J. Soriano, A. Mercier, R. Planet, A. Hern\'andez-Machado, M. A.
  Rodr\'{\i}guez, and J. Ort\'{\i}n, Phys. Rev. Lett. \textbf{95}, 104501
  (2005). The authors analyze the experimental data for the non-equilibrium
  surface interfacial roughening using an anomolous dynamic
  scaling}\BibitemShut {NoStop}%
\bibitem [{\citenamefont {Martys}(1994)}]{PhysRevE.50.335}%
  \BibitemOpen
  \bibfield  {author} {\bibinfo {author} {\bibfnamefont {N.~S.}\ \bibnamefont
  {Martys}},\ }\href {\doibase 10.1103/PhysRevE.50.335} {\bibfield  {journal}
  {\bibinfo  {journal} {Phys. Rev. E}\ }\textbf {\bibinfo {volume} {50}},\
  \bibinfo {pages} {335} (\bibinfo {year} {1994})}\BibitemShut {NoStop}%
\bibitem [{len()}]{lenormand}%
  \BibitemOpen
  \href@noop {} {}\bibinfo {note} {R. Lenormand, E. Touboul and C. Zarcone, J.
  Fluid. Mech. \textbf{189}, 165 (1988)}\BibitemShut {NoStop}%
\bibitem [{\citenamefont {Frette}\ \emph {et~al.}(1997)\citenamefont {Frette},
  \citenamefont {M{\aa}l{\o}y}, \citenamefont {Schmittbuhl},\ and\
  \citenamefont {Hansen}}]{PhysRevE.55.2969}%
  \BibitemOpen
  \bibfield  {author} {\bibinfo {author} {\bibfnamefont {O.~I.}\ \bibnamefont
  {Frette}}, \bibinfo {author} {\bibfnamefont {K.~J.}\ \bibnamefont
  {M{\aa}l{\o}y}}, \bibinfo {author} {\bibfnamefont {J.}~\bibnamefont
  {Schmittbuhl}}, \ and\ \bibinfo {author} {\bibfnamefont {A.}~\bibnamefont
  {Hansen}},\ }\href {\doibase 10.1103/PhysRevE.55.2969} {\bibfield  {journal}
  {\bibinfo  {journal} {Phys. Rev. E}\ }\textbf {\bibinfo {volume} {55}},\
  \bibinfo {pages} {2969} (\bibinfo {year} {1997})}\BibitemShut {NoStop}%
\bibitem [{\citenamefont {M{\'{e}}heust}\ \emph {et~al.}(2002)\citenamefont
  {M{\'{e}}heust}, \citenamefont {L{\o}voll}, \citenamefont {M{\aa}l{\o}y},\
  and\ \citenamefont {Schmittbuhl}}]{PhysRevE.66.05}%
  \BibitemOpen
  \bibfield  {author} {\bibinfo {author} {\bibfnamefont {Y.}~\bibnamefont
  {M{\'{e}}heust}}, \bibinfo {author} {\bibfnamefont {G.}~\bibnamefont
  {L{\o}voll}}, \bibinfo {author} {\bibfnamefont {K.~J.}\ \bibnamefont
  {M{\aa}l{\o}y}}, \ and\ \bibinfo {author} {\bibfnamefont {J.}~\bibnamefont
  {Schmittbuhl}},\ }\href {\doibase 10.1103/PhysRevE.66.051603} {\bibfield
  {journal} {\bibinfo  {journal} {Phys. Rev. E}\ }\textbf {\bibinfo {volume}
  {66}},\ \bibinfo {pages} {051603} (\bibinfo {year} {2002})}\BibitemShut
  {NoStop}%
\bibitem [{\citenamefont {Tallakstad}\ \emph
  {et~al.}(2009{\natexlab{a}})\citenamefont {Tallakstad}, \citenamefont
  {Knudsen}, \citenamefont {Ramstad}, \citenamefont {L\o{}voll}, \citenamefont
  {M\aa{}l\o{}y}, \citenamefont {Toussaint},\ and\ \citenamefont
  {Flekk\o{}y}}]{PhysRevLett.102.074502}%
  \BibitemOpen
  \bibfield  {author} {\bibinfo {author} {\bibfnamefont {K.~T.}\ \bibnamefont
  {Tallakstad}}, \bibinfo {author} {\bibfnamefont {H.~A.}\ \bibnamefont
  {Knudsen}}, \bibinfo {author} {\bibfnamefont {T.}~\bibnamefont {Ramstad}},
  \bibinfo {author} {\bibfnamefont {G.}~\bibnamefont {L\o{}voll}}, \bibinfo
  {author} {\bibfnamefont {K.~J.}\ \bibnamefont {M\aa{}l\o{}y}}, \bibinfo
  {author} {\bibfnamefont {R.}~\bibnamefont {Toussaint}}, \ and\ \bibinfo
  {author} {\bibfnamefont {E.~G.}\ \bibnamefont {Flekk\o{}y}},\ }\href
  {\doibase 10.1103/PhysRevLett.102.074502} {\bibfield  {journal} {\bibinfo
  {journal} {Phys. Rev. Lett.}\ }\textbf {\bibinfo {volume} {102}},\ \bibinfo
  {pages} {074502} (\bibinfo {year} {2009}{\natexlab{a}})}\BibitemShut
  {NoStop}%
\bibitem [{\citenamefont {Tallakstad}\ \emph
  {et~al.}(2009{\natexlab{b}})\citenamefont {Tallakstad}, \citenamefont
  {L\o{}voll}, \citenamefont {Knudsen}, \citenamefont {Ramstad}, \citenamefont
  {Flekk\o{}y},\ and\ \citenamefont {M\aa{}l\o{}y}}]{PhysRevE.80.036308}%
  \BibitemOpen
  \bibfield  {author} {\bibinfo {author} {\bibfnamefont {K.~T.}\ \bibnamefont
  {Tallakstad}}, \bibinfo {author} {\bibfnamefont {G.}~\bibnamefont
  {L\o{}voll}}, \bibinfo {author} {\bibfnamefont {H.~A.}\ \bibnamefont
  {Knudsen}}, \bibinfo {author} {\bibfnamefont {T.}~\bibnamefont {Ramstad}},
  \bibinfo {author} {\bibfnamefont {E.~G.}\ \bibnamefont {Flekk\o{}y}}, \ and\
  \bibinfo {author} {\bibfnamefont {K.~J.}\ \bibnamefont {M\aa{}l\o{}y}},\
  }\href {\doibase 10.1103/PhysRevE.80.036308} {\bibfield  {journal} {\bibinfo
  {journal} {Phys. Rev. E}\ }\textbf {\bibinfo {volume} {80}},\ \bibinfo
  {pages} {036308} (\bibinfo {year} {2009}{\natexlab{b}})}\BibitemShut
  {NoStop}%
\bibitem [{\citenamefont {Toussaint}\ \emph {et~al.}(2012)\citenamefont
  {Toussaint}, \citenamefont {M{\aa}l{\o}y}, \citenamefont {Meheust},
  \citenamefont {L{\o}voll}, \citenamefont {Jankov}, \citenamefont {Schaefer},\
  and\ \citenamefont {Schmittbuhl}}]{vadose}%
  \BibitemOpen
  \bibfield  {author} {\bibinfo {author} {\bibfnamefont {R.}~\bibnamefont
  {Toussaint}}, \bibinfo {author} {\bibfnamefont {K.~J.}\ \bibnamefont
  {M{\aa}l{\o}y}}, \bibinfo {author} {\bibfnamefont {Y.}~\bibnamefont
  {Meheust}}, \bibinfo {author} {\bibfnamefont {G.}~\bibnamefont {L{\o}voll}},
  \bibinfo {author} {\bibfnamefont {M.}~\bibnamefont {Jankov}}, \bibinfo
  {author} {\bibfnamefont {G.}~\bibnamefont {Schaefer}}, \ and\ \bibinfo
  {author} {\bibfnamefont {J.}~\bibnamefont {Schmittbuhl}},\ }\href {\doibase
  {10.2136/vzj2011.0123}} {\bibfield  {journal} {\bibinfo  {journal} {{Vadose
  Zone J.}}\ }\textbf {\bibinfo {volume} {{11}}} (\bibinfo {year} {{2012}}),\
  {10.2136/vzj2011.0123}},\ \bibinfo {note} {and references
  therein}\BibitemShut {NoStop}%
\bibitem [{\citenamefont {Dong}\ \emph {et~al.}(2011)\citenamefont {Dong},
  \citenamefont {Yan},\ and\ \citenamefont {Li}}]{dong2011}%
  \BibitemOpen
  \bibfield  {author} {\bibinfo {author} {\bibfnamefont {B.}~\bibnamefont
  {Dong}}, \bibinfo {author} {\bibfnamefont {Y.}~\bibnamefont {Yan}}, \ and\
  \bibinfo {author} {\bibfnamefont {W.}~\bibnamefont {Li}},\ }\href {\doibase
  10.1007/s11242-011-9740-y} {\bibfield  {journal} {\bibinfo  {journal}
  {Transp. Porous Media}\ }\textbf {\bibinfo {volume} {88}},\ \bibinfo {pages}
  {293} (\bibinfo {year} {2011})}\BibitemShut {NoStop}%
\bibitem [{\citenamefont {Birovljev}\ \emph {et~al.}(1991)\citenamefont
  {Birovljev}, \citenamefont {Furuberg}, \citenamefont {Feder}, \citenamefont
  {J\o{}ssang}, \citenamefont {M\aa{}l\o{}y},\ and\ \citenamefont
  {Aharony}}]{PhysRevLett.67.584}%
  \BibitemOpen
  \bibfield  {author} {\bibinfo {author} {\bibfnamefont {A.}~\bibnamefont
  {Birovljev}}, \bibinfo {author} {\bibfnamefont {L.}~\bibnamefont {Furuberg}},
  \bibinfo {author} {\bibfnamefont {J.}~\bibnamefont {Feder}}, \bibinfo
  {author} {\bibfnamefont {T.}~\bibnamefont {J\o{}ssang}}, \bibinfo {author}
  {\bibfnamefont {K.~J.}\ \bibnamefont {M\aa{}l\o{}y}}, \ and\ \bibinfo
  {author} {\bibfnamefont {A.}~\bibnamefont {Aharony}},\ }\href {\doibase
  10.1103/PhysRevLett.67.584} {\bibfield  {journal} {\bibinfo  {journal} {Phys.
  Rev. Lett.}\ }\textbf {\bibinfo {volume} {67}},\ \bibinfo {pages} {584}
  (\bibinfo {year} {1991})}\BibitemShut {NoStop}%
\bibitem [{Note1()}]{Note1}%
  \BibitemOpen
  \bibinfo {note} {The codes we use were built from \protect \url
  {http://www.lmm.jussieu.fr/\~zaleski/latgas.html}; see Ref. \cite
  {Rothman_book}.}\BibitemShut {Stop}%
\bibitem [{\citenamefont {Family}(1986)}]{RDBwS}%
  \BibitemOpen
  \bibfield  {author} {\bibinfo {author} {\bibfnamefont {F.}~\bibnamefont
  {Family}},\ }\href@noop {} {\bibfield  {journal} {\bibinfo  {journal} {J.
  Phys. A}\ }\textbf {\bibinfo {volume} {19}},\ \bibinfo {pages} {L441}
  (\bibinfo {year} {1986})}\BibitemShut {NoStop}%
\bibitem [{\citenamefont {Edwards}\ and\ \citenamefont
  {Wilkinson}(1982)}]{Edwards08051982}%
  \BibitemOpen
  \bibfield  {author} {\bibinfo {author} {\bibfnamefont {S.~F.}\ \bibnamefont
  {Edwards}}\ and\ \bibinfo {author} {\bibfnamefont {D.~R.}\ \bibnamefont
  {Wilkinson}},\ }\href {\doibase 10.1098/rspa.1982.0056} {\bibfield  {journal}
  {\bibinfo  {journal} {Proc. R. Soc. London, Sect. A}\ }\textbf {\bibinfo
  {volume} {381}},\ \bibinfo {pages} {17} (\bibinfo {year} {1982})}\BibitemShut
  {NoStop}%
\end{thebibliography}%

\end{document}